\documentclass[12pt]{article}

\usepackage[utf8]{inputenc}
\usepackage{epstopdf}
\usepackage{bm}
\usepackage{amsmath}
\usepackage[font=small,labelfont=bf]{caption}
\usepackage{amssymb}
\usepackage{pifont}
\usepackage{xtab}
\usepackage{mathrsfs}
\usepackage{subeqnarray}
\usepackage{cases}
\usepackage{booktabs}
\usepackage{multirow}
\usepackage{mathtools}
\usepackage{enumitem}
\usepackage{algpseudocode,algorithm}
\usepackage{algcompatible}
\usepackage{array}
\usepackage{subcaption}
\usepackage[overload]{empheq}
\usepackage{hyperref}
\usepackage[export]{adjustbox}
\usepackage{float}
\usepackage{makecell}
\usepackage[percent]{overpic}
\usepackage[flushleft]{threeparttable}
\usepackage[square,numbers]{natbib}
\usepackage{lineno}
\usepackage{titletoc,tocloft}

\hypersetup{colorlinks,linkcolor={black},citecolor={black},urlcolor={blue}}

\algblockdefx{FORALLP}{ENDFAP}[1]%
  {\textbf{for all }#1 \textbf{do in parallel}}%
  {\textbf{end for}}

\newcommand{\blue}[1]{{\color{blue}{#1}}}

\newtheorem{condition}{\textbf{Condition}}
\newtheorem{theorem}{\textbf{Theorem}}

\newtheorem{lemma}{\textbf{Lemma}}

\newcommand{\reviseOne}[1]{{\textcolor{black}{#1}}}
\newcommand{\reviseTwo}[1]{{\textcolor{black}{#1}}}
\newcommand{\reviseThree}[1]{{\textcolor{black}{#1}}}

\newcommand{\reviseFive}[1]{{\textcolor{black}{#1}}}

\newcommand{\reviseSix}[1]{{\textcolor{black}{#1}}}

\newcommand{\reviseTen}[1]{#1}

\newcommand{\norm}[1]{\left\lVert#1\right\rVert}
\newcommand{\abs}[1]{\lvert#1\rvert}
\allowdisplaybreaks

\usepackage{times}

\newenvironment{sciabstract}{%
\begin{quote} \bf}
{\end{quote}}

\renewcommand\refname{References}

\makeatletter
\renewcommand*{\@biblabel}[1]{\hfill#1.}
\makeatother

\newcommand{\mytitle}{{Prevalence and scalable control of localized networks}}
\title{\mytitle}

\author{
\hspace{-6mm}Chao Duan,$^{1,2}$ Takashi Nishikawa,$^{2,3}$  Adilson E. Motter$^{2,3}$\\
\\
\hspace{-6mm}\normalsize{$^{1}$School of Electrical Engineering, Xi'an Jiaotong University, Xi'an 710049, China}\\
\hspace{-6mm}\normalsize{$^{2}$Department of Physics and Astronomy, Northwestern University, Evanston, IL 60208, USA}\\
\hspace{-6mm}\normalsize{$^{3}$Northwestern Institute on Complex Systems, Northwestern University, Evanston, IL 60208, USA}\\
}

\date{}

\begin{document} 
\nolinenumbers


\maketitle 

\baselineskip17pt



\vspace{-0.6cm}
\begin{sciabstract}
The ability to control network dynamics is essential for ensuring desirable functionality of many technological, biological, and social systems. Such systems often consist of a large number of network elements, and controlling large-scale networks remains challenging because the computation and communication requirements increase prohibitively fast with network size.
\reviseTen{Here, we introduce a notion of {\em network locality} ~that can be exploited to make the control of networks scalable even when the dynamics are nonlinear}. We show that network locality is captured by an information metric and is almost universally observed across real and model networks. In localized networks, the optimal control actions and system responses are both shown to be necessarily concentrated in small neighborhoods induced by the information metric. This allows us to develop localized algorithms for determining network controllability and optimizing 
the placement of driver nodes. This also allows us to develop a localized algorithm for designing local feedback controllers that approach the performance of the corresponding best global controllers while incurring a computational cost orders-of-magnitude lower. We validate the locality, performance, and efficiency of the algorithms in \reviseTen{Kuramoto oscillator networks as well as}
three large empirical networks: synchronization dynamics in the Eastern U.S. power grid, epidemic spreading mediated by the global air transportation network, and Alzheimer's disease dynamics in a human brain network. Taken together, our results establish that large networks can be controlled with computation and communication costs comparable to those for small networks.

DOI: \href{https://doi.org/10.1073/pnas.2122566119}{10.1073/pnas.2122566119} 
\end{sciabstract}

\baselineskip24pt

\clearpage

Many complex networks derive their functionalities from the dynamical processes they host \cite{newman2006structure,barrat2008dynamical,albert2002statistical}, such as synchronization of generators in power grids \cite{motter2013spontaneous,skardal2015control}, coordination dynamics in robotic networks \cite{bullo2009distributed}, production and distribution of goods in supply chain networks \cite{nagurney2006supply},
species interactions in biochemical \cite{feinberg2019foundations,wuchty2014controllability} and ecological \cite{lessard2005should,sahasrabudhe2011rescuing} networks, and exchange of assets and other transactions in financial networks \cite{galbiati2013power}. The control of the dynamics of such networks for desirable outcomes is a fundamental problem in network science \cite{cornelius2013realistic}. Crucially, the dynamics of large real networks are high dimensional. This calls 
for the integration of control theory and network science in order to solve both the analysis problem (whether a network is controllable) and the synthesis problem (how to control the network), so that network properties can be exploited to avoid computation and communication intractability \cite{liu2016control,motter2015networkcontrology,scholl2016control}. 
A promising line of research has been developed by focusing on structure-based approaches \cite{lin1974structural,liu2011controllability,zanudo2017structure,menara2018structural,montanari2021functional}, in which the nodes that need to be controlled are determined using network-topological information only.
\reviseOne{For instance, based on} the graph-theoretic characterizations of the Kalman \cite{kalman1960general} or Popov-Belevitch-Hautus \cite{hautus1970stabilization} rank conditions for controllability, efficient algorithms have been designed to identify the minimal set of driver nodes for a network to be controllable \reviseOne{\cite{liu2016control}}. This qualitative notion of controllability has proved to be insightful and broadly applicable. However, this concept is not designed to characterize the difficulty in actually carrying out the control actions nor to inform the design of control \reviseOne{laws}. This is important because the control energy \reviseTwo{needed to steer the system (i.e., the amount of physical, human, social, or economic resources required for control)} may increase exponentially as one reduces the fraction of nodes controlled, even when \reviseTwo{the system is controllable in principle} \cite{yan2012controlling,sun2013controllability,yan2015spectrum}. To enable network control in practice, numerous studies have shifted focus from qualitative 
to quantitative controllability \cite{yan2012controlling,sun2013controllability,pasqualetti2014controllability}, from controlling the entire network to controlling a target subset of nodes \cite{gao2014target,klickstein2017energy}, and from centralized to decentralized control designs \cite{li2018enabling,sanchez2021nonlinear}. 

In this Article, we develop a theory and an associated computational approach for controlling large complex dynamical networks by exploring the concept of \emph{locality} \reviseOne{(defined below)}, and we show that empirical networks are \reviseTen{most often localized}. Our study uncovers a dichotomy in controlling localized networks: \reviseOne{even} though a significant fraction of nodes need to be directly controlled to make the system \reviseOne{controllable in practice}, analysis and control is possible using only local computation and communication, while keeping the control performance near the optimal \reviseOne{achieved by} global control.

\begin{figure}[t]
    \centering
\begin{overpic}[width=0.66\textwidth,tics=10]{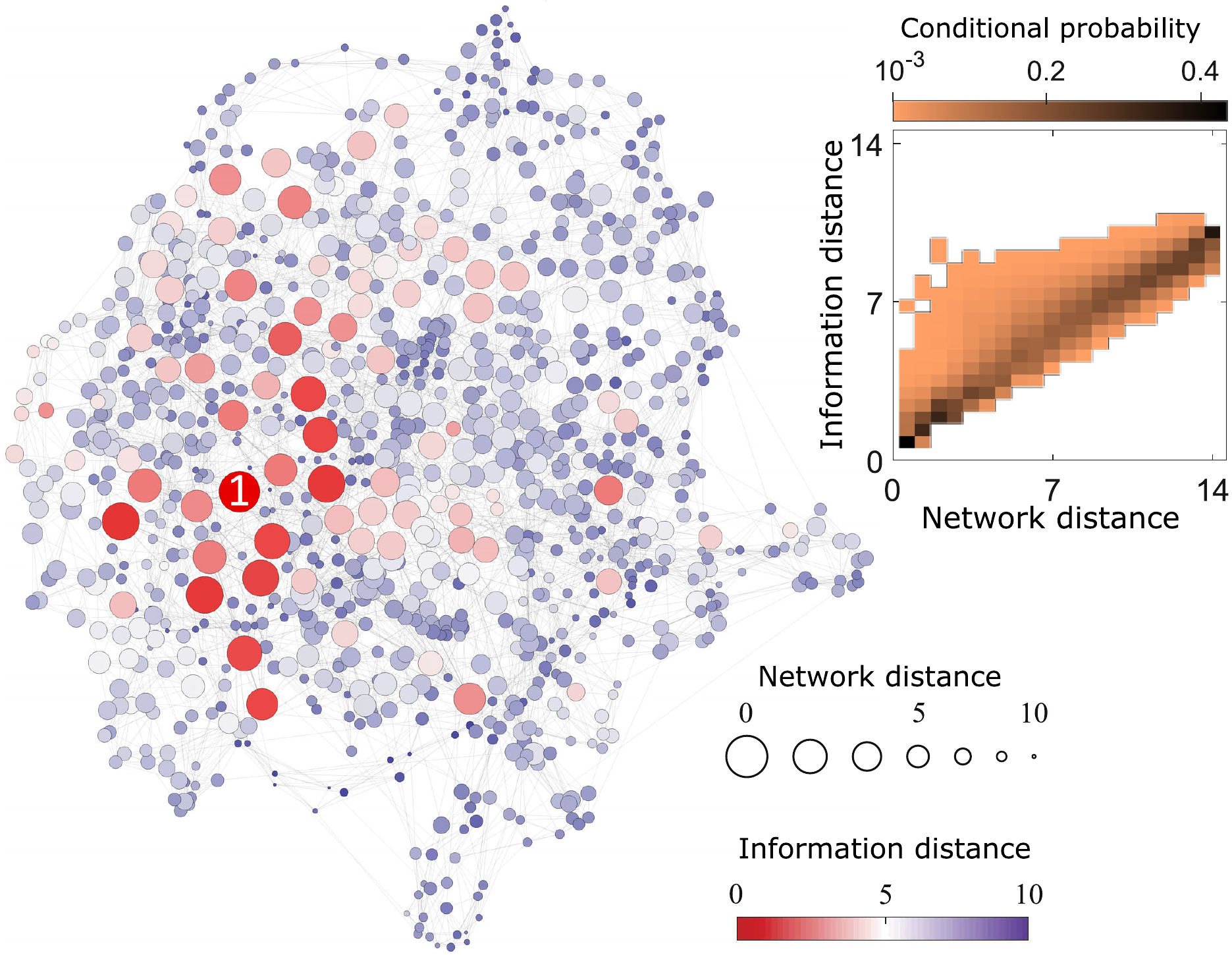}
\put (4,72) {\textbf{A}}
\put (64,72) {\textbf{B}}
\end{overpic}
        \caption{Information distance v.s.\ network distance on a weighted Watts-Strogatz (WS) network of $N=1000$ nodes with average degree $\bar{d}=20$ and rewiring probability $p=0.1$. \reviseFive{The network distance is the geodesic distance on the network with edge lengths defined as the reciprocal of coupling strengths.} (A) Information distances and network distances to a reference node (labeled ``1''), visualized on the network by node colors and sizes, respectively. (B) Information distances vs.\ network distances for each pair of nodes. The color indicates the conditional probability density for the information distance given the network distance.}
        \label{smallworldfig}
         
\end{figure}

Intuitively, a network is localized if each node is associated with a small group of other nodes and interacts significantly more strongly with the nodes \reviseOne{within} this group than outside \reviseOne{it}. This notion of locality can be seen as a generalization of sparsity, defined as the property \reviseOne{in which} each node 
is connected with
only a small subset of other nodes. Since locality additionally accounts for interaction strengths, a network can be localized even if all pairs of nodes are connected. In addition, for the concept of locality to be useful in network control, the locality property should be preserved in the dynamical responses of the network, which we formalize by introducing a metric space on the network. We characterize network locality by how the interaction strengths decay with a metric that we call the \emph{information distance}, such that each node in a network interacts strongly only with its so-called \emph{information neighborhood} (see Fig.~\ref{smallworldfig} for an example and the next section for precise definitions). We present an efficient algorithm to construct the information distance from the given network data, and we show that locality is observed in a broad class of real and model networks. We emphasize that network locality is different from the presence of a community structure \cite{girvan2002community}, since the information neighborhood of a node can be different from that of another node in that neighborhood, whereas a community is generally shared by all of its member nodes. For example, a ring network in which each node is connected to its two nearest neighbors, does not have a community structure and yet is localized.

To address the \emph{analysis problem} in network control, we prove that locality allows both the construction of the controllability Gramian and the approximation of its smallest eigenvalue (a measure of controllability, \reviseTen{see} \textcolor{blue}{\emph{SI Text 3}}) to be performed
using only local information and computation.
Based on this observation, we develop a highly scalable algorithm that can compute a near-optimal solution of the driver placement problem, in which the \reviseOne{smallest} eigenvalue of the Gramian is maximized.
Moreover, the locality of the Gramian implies that a driver can efficiently control only the nodes in its information neighborhood and that the energy needed to control a distant node becomes prohibitively large as the information distance increases. Incidentally, this provides a theoretical explanation for the observation in \reviseOne{\cite{sun2013controllability,yan2015spectrum}} that\reviseOne{, in the worst-case scenario, the} control energy increases exponentially when the number of driver nodes decreases.

To address the \emph{synthesis problem}, we show that 
network locality can enable stable and near-optimal control of large networks. This follows from showing that the (globally) optimal control actions and the corresponding system responses are both localized in the information neighborhoods of the disturbed nodes, implying that the optimal feedback matrix is also localized. Taking advantage of this, we develop a decentralized algorithm for calculating a sparse approximation of the optimal feedback law, in which the state measurements of each node are used only by the drivers in the information neighborhood of that node. 

\reviseTen{Our theory and methods are applicable to the control of nonlinear networks. This is achieved by allowing the local control laws to be time dependent and is demonstrated using four concrete examples:} 
control of synchronization in Kuramoto oscillator networks, stability control for the Eastern U.S.\ power-grid network, suppression of epidemic spreading \reviseOne{mediated by} the global air transportation network, and control of whole-brain network dynamics associated with a neurological disease.
\reviseOne{These} examples illustrate the methods' applicability to diverse domains\reviseOne{---infrastructural, epidemiological, and biomedical---}and to control tasks \reviseOne{ranging from} synchronization and stabilization \reviseOne{to} trajectory tracking and command following.
Thus, by exploiting network locality, the developed method successfully addresses existing computation and communication scalability issues in controlling large complex networks.

\section*{Locality in Dynamical Networks}


\subsection*{Definitions and Basic Implications}
\reviseTen{While our results will apply to nonlinear networks, to develop our theory we first consider networks described by}
\begin{equation} \label{netsys}
    \dot{\bm{x}}_i = \bm{C}_{ii}\bm{x}_{i}+ \sum_{j=1,j\neq i}^N \bm{C}_{ij}\bm{x}_j,\ i=1,\cdots,N,
\end{equation}
where $\bm{x}_i\in \mathbb{R}^{n_i}$ is the state vector for node $i$
and the dimension $n_i$ can in principle be different for different nodes. In compact form, Eq.~\textbf{\ref{netsys}} reads $\dot{\bm{x}} = \bm{C} \bm{x}$, where $\bm{x}\in \mathbb{R}^{m}$, $\bm{C}\in \mathbb{R}^{m \times m}$, and $m=\sum_{i=1}^{N}n_i$. Here, $\bm{C}$ represents the Jacobian matrix of a general network system of $N$ nodes, which can be directed and weighted. In the case of an adjacency-like matrix $\bm{C}$, \reviseOne{the block $\bm{C}_{ij}$ represents the coupling from node $j$ to node $i$ if $i \neq j$, whereas} $\bm{C}_{ii}$ captures the \reviseThree{nodal dynamics and self-links}, collectively referred to as the self-interaction of node $i$. 
As a scalar measure of the coupling strength from node $j$ to $i$, we use the matrix norm $\norm{\bm{C}_{ij}}$ induced by the \reviseOne{given} vector norms for $\mathbb{R}^{n_i}$ and $\mathbb{R}^{n_j}$ (the notation $\norm{\cdot}$ is used throughout to indicate these norms for any vector and matrix). \reviseTen{The theory presented below is applicable to arbitrary matrices $\bm{C}$ and is explicitly illustrated for systems with multi-dimensional node dynamics. 
However, except when noted otherwise, our numerical simulations assume for concreteness that $\bm{C}=-\bm{L}$, where $\bm{L}$ is the Laplacian matrix of a network. Given a network with adjacency matrix $\bm{A}$, the Laplacian matrix is defined as $L_{ij} = -A_{ij}$ for $i\neq j$ and $L_{ii} = \sum_{j\neq i} A_{ij}$.}

To define a notion of \emph{locality} for dynamical networks, we use the algebra of matrices with off-diagonal decay \cite{grochenig2006symmetry}. 
The system matrix $\bm{C}$ is said to be \emph{localized} with respect to \reviseOne{a} characteristic function $v:\mathbb{R^{+}}\rightarrow \mathbb{R^{+}}$ and metric $\rho:\mathbb{Z}\times\mathbb{Z}\rightarrow \mathbb{R^{+}}$ provided \reviseOne{that}
\begin{equation}\label{localmatrix}
    \norm{\bm{C}_{ij}}\leq \kappa \cdot v\big(\rho(i,j)\big)^{-1}, \  i,j=1,2,\ldots,N,
\end{equation}
for some positive real constant $\kappa$. A network with system matrix $\bm{C}$ is localized if, in addition, the resulting information neighborhoods defined below are small for a tight choice of the bound in Eq. \textbf{\ref{localmatrix}}. The characteristic function $v(\cdot)$ is required to (i) be monotonically increasing, (ii) satisfy $v(0)=1$ and $v(\infty)=\infty$, and (iii) be sub-multiplicative \reviseOne{(i.e.,\ $v(z+y)\leq v(z)v(y)$)}. As a metric, $\rho(\cdot,\cdot)$ is required to satisfy (i\reviseOne{$'$}) the identity of indiscernibles, i.e., $\rho(i,j)= 0$ if and only if $i=j$, (ii\reviseOne{$'$}) the symmetry relation $\rho(i,j)=\rho(j,i)$, and (iii\reviseOne{$'$}) the triangle inequality $\rho(i,j)+\rho(j,k)\geq \rho(i,k)$. We refer to $\rho(\cdot,\cdot)$ as the \emph{information distance} associated with the network system in Eq. \textbf{\ref{netsys}}, as it \reviseTwo{measures the distance between two nodes in terms of information exchange: the farther apart two nodes are, the less information they exchange.} The reciprocal of the function $v(\cdot)$ in Eq. \textbf{\ref{localmatrix}} characterizes how the coupling strength in matrix $\bm{C}$ decays as the information distance grows. We define
$
    \mathcal{N}_i(\tau) = \{1 \leq j\leq N\ |\ \rho(i,j)\leq \tau\}
$
to be the \emph{information neighborhood} of radius $\tau$ centered at node $i$. Thus, Eq. \textbf{\ref{localmatrix}} ensures that the coupling from node $j$ to $i$ is weaker than $\kappa \cdot v(\tau)^{-1}$ for all nodes $j\notin\mathcal{N}_i(\tau)$, whereas all nodes $j\in\mathcal{N}_i(\tau)$ can have coupling with $i$ stronger than $\kappa \cdot v(\tau)^{-1}$. This coincides with the intuitive idea of network locality mentioned above. 
The rest of this Article will establish the legitimacy of this formal definition by demonstrating its explanatory and predictive power for analyzing and designing the control of dynamical networks.

For this purpose, it is instructive to first consider some basic implications of the notion of locality just introduced. Given a characteristic function $v(\cdot)$ and a metric $\rho(\cdot,\cdot)$, the set $\mathcal{L}_{v,\rho}$ of all block matrices $\bm{M}$ of block sizes $n_1,n_2,\cdots,n_N$  satisfying the locality property in Eq. \textbf{\ref{localmatrix}} (for $\bm{C}_{ij}$ replaced by $\bm{M}_{ij}$) forms a Banach algebra \cite{grochenig2006symmetry,motee2008decentralized}. That is, the set $\mathcal{L}_{v,\rho}$, which can include the system matrix $\bm{C}$, is closed under matrix arithmetics and contains its limit elements. In addition, if we choose a characteristic function $v(\cdot)$ satisfying the Gelfand-Raikov-Shilov (GRS) condition, $\lim_{{n}\rightarrow \infty} v(n{z})^{1/n}=1$, then the set $\mathcal{L}_{v,\rho}$ is inverse-closed, i.e., $\bm{M}^{-1} \in \mathcal{L}_{v,\rho}$ if $\bm{M}$ is an invertible element in $\mathcal{L}_{v,\rho}$ \cite{grochenig2006symmetry}. A special class of functions satisfying the GRS condition consists of the sub-exponential functions $v({z})=e^{\alpha {{z}}^{\beta}} (1+z)^q$ with $\alpha>0$, $0<\beta<1$, and $q>1$. When $\mathcal{L}_{v,\rho}$ is an inverse-closed Banach algebra, the locality defined above is invariant under various operations on the system matrix $\bm{M}$ and hence is preserved in key matrices for system analysis and control, such as the controllability and observability Gramians.
If the algebra is inverse-closed, locality is also preserved in the solutions of linear equations of the form $\bm{M}\bm{x}=\bm{b}$, the Riccati equation 
$
    \bm{C}^{T}\bm{P}+\bm{P}\bm{C}-\bm{P}\bm{B}\bm{R}^{-1}\bm{B}^{T}\bm{P}+\bm{Q}=\bm{0},
$
and the Lyapunov equation
$
    \bm{C}^{T}\bm{P}'+\bm{P}'\bm{C}+\bm{Q}'=\bm{0}
$
(assuming that $\bm{b}$ is localized around a given node $i$ and that the matrices $\bm{Q}$, $\bm{B}\bm{R}^{-1}\bm{B}^T$, and $\bm{Q}'$ belong to $\mathcal{L}_{v,\rho}$) \cite{motee2008decentralized,curtain2011riccati}. For localized networks, this leads to localized feedback matrices that solve the linear-quadratic optimal control problem. \reviseTen{These properties are derived in \textcolor{blue}{\emph{SI Text 1}} and used in our theory below.}

\subsection*{Constructing the Information Distance and Locality Measures}

To systematically construct an information distance, we note that given a function $v(\cdot)$ satisfying the conditions (i)-(iii) above, there is always a function $\rho(\cdot,\cdot)$ satisfying Eq. \textbf{\ref{localmatrix}} for some $\kappa>0$ whose explicit identification is presented below. For simplicity, we use the characteristic function $v({z})=e^{\alpha {{z}}^{\beta}} (1+{z})^q$ with $\alpha=1$, \reviseTen{$\beta=0.9$, and $q=1.2$} throughout. The locality of a given network is then characterized solely by $\rho(\cdot,\cdot)$, which is unknown \textit{a priori} and thus needs to be constructed from the system matrix $\bm{C}$. Since $v(\cdot)$ is monotonically increasing, it has an inverse function, which we denote by ${w}(\cdot)$. {Let $\widetilde{G}$ denote the graph in which an undirected edge exists between nodes $i$ and $j$ if and only if $\max \{\norm{\bm{C}_{ij}},\norm{\bm{C}_{ji}}\}>0$ and define the length of each edge to be $\widetilde{\rho}_{ij} = \max \{ {w}\left(\max_{1\leq i',j'\leq N}\norm{\bm{C}_{i'j'}}/ {\max \{\norm{\bm{C}_{ij}},\norm{\bm{C}_{ji}}\} } \right),\epsilon \}$.} Here, we introduce \reviseThree{a small number $\epsilon>0$} to ensure that $\rho(\cdot,\cdot)$ to be constructed will be a metric and is set as $\epsilon = 10^{-12}$ throughout this Article. 
While it is straightforward to verify that any function $\rho(i,j)\leq \widetilde{\rho}_{ij}$ satisfies Eq.~\textbf{\ref{localmatrix}} with constant $\kappa =\max_{1\leq i,j\leq N}\norm{\bm{C}_{ij}}\cdot  v(\epsilon)$, such a function is generally not a metric. We thus choose $\rho(i,j)$ to be instead the geodesic distance over the weighted graph $\widetilde{G}$, i.e., the smallest sum of edge lengths $\widetilde{\rho}_{ij}$ along a path between nodes $i$ and $j$ ($\rho(i,j)=+\infty$ if no such path exists). This guarantees that $\rho(\cdot,\cdot)$ is a metric, in addition to being upper-bounded by $\widetilde{\rho}_{ij}$, and satisfies all the conditions required for an information distance. 

Note that the information distance defined above is generally different from the (conventional) network distance on networks, where the edge length is taken to be the inverse of the coupling strength. For the characteristic function $v(\cdot)$ adopted above, the only case in which these two notions of distances coincide is when the off-diagonal of $\bm{C}$ is given by an undirected and uniformly weighted adjacency matrix and the diagonal satisfies $ \norm{\bm{C}_{kk}}> \max_{i\neq j} \norm{\bm{C}_{ij}}$ for all $k$. If instead we take a linear function $v(\rho(i,j)) = (\max_{1\leq i',j'\leq N}\norm{\bm{C}_{i'j'}}) \rho(i,j)$, by our construction, $\widetilde{\rho}_{ij} = {(\max \{\norm{\bm{C}_{ij}},\norm{\bm{C}_{ji}}\})^{-1}}$ and the geodesic distance on $\widetilde{G}$ coincides with the network distance if the networks are undirected. However, the linear function $v(\cdot)$ does not satisfy the GRS condition and the properties (ii)-(iii) required to be a characteristic function. \reviseTen{We refer to Fig.\ S1 for an illustration of the necessity of the GRS condition.}

The problem of constructing the information distance $\rho(\cdot,\cdot)$ above has been reduced to that of determining the geodesic distances on the graph $\widetilde{G}$, which is a classical problem that is solvable, for example, by Dijkstra's algorithm \cite{dijkstra1959note}. For our purpose, a variant of Dijkstra's algorithm, called the Uniform Cost Search (UCS) \cite{felner2011position} (Algorithm 1 in \textcolor{blue}{\emph{Materials and Methods}}), is suitable because the algorithm itself can then be implemented in a distributed way. That is, the calculations can be parallelized and performed at each node using only local information. Since the algorithm starts from a given node and sequentially visits its geodesic neighbors from the nearest to the farthest, it can be terminated once the desired distances are calculated. These features make the algorithm scalable to large networks.

The information distance constructed above is visualized in Fig.~\ref{smallworldfig} for a weighted variant of the WS small-world model \cite{watts1998collective}.  
The figure shows that a short network distance between two nodes does not necessarily imply a short information distance between them. Fundamentally, the difference between the two distances arises because the information distance is a metric and captures not only the node-to-node interactions but also the self-interactions. In contrast with the network distance, the information distance is based on a characteristic function $v(\cdot)$ that satisfies the sub-multiplicativity and GRS conditions, which guarantees that the locality associated with the information distance is inherited by the dynamical responses in the network. It follows that the information distance is the appropriate distance representing the dynamical interaction strengths among nodes in a network.
For the example in Fig.~\ref{smallworldfig} and other model networks considered, the (scalar) edge weights $A_{ij}$ are drawn randomly from the uniform distribution in $[0,1]$, but we note that our main conclusions do not depend sensitively on this choice. 
To quantify the locality of a network, we now introduce two measures, which we will refer to as the \emph{$\gamma$-locality} and the \emph{$L$-neighborhood reduction rate}, using the information distance constructed above. Here, the $\gamma$-locality will quantify the size of the neighborhood given a fixed reduction rate $\gamma$, whereas the $L$-neighborhood reduction rate will measure the reduction rate given a neighborhood size $L$.

\begin{figure}[!t]
    \centering
    \begin{overpic}[width=0.8\textwidth,tics=10]{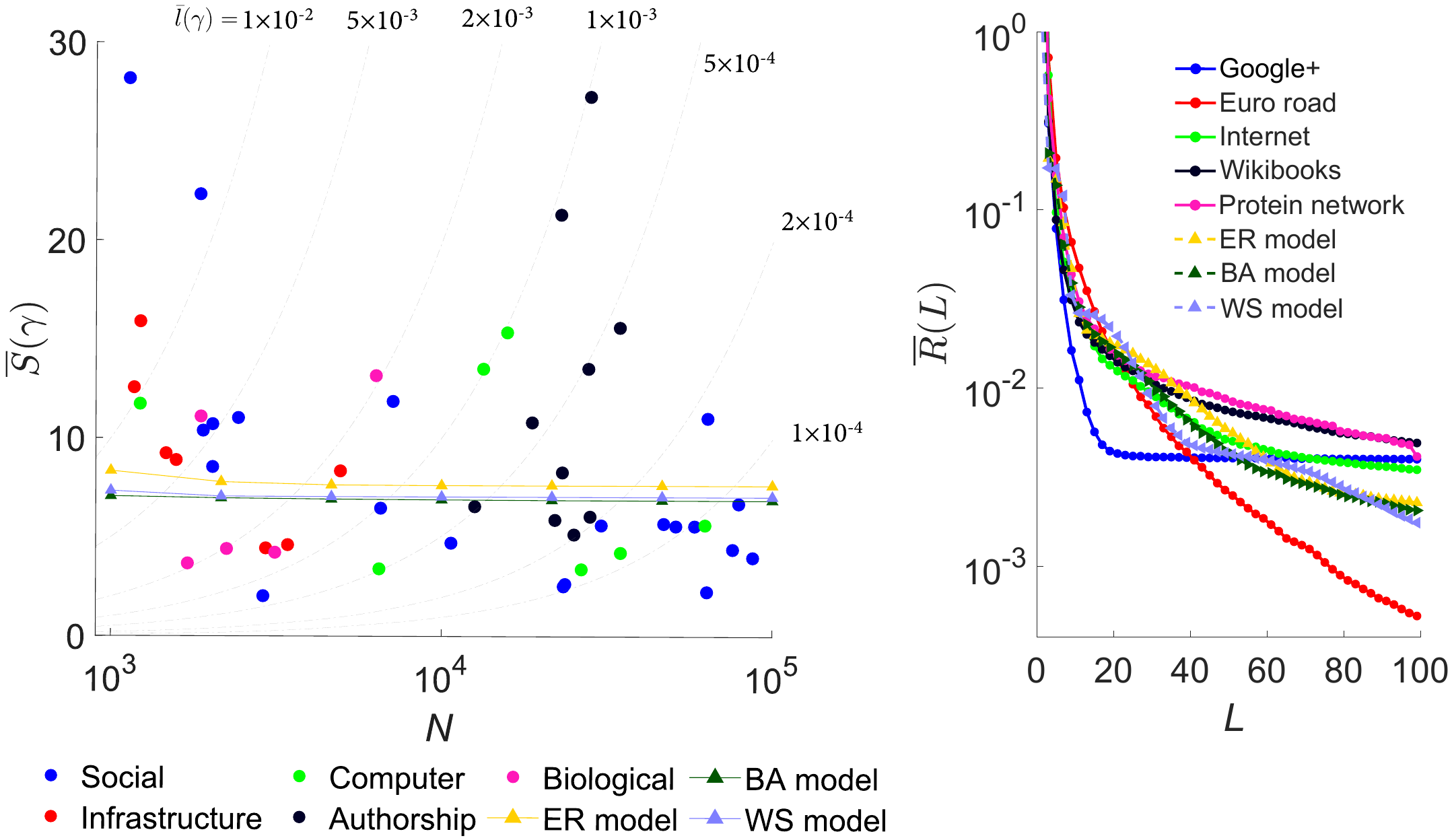}
\put (0,59) {\textbf{A}}
\put (59,59) {\textbf{B}}
\end{overpic}
         \caption{Locality measures for both empirical and model networks. (A) Average $\gamma$-neighborhood size $\bar{S}(\gamma)$ with \reviseTen{$\gamma = 0.05$} vs.\ the number of nodes $N$ for $50$ empirical networks from the KONECT dataset \cite{konect} (color-coded dots) and model networks generated by the Erd\H{o}s--R\'{e}yni (ER) model \cite{erdHos1960evolution}, Barab{\'a}si--Albert (BA) model \cite{barabasi1999emergence}, and WS model \cite{watts1998collective}. For the model networks (color-coded triangles), each data point represents an average over $20$ realizations. The gray dot-dash lines represent contour curves of the average $\gamma$-{locality} $\bar{l}({\gamma})$.  (B)~Average $L$-neighborhood reduction rate $\bar{R}(L)$ vs.\ the neighborhood size $L$ for a representative subset of empirical networks and the model networks with $N=1000$. 
         The model networks are set to have average degree $\bar{d} = 6$ using a seed network size of $m_0=3$ for the BA networks and a rewiring probability of $p = 0.2$ for the WS networks. The empirical network data is described in \textcolor{blue}{\emph{SI Text 2}}.}
         \label{gammalocality}
        
\end{figure}

For this purpose, we first define the $\gamma$-\emph{neighborhood} $\widetilde{\mathcal{N}}_i(\gamma)$ of node $i$ for a given constant $0<\gamma<1$ as the set of nodes $j$ for which the upper bound $\kappa\cdot v(\rho(i,j))^{-1}$ in Eq.~\textbf{\ref{localmatrix}} is larger than $\gamma \mu_i$, where $\mu_i = \max_{1\leq j \leq N} \max \{\norm{\bm{C}_{ij}},\norm{\bm{C}_{ji}}\}$.
Thus, the strength of the interaction between any node outside this $\gamma$-neighborhood and node $i$ is weaker than $\gamma$ times the maximum interaction strength involving node $i$.
In terms of the radius in information distance, we can write $\widetilde{\mathcal{N}}_i(\gamma) = \mathcal{N}_i\big({w}\left(\kappa/(\gamma \mu_i)\right)\big)$, and thus the $\gamma$-neighborhoods can themselves be referred to as information neighborhoods.
With this definition, we can now quantify the degree to which node $i$ is localized by the neighborhood size \reviseThree{$S_i(\gamma)=\abs{\widetilde{\mathcal{N}}_i(\gamma)}$ and its normalized version, $l_i(\gamma) = S_i(\gamma)/N$,} where $\abs{\cdot}$ denotes the number of elements in the set (when applied to numbers, the notation $\abs{\cdot}$ will denote absolute value). We call $l_i(\gamma)$ the $\gamma$-\emph{locality} of node $i$. To measure the locality of the entire network, we use the average $\gamma$-{locality}, $ \bar{l}({\gamma}) =\sum_{1\leq i\leq N} S_i(\gamma)/N^2$, which is a number between $0$ and $1$. In the extreme case of completely isolated nodes (i.e., $\norm{\bm{C}_{ij}}=0$ for all $i\neq j$), the {information} distance is given by $\rho(i,j)=+\infty$ for all $i\neq j$ and $\rho(i,i)=0$ for all $i$, yielding $\widetilde{\mathcal{N}}_i(\gamma)=\{i\}$ and $\bar{l}({\gamma})=\frac{1}{N}$, which approaches zero in the large-network limit. In the other extreme of all-to-all uniform coupling (i.e., $\norm{\bm{C}_{ij}}=\norm{\bm{C}_{i'j'}}>0$ for all $i\neq j$ and $i'\neq j'$), the distance is given by $\rho(i,j)= \epsilon$ for all $i \neq j$ and $\rho(i,i)=0$, yielding $\bar{l}({\gamma})=1$. For typical networks, $\bar{l}({\gamma})$ takes values intermediate between these two extremes. We note that the calculation of the $\gamma$-{locality} does not require obtaining $\rho(\cdot,\cdot)$ in advance; instead, the UCS algorithm can be run in parallel to efficiently construct $\widetilde{\mathcal{N}}_i(\gamma)$ for all nodes $i$. Fig.~\ref{gammalocality}A shows the average of \reviseThree{$S_i(\gamma)$ among all nodes, denoted as $\bar{S}(\gamma)$,} for \reviseTen{$\gamma = 0.05$} as a function of the network size for various empirical and model networks. We observe $\bar{l}({\gamma})<0.05$ for all networks considered, with nearly \reviseTen{$90\%$} of them showing $\bar{l}({\gamma})<0.01$, which suggests that the locality in the sense defined here is pervasive across both real and model networks. In addition, the average neighborhood size $\bar{S}(\gamma)$ for model networks does not grow with the network size, indicating that larger networks may not be more difficult to analyze and control if locality is properly exploited.

The upper bound $\kappa\cdot v(\rho(i,j))^{-1}$ on the strength of coupling from a given node $j$ to a given node $i$ in Eq.~\textbf{\ref{localmatrix}} reduces as the information distance $\rho(i,j)$ increases. Thus, the locality of node $i$ can also be measured by the reduction of this bound achieved at the boundary of the $L$-neighborhood $\widehat{\mathcal{N}}_i(L)$, which we define as the
set of $L$ nodes closest to the node $i$ according the information distance. This neighborhood includes node $i$ itself, and nodes at equal information distances are ordered randomly.
 The interaction strength reduction achieved at the boundary of the $L$-neighborhood is then given 
 by $R_i(L):= \kappa\cdot v\big(\max_{k\in \widehat{\mathcal{N}}_i(L)} \rho(i,k)\big)^{-1} / \mu_i$, which we call the $L$-neighborhood reduction rate of node $i$. This implies that $\norm{\bm{C}_{ik}} \leq  R_i(L) \mu_i$ and that any node farther away must couple more weakly to node $i$ than $R_i(L) \mu_i$. Fig.~\ref{gammalocality}B shows the average $L$-neighborhood reduction rate as a function of $L$ for several real and model networks. The average reduction rate $ \bar{R}({L})  = \sum_{i=1}^N R_i(L)/N$ exhibits a sharp initial decrease for a small $L$ on various networks (note the logarithmic scale), further suggesting that local control may be possible with small information neighborhoods. Below, we show that this is indeed the case by establishing that the controllability Gramian and optimal control actions inherit the network locality.

\section*{Controllability of Localized Networks}

\subsection*{Locality of the Controllability Gramian and Control Effort}

We now examine the network system described by Eq. \textbf{\ref{netsys}} in a control-theoretic context. In this analysis, we use the notion of driver node to refer to a node that is directly actuated by an independent control input, which is in turn referred to as a driver. By selecting driver nodes as a subset of nodes $\mathcal{D}\subseteq \mathcal{N}:=\{1,\ldots,N\}$, the system
dynamics can be expressed as
\begin{equation}\label{consys}
\begin{aligned}
   & \dot{\bm{x}} = \bm{C}\bm{x}+\bm{B}\bm{u}, \\
    \end{aligned}
\end{equation}
where $\bm{B}\in \mathbb{R}^{m\times r}$ is comprised of $[\bm{e}_{1}\bm{b}_{1}, \bm{e}_{2}\bm{b}_{2},\cdots,\bm{e}_{N}\bm{b}_{N}]$, $\bm{b}_i \in \mathbb{R}^{n_i\times r_i}$ is the input matrix of node $i$, and $\bm{e}_i^T = [\bm{0}_{n_i\times n_1},\cdots,\bm{I}_{n_i},\cdots,\bm{0}_{n_i\times n_N}] \in \mathbb{R}^{n_i \times m}$ is the projection from the entire state space to the subspace of node $i$. The matrix $\bm{b}_i$ is zero if $i \notin \mathcal{D}$, and we define $\eta = \max_{i} \norm{\bm{b}_{i}}$. The total input dimension of the system is $r=\sum_{i=1}^{N}r_i$ and we denote by $\bm{f}_i^T =[\bm{0}_{r_i\times r_1},\cdots,\bm{I}_{r_i},\cdots,\bm{0}_{r_i\times r_N}]\in \mathbb{R}^{r_i \times r}$ the projection from the entire input space to the input subspace of node $i$. The dynamical system in Eq. \textbf{\ref{consys}} is controllable if, for any given initial state $\bm{x}_0$, final state $\bm{x}_1$, and finite time $t_1>0$, there exists an input $\bm{u}$ such that the system state 
is steered from $\bm{x}=\bm{x}_0$ at time $t=0$ to $\bm{x}=\bm{x}_1$ at $t=t_1$. 
It can be shown that this controllability condition is satisfied
 if the controllability Gramian matrix
\begin{equation}\label{Wc}
 { \bm{W}_{\text{c}}^t =\sum_{i\in \mathcal{D}}    \bm{W}_{\text{c}i}^t= \sum_{i\in \mathcal{D}}    \int_{0}^t e^{\bm{C}t'} \bm{e}_{i}\bm{b}_{i}\bm{b}_{i}^T\bm{e}_{i}^Te^{\bm{C}^Tt'} dt'}
\end{equation}
is positive definite for any $t>0$ \cite{controltheory}, where the component $\bm{W}_{\text{c}i}^t$ represents the contribution of the $i$th driver.
Since the matrix exponential $e^{\bm{C}t}$ is localized if the network system is localized (see \textcolor{blue}{\emph{SI Text 1}} for a proof), we have $\norm{[e^{\bm{C}t}]_{ij}}\leq \kappa_{t} v(\rho(i,j))^{-1}$ for some constant $\kappa_{t}>0$, where $[e^{\bm{C}t}]_{ij}$ denotes the $(i,j)$ block of matrix $e^{\bm{C}t}$ (according to the same block partition as in matrix $\bm{C}$). Thus, $ \norm{ [e^{\bm{C}t} \bm{e}_{i}\bm{b}_{i}(e^{\bm{C}t} \bm{e}_{i}\bm{b}_{i})^T]_{jk}} \leq \kappa_{t}^2  \eta^2  v(\rho(i,j))^{-1} v(\rho(i,k))^{-1}
\leq \kappa_{t}^2\eta^2  v\big(\rho(i,j)+\rho(i,k)\big)^{-1} $, where the second inequality comes from the submuliplicative property of the characteristic function. This decay pattern is preserved under integration: $\norm{ [ {\bm{W}_{\text{c}i}^t}]_{jk}} \leq  {\breve{\kappa}_t}\eta^2 v\big(\rho(i,j)+\rho(i,k)\big)^{-1}$, where {$\breve{\kappa}_t = \int_{0}^t  \kappa_{t'}^2 dt'$}. This shows that the contribution to the controllability Gramian from the driver at node $i$ concentrates around the $(i,i)$ block and decays as one moves away vertically or horizontally from that block. Combining the contributions from all driver nodes and using the triangle inequality, we have $\norm{ [ {\bm{W}_{\text{c}}^t}]_{jk} } \leq {\breve{\kappa}_t}\eta^2 \abs{\mathcal{D}}  v\big(\rho(j,k)\big)^{-1}$, which implies that the Gramian is also localized and belongs to the algebra $\mathcal{L}_{v,\rho}$ according {to} the block partition of the system matrix $\bm{C}$.

{The quadratic integral $\int_{0}^{\infty}\bm{u}(t)^T\bm{u}(t)dt$ is usually referred to as the energy of the control input and serves as an quantitative measure of the control effort required to achieve certain task.}
It is known that the worst-case minimum energy needed to drive a system to a target state is inversely proportional to {$\lambda_{\text{min}} ({\bm{W}_{\text{c}}^t})$, the smallest eigenvalue of the controllability Gramian \cite{sun2013controllability,pasqualetti2014controllability}.} In general, $\lambda_{\text{min}} ({\bm{W}_{\text{c}}^t})$ is upper-bounded by the smallest diagonal element of ${\bm{W}_{\text{c}}^t}$. Thus, if a localized network system is equipped with only one driver at node $i$, we have 
$
        \lambda_{\text{min}} ({\bm{W}_{\text{c}}^t}) \leq  {\breve{\kappa}_t}\eta^2 v\big(\max_{j} \rho(i,j)\big)^{-2},
$
{which implies that the worst-case control energy grows at least near-exponentially with the information distance between the driver and the farthest node in the network.}
{A system that} has nodes far from node $i$ in the metric $\rho(\cdot,\cdot)$, as in the case of a highly localized network, would be uncontrollable in practice by just placing one driver at node $i$. That is, even if the system is {theoretically controllable (i.e., $\bm{W}_{\text{c}}^t$ has full rank)}, the control energy needed to drive the system would be prohibitively high. Moreover, the analysis above extends to the case of multiple driver nodes, providing a more general upper bound on the smallest eigenvalue of the controllability Gramian:
\begin{equation}\label{upp_bnd_min_gram}
     \lambda_{\text{min}} ({\bm{W}_{\text{c}}^t}) \leq {\breve{\kappa}_t} \eta^2 \cdot\abs{\mathcal{D}} \cdot  v\big({\rho_{\text{H}}} (\mathcal{D},\mathcal{N})\big)^{-2},
\end{equation}
where $\mathcal{D} := \{i_1, i_2,\cdots,i_{\abs{\mathcal{D}}}\}$ is the set of driver nodes, and ${\rho_{\text{H}}} (\mathcal{D},\mathcal{N}) := \max_{j\in \mathcal{N}} \, \min_{i\in \mathcal{D}}\,\rho(i,j)$ is the directed Hausdorff distance between the sets $\mathcal{D}$ and $\mathcal{N}$ induced by the metric $\rho(\cdot,\cdot)$. 
Eq.~\textbf{\ref{upp_bnd_min_gram}} establishes a locality requirement for energy-efficient control: to ensure that the network is controllable in practice, every node in the network must be within a small information neighborhood of a node in $\mathcal{D}$ (i.e., $\min_{i\in \mathcal{D}}\,\rho(i,j)$ is small for all $j\in\mathcal{N}$, so that ${\rho_{\text{H}}} (\mathcal{D},\mathcal{N})$ is small). This condition usually means that a significant portion of the nodes need to be directly controlled. This analysis provides a theoretical explanation for the empirical observations in \cite{gao2014target,yan2015spectrum} that the worst-case control energy increases drastically as the number of driver nodes is reduced. 

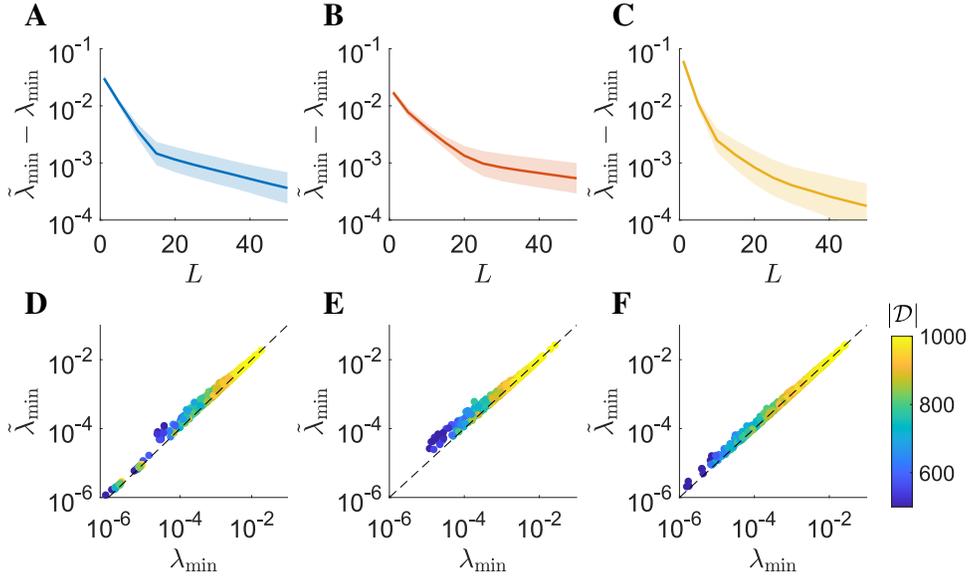
\begin{figure}[!t]
    \centering
        \begin{overpic}[width=0.8\textwidth,tics=5]{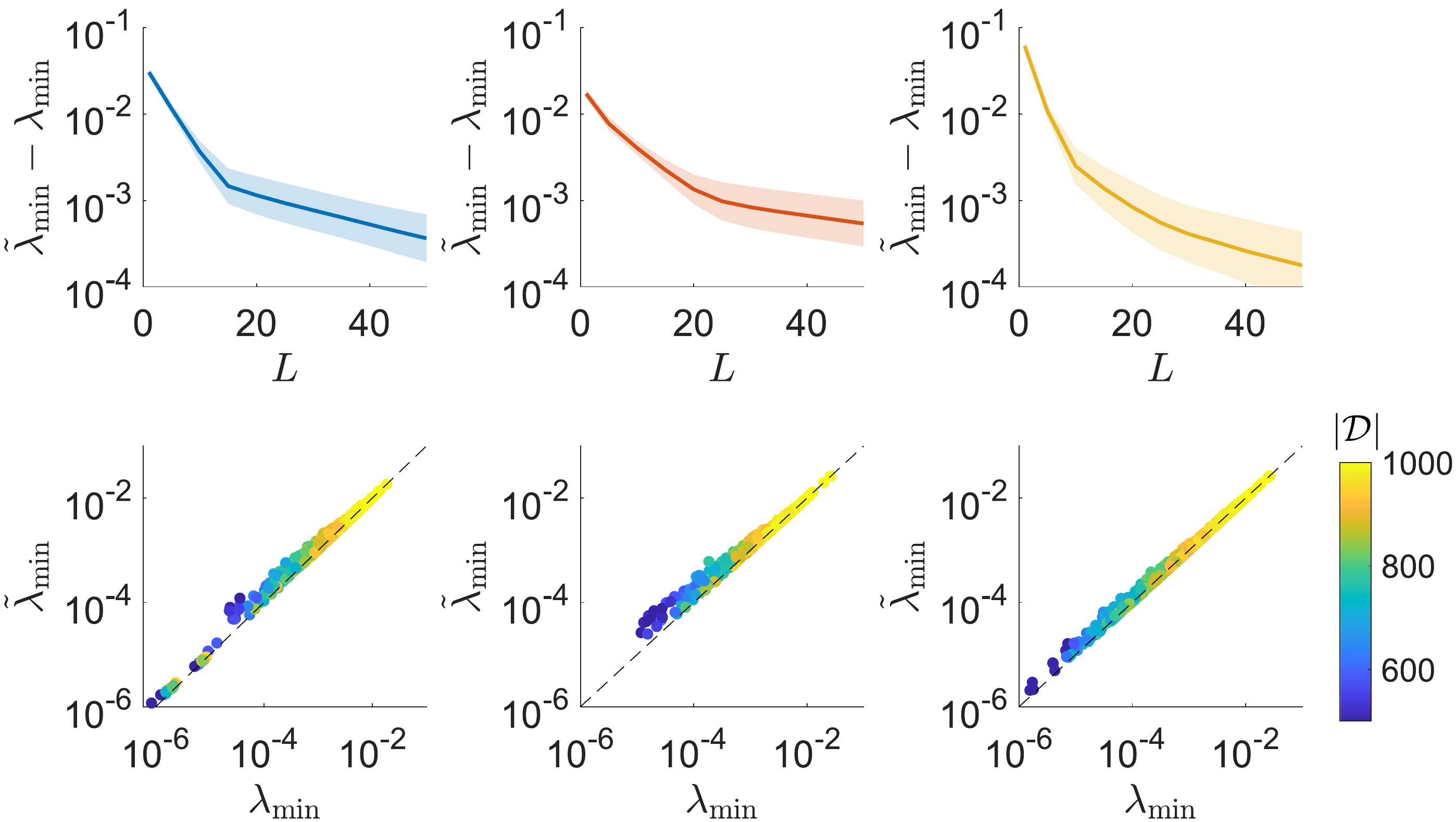}
\put (2,57) {\textbf{A}}
\put (33,57) {\textbf{B}}
\put (63,57) {\textbf{C}}
\put (2,27) {\textbf{D}}
\put (33,27) {\textbf{E}}
\put (63,27) {\textbf{F}}
\end{overpic}
        \caption{Local approximability of the smallest eigenvalue of the controllability Gramian. (A)--(C) Approximation error ${\widetilde{\lambda}_{\text{min}}-\lambda_{\text{min}}}$ vs.\ the information neighborhood size $L$ for the ER, BA, and WS models, respectively. The networks are generated for $N=1000$ with the other parameters set as in Fig.~\ref{gammalocality}. For each $L$, we choose as drivers $950$ randomly selected nodes, specifying $\bm{B}$ in Eq.~\textbf{\ref{consys}} as the diagonal matrix whose diagonal elements equal $1$ for the selected nodes and $0$ for the others. The curves and shaded areas represent the mean and standard deviation of the approximation error over $100$ realizations of driver placement. (D)--(F) Exact vs.\ estimated smallest eigenvalue for the network models respectively used in (A)--(C) for $1000$ realizations of a random number of drivers $\abs{\mathcal{D}} = N-\xi$ (each placed randomly). Here, $\xi$ is drawn from the Poisson distribution with mean $\mu = 100$, the information neighborhood size is fixed at $L=50$, and each realization is color coded by $\abs{\mathcal{D}}$.}
         \label{localappeig}
         
\end{figure}

\subsection*{Localized Approximation of the Controllability Measure}

\begin{figure}[h]
\centering
      \begin{overpic}[width=0.33\textwidth,tics=5]{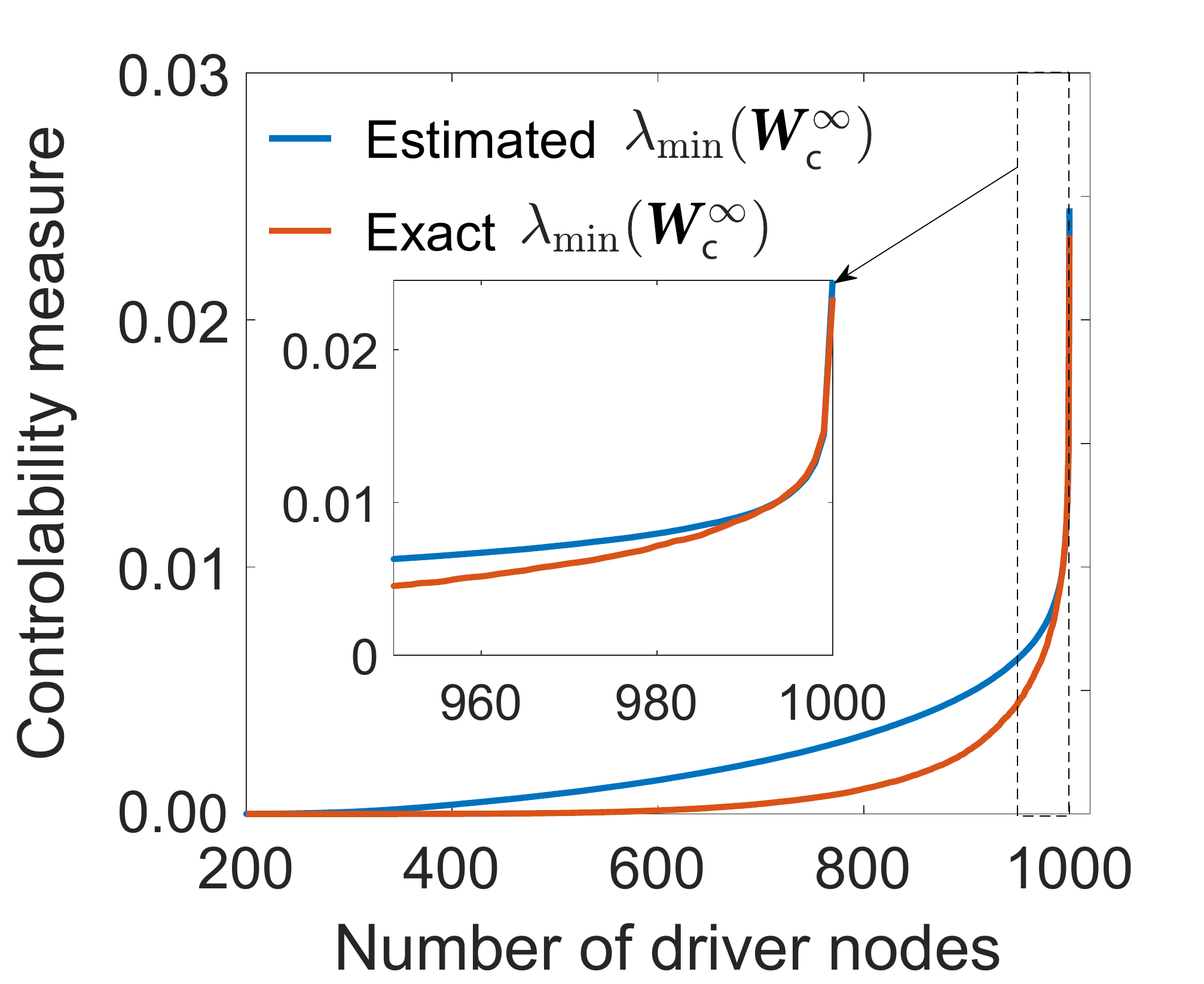}
\put (0,80) {\textbf{A}}
\end{overpic}
      \begin{overpic}[width=0.33\textwidth,tics=5]{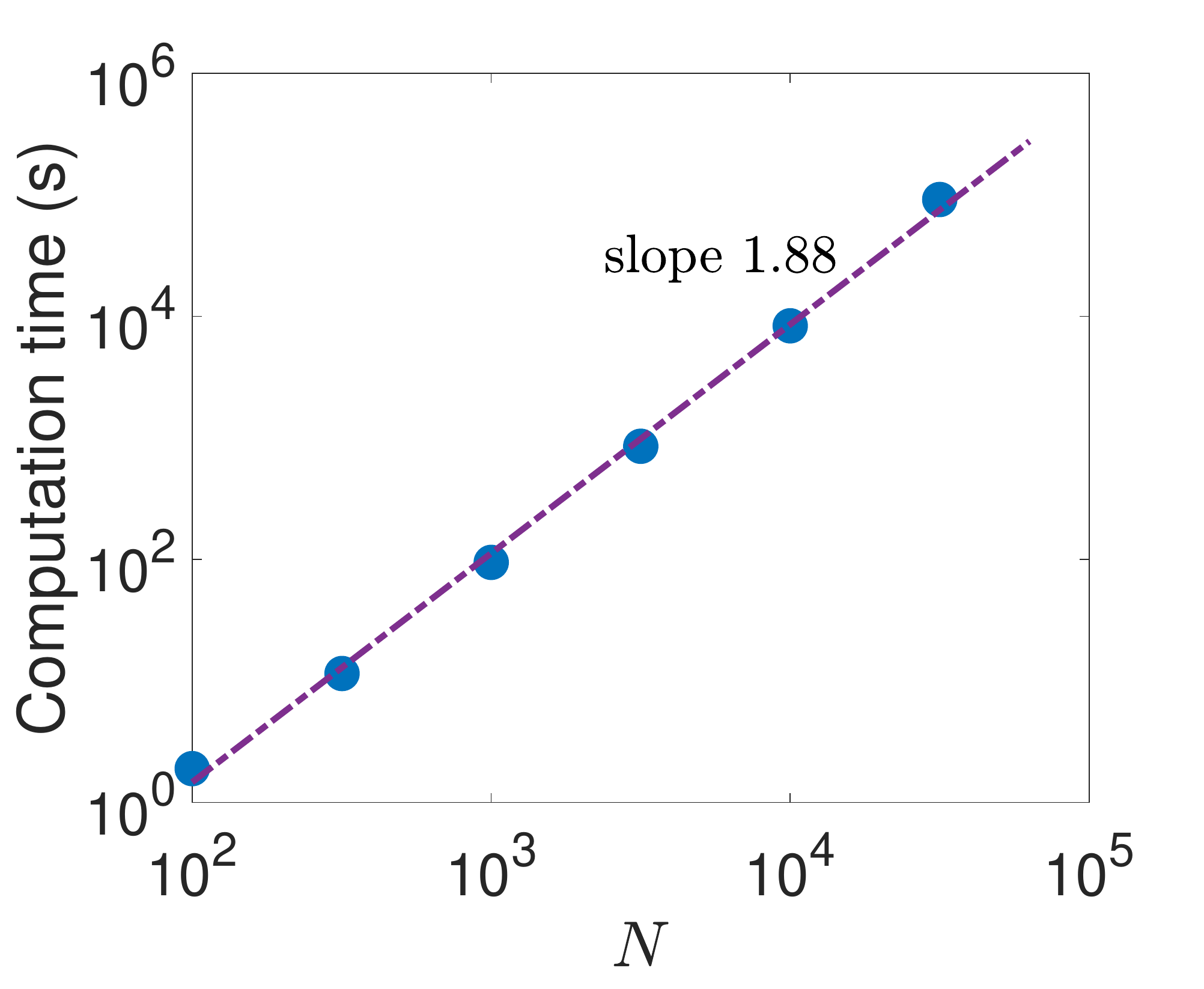}
\put (0,80) {\textbf{B}}
\end{overpic}
      \begin{overpic}[width=0.33\textwidth,tics=5]{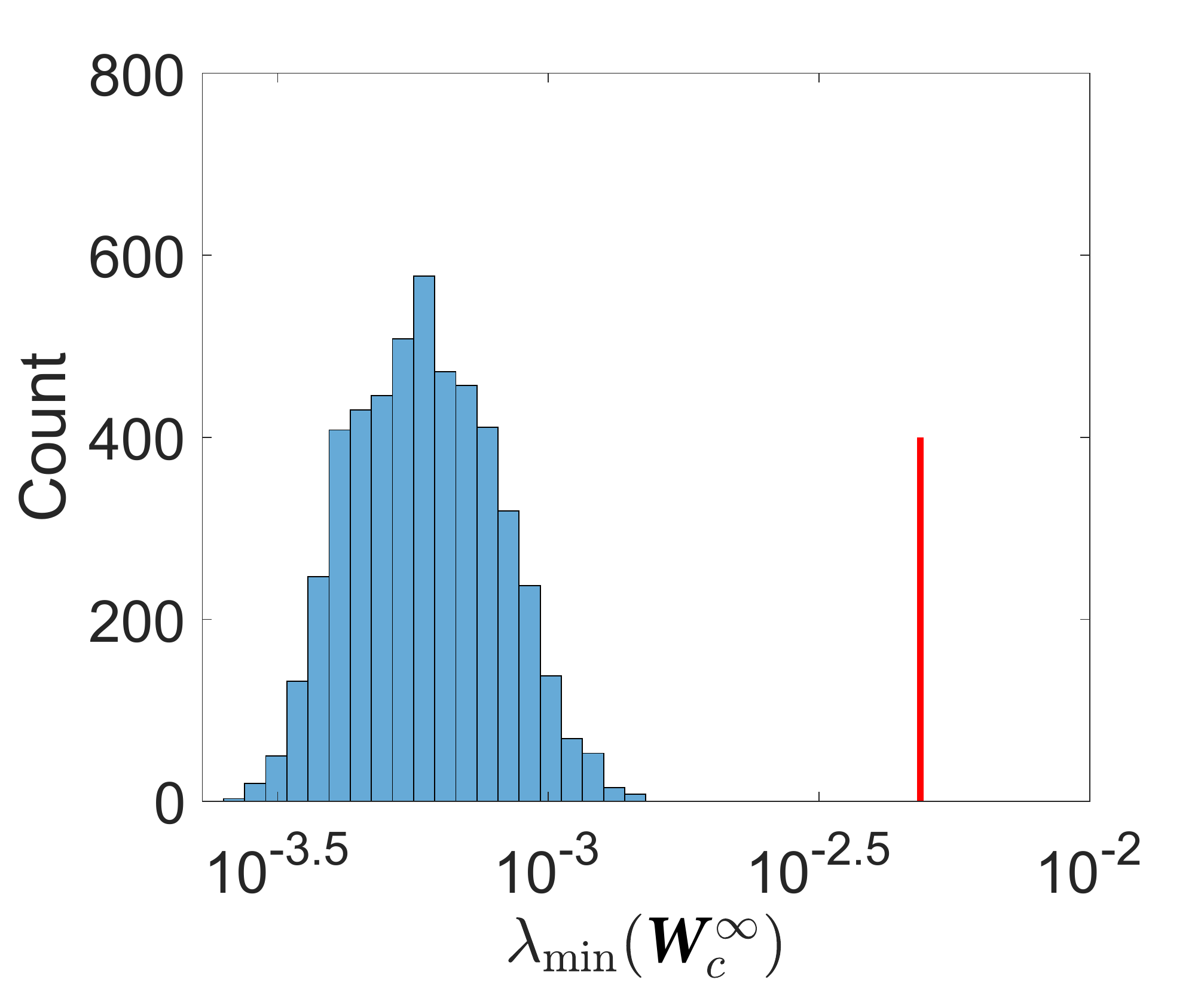}
\put (0,80) {\textbf{C}}
\end{overpic}
      \begin{overpic}[width=0.33\textwidth,tics=5]{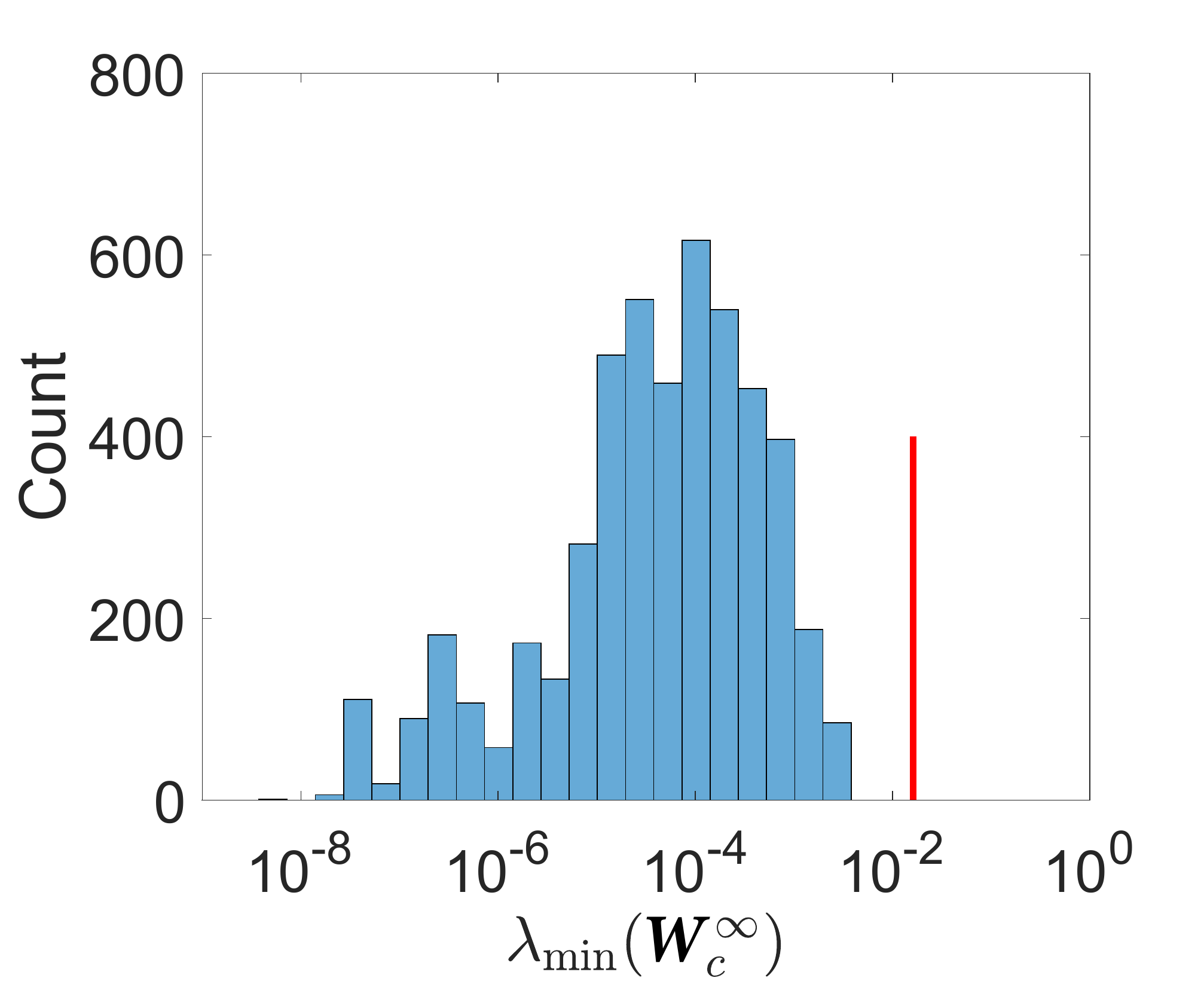}
\put (0,80) {\textbf{D}}
\end{overpic}
      \begin{overpic}[width=0.33\textwidth,tics=5]{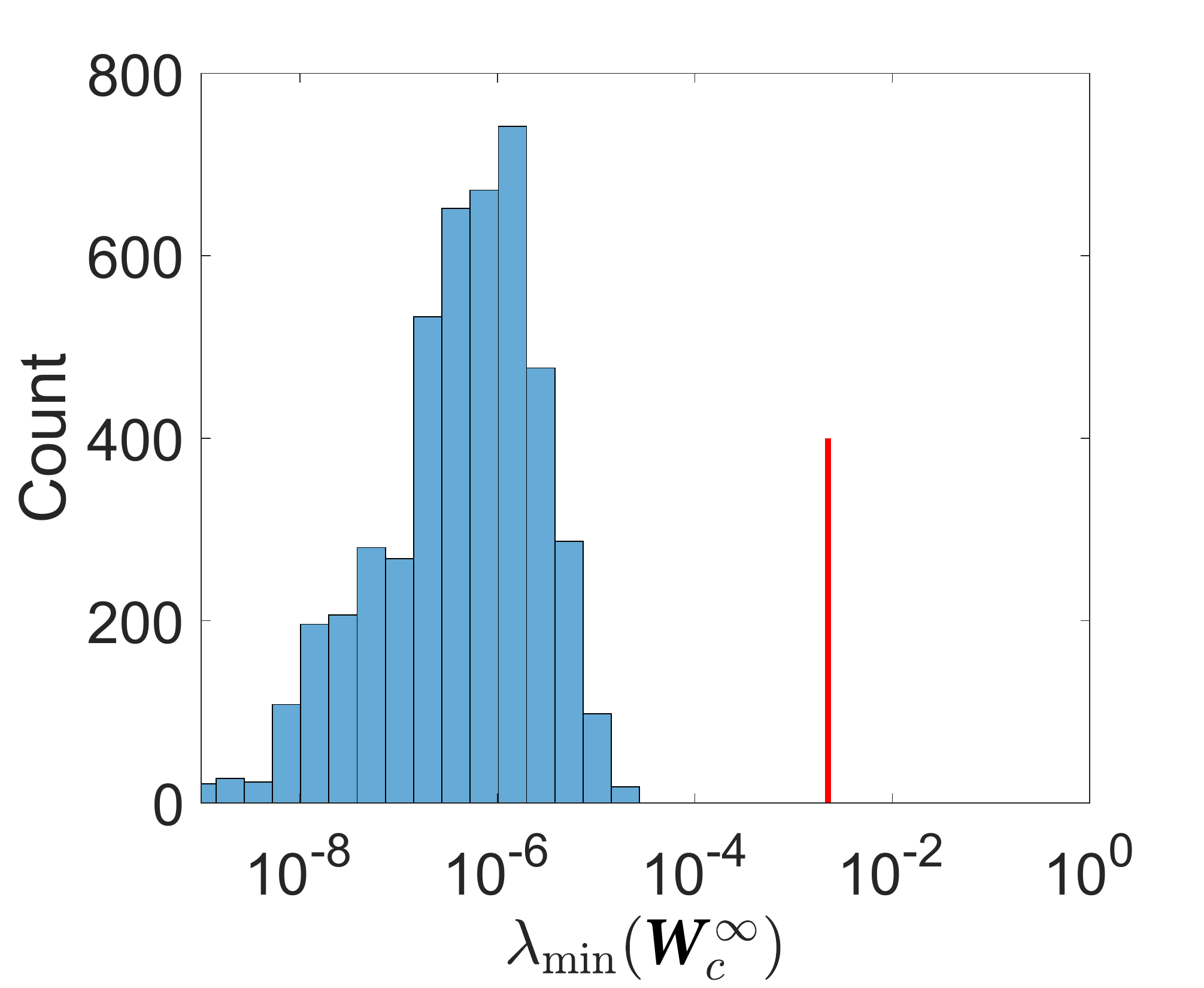}
\put (0,80) {\textbf{E}}
\end{overpic}
      \begin{overpic}[width=0.33\textwidth,tics=5]{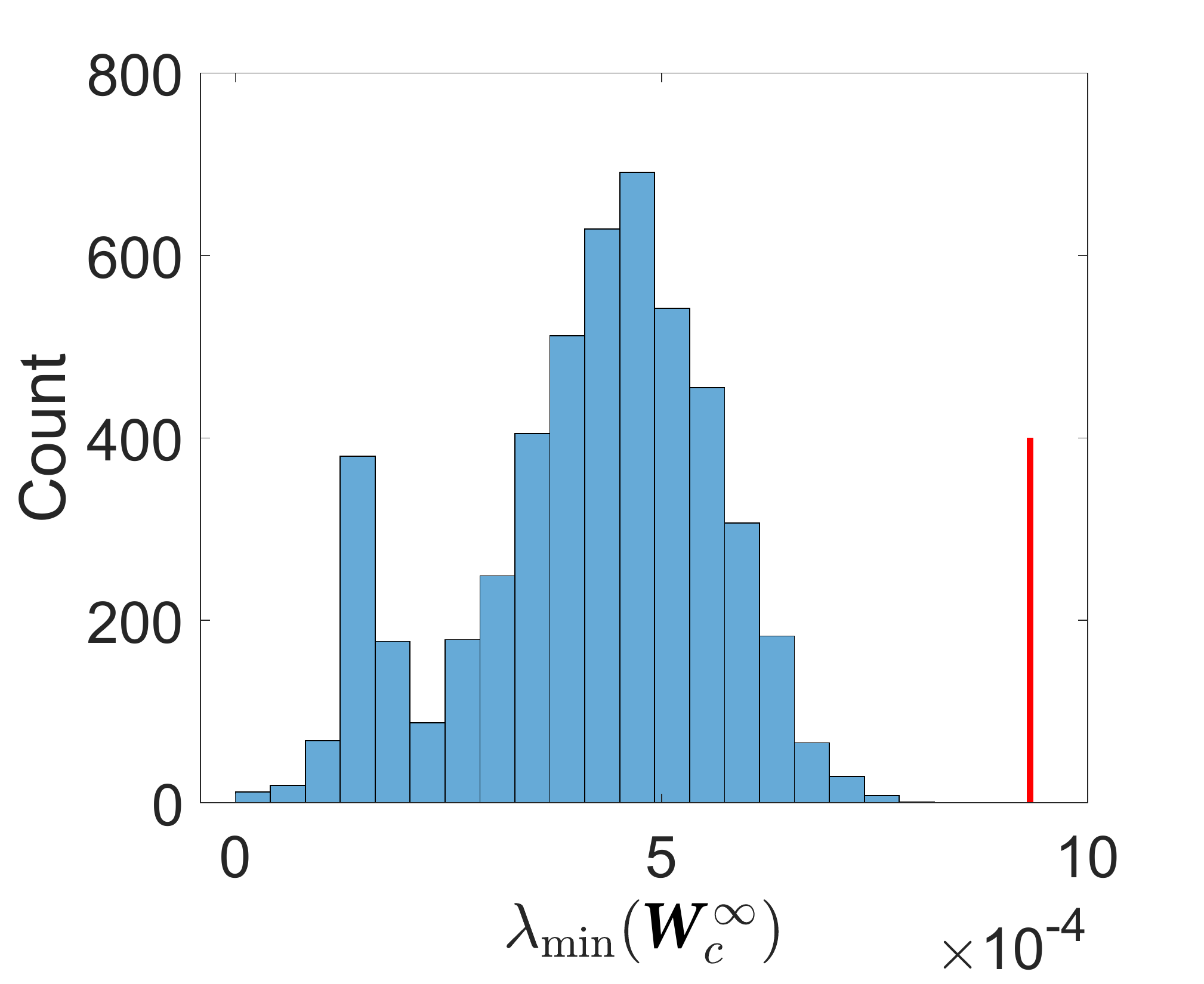}
\put (0,80) {\textbf{F}}
\end{overpic}
\caption{Performance of Algorithm 2 for driver placement. (A) Exact controllability measure $\lambda_{\text{min}}(\bm{W}_{\text{c}}^{\infty})$ and its estimate $\widetilde{\lambda}_{\text{min}}(\widetilde{\bm{W}}_\text{c})$ vs.\ the number of drivers $\abs{\mathcal{D}}$ for the optimal driver placement identified by the algorithm. The curves are averages over 100 realizations of (weighted) BA networks of Kuramoto oscillators (governed by Eq.~\textbf{\ref{kuramoto}} in \textcolor{blue}{\emph{Materials and Methods}}) with $N=1000$ and average degree $\bar{d}=10$ for $L=20$. (B)~Computational time of the algorithm for the network used in (A) as the number of nodes $N$ is varied (obtained using 12 cores of an Intel Xeon E7-8867v4 processor). (C)~Performance of the algorithm for the network model in (A) with $\abs{\mathcal{D}}=950$, where the red line indicates $\lambda_{\text{min}}(\bm{W}_{\text{c}}^{\infty})$ for the optimal driver placement identified by the algorithm and the histogram shows $\lambda_{\text{min}}(\bm{W}_{\text{c}}^{\infty})$ 
for $5000$ random placements. (D)~Performance of the algorithm with $\abs{\mathcal{D}}=2500$ for the dynamics of the $3907$ generators in the Eastern U.S.\ power grid used in Fig.~\ref{WcWcest}. (E)~Performance of the algorithm with $\abs{\mathcal{D}}=2200$ for epidemic spreading over the global air transportation network between $2290$ major cities. (F)~Performance of the algorithm with $\abs{\mathcal{D}}=600$ for the neuronal dynamics in the brain network of $638$ cortical areas. The results in (D)--(F) are visualized as in (C). 
The empirical network systems in (D)--(F) are constructed from data as described in \textcolor{blue}{\emph{SI Text 2}}, with Eq.~\textbf{\ref{consys}} for each network specified in \textcolor{blue}{\emph{Materials and Methods}}.
}
\label{driverpalceperform}
\end{figure}

The Gramian eigenvalue $\lambda_{\text{min}} (\bm{W}_{\text{c}}^t)$, being a quantitative measure of the system's controllability, can be used as an objective function to guide the selection of driver nodes \cite{pasqualetti2014controllability}. The exact computation of $\lambda_{\text{min}} (\bm{W}_{\text{c}}^t)$, which requires solving the eigenvalue problem for the entire system, is inefficient and can be prohibitive for large-scale networks. However, when the network is localized, $\lambda_{\text{min}} (\bm{W}_{\text{c}}^t)$ can be well approximated by $\lambda_{\text{min}} \big(\bm{W}_{\text{c}}^t(\mathcal{N}_i(\tau),\mathcal{N}_i(\tau))\big)$, where $\bm{W}_{\text{c}}^t(\mathcal{N}_i(\tau),\mathcal{N}_i(\tau))$ denotes the submatrix of $\bm{W}_{\text{c}}^t$ induced by an information neighborhood $\mathcal{N}_i(\tau)$ of radius $\tau$ around a  certain node $i$. Indeed, we show that in a localized network there exists an $i\in \mathcal{N}$ such that ${\lambda_{\text{min}} \big(\bm{W}_{\text{c}}^t(\mathcal{N}_i(\tau),\mathcal{N}_i(\tau))\big)-\lambda_{\text{min}} (\bm{W}_{\text{c}}^t)}=\mathcal{O}\big(v(\tau)^{-1}\big)$ (\textcolor{blue}{\emph{SI Text 4}}). Since identifying such a node $i$ may be difficult in practice, we consider the smallest eigenvalue over all nodes:
\begin{equation}\label{minlam}
   {\widetilde{\lambda}_{\text{min}} (\bm{W}_{\text{c}}^t) = \min_{1\leq k \leq N}\ \lambda_{\text{min}} \big(\bm{W}_{\text{c}}^t(\mathcal{N}_k(\tau),\mathcal{N}_k(\tau))\big).}
\end{equation}
{It follows from the existence of the node $i$ with the property above that} $\widetilde{\lambda}_{\text{min}} (\bm{W}_{\text{c}}^t)$ converges to $\lambda_{\text{min}} (\bm{W}_{\text{c}}^t)$ as $\mathcal{O}\big(v(\tau)^{-1}\big)$. If the sizes of the information neighborhoods do not grow with the network size $N$, the cost of computing the smallest eigenvalue of each ``sub-Gramian'' in Eq. \textbf{\ref{minlam}} would remain constant, and hence the cost of computing $\widetilde{\lambda}_{\text{min}}$ would scale linearly with $N$. {This analysis also applies to the infinite-horizon controllability Gramian $\bm{W}_{\text{c}}^{\infty}$ when the system matrix $\bm{C}$ is stable (i.e., all its eigenvalues have negative real parts) since the integration in Eq. \textbf{\ref{Wc}} converges as $t\to\infty$ and the algebra $\mathcal{L}_{v,\rho}$ is complete.} Fig.~\ref{localappeig} demonstrates the ability of $\widetilde{\lambda}_{\text{min}}(\bm{W}_{\text{c}}^{\infty})$ to approximate the exact $\lambda_{\text{min}} (\bm{W}_{\text{c}}^{\infty})$ for model networks. As the neighborhood size $L=\abs{\mathcal{N}_i(\tau)}$ increases, the estimate $\widetilde{\lambda}_{\text{min}}(\bm{W}_{\text{c}}^{\infty})$ quickly approaches the true value $\lambda_{\text{min}} (\bm{W}_{\text{c}}^{\infty})$, as shown in Fig.~\ref{localappeig}A--C. For a fixed $L$, the estimate $\widetilde{\lambda}_{\text{min}}(\bm{W}_{\text{c}}^{\infty})$ provides an upper bound for $\lambda_{\text{min}} (\bm{W}_{\text{c}}^{\infty})$, as verified in Fig.~\ref{localappeig}D--F. Moreover, placing additional drivers decreases the relative differences between $\widetilde{\lambda}_{\text{min}}(\bm{W}_{\text{c}}^{\infty})$ and $\lambda_{\text{min}} (\bm{W}_{\text{c}}^{\infty})$, as shown in Fig.~\ref{localappeig}D--F (this point will be further illustrated in the context of driver placement in Fig.~\ref{driverpalceperform}A). Interestingly, it follows that the higher the degree of controllability, the more accurate is the localized approximation of the controllability measure $\lambda_{\text{min}} (\bm{W}_{\text{c}}^{\infty})$.

\clearpage

\subsection*{Localized Approximation of the Gramian}

In the limit $t\rightarrow \infty$, the controllability Gramian can be obtained by solving the algebraic Lyapunov equation:
$
    \bm{C} \bm{W}_{\text{c}}^{\infty} + \bm{W}_{\text{c}}^{\infty}\bm{C}^T + \bm{B}\bm{B}^T =\bm{0}
$. 
When $\bm{C}$ is the Laplacian matrix, it always has a zero eigenvalue due to the system's translational invariance, but we show that the Lyapunov equation is valid
after eliminating the trivial eigenspace associated with the zero eigenvalue (\textcolor{blue}{\emph{SI Text 5}}). In such cases, the notation $\bm{W}_{\text{c}}^{\infty}$ should always be interpreted as the Gramian after this elimination. Since the Lyapunov equation is linear in $\bm{B}\bm{B}^T$, its solution can be decomposed as $ \bm{W}_{\text{c}}^{\infty}  = \sum_{i=1}^N \bm{W}_{\text{c}i}^{\infty}$, where each $\bm{W}_{\text{c}i}^{\infty}$ solves
\begin{equation}\label{lyacon}
     \bm{C} \bm{W}_{\text{c}i}^{\infty} + \bm{W}_{\text{c}i}^{\infty}\bm{C}^T + \bm{e}_{i}\bm{b}_{i}\bm{b}_{i}^T\bm{e}_{i}^T =\bm{0}.
\end{equation}
The components $\bm{W}_{\text{c}i}^{\infty}$ are exactly the limits of the individual integral terms of the sum in Eq.~\textbf{\ref{Wc}} as $t\to\infty$ and hence inherit the locality property of $\bm{W}_{\text{c}i}^t$. Thus, for a localized network system, each $\bm{W}_{\text{c}i}^{\infty}$ is concentrated around the $(i,i)$ block, with a rapid decay away from that block, implying that there is a $\tau_i$-information neighborhood $\mathcal{N}_i(\tau_i)$ of node $i$ that captures the most significant matrix elements of $\bm{W}_{\text{c}i}^{\infty}$.
If we denote by $\bm{N}_i$ the matrix of projection from the entire state space to the subspace of the nodes in ${\mathcal{N}_i(\tau_i)}$, it follows from the locality properties analyzed above that $\norm{\bm{N}_i^T\bm{N}_i\bm{W}_{\text{c}i}^{\infty}\bm{N}_i^T\bm{N}_i-\bm{W}_{\text{c}i}^{\infty}}_{\infty} \leq {\breve{\kappa}_t}\norm{\bm{b}_{i}}^2v(\tau_i)^{-2}$. Here, we used the induced infinity norm of a matrix $\bm{M}\in \mathbb{R}^{m\times m}$ given by $\norm{\bm{M}}_{\infty}=\max_{1\leq i,j\leq N} \norm{\bm{M}_{ij}}$, where $\bm{M}_{ij}$ is the $(i,j)$th block of matrix $\bm{M}$ following the same partition of the system matrix $\bm{C}$. 

\begin{figure}[ht]
    \centering
      \begin{overpic}[width=0.66\textwidth,tics=5]{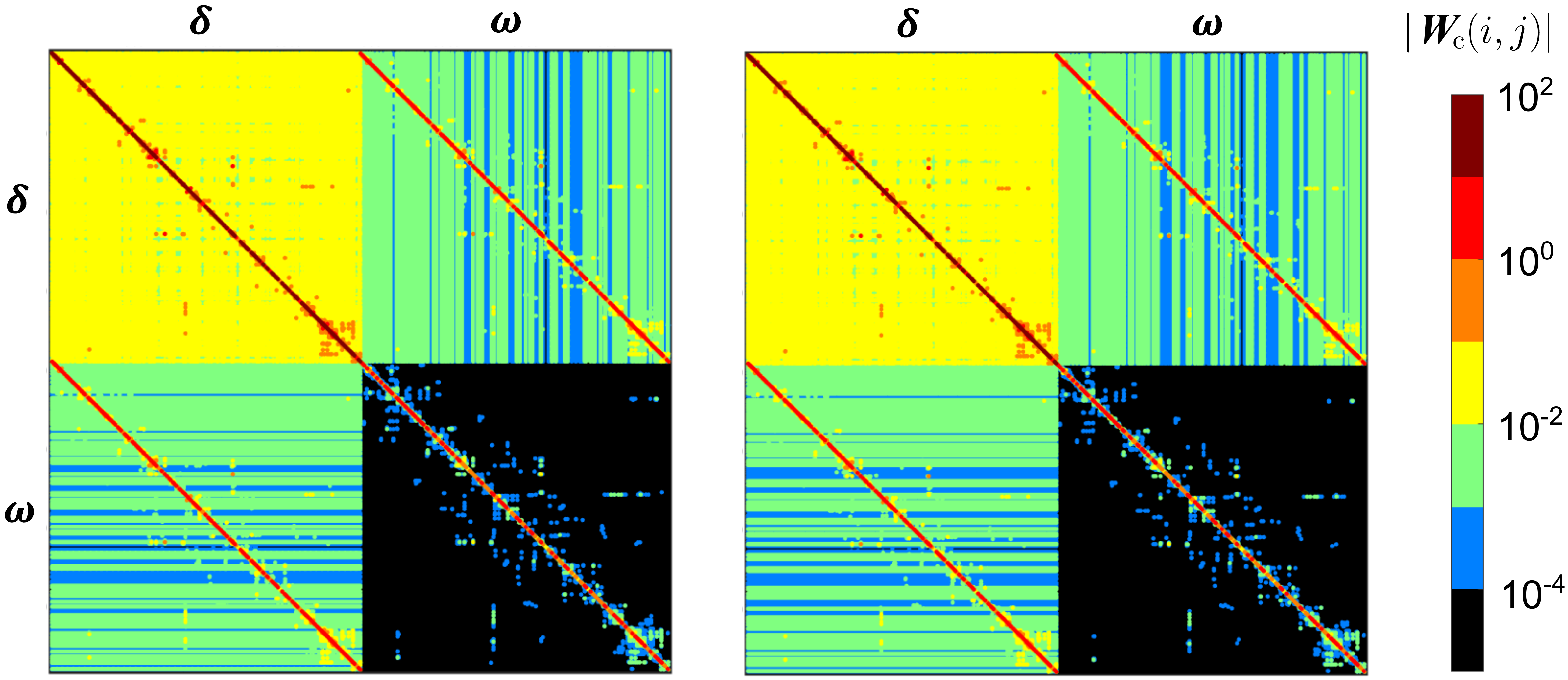}
\put (0,43) {\textbf{A}}
\put (46,43) {\textbf{B}}
\end{overpic}
         \caption{Controllability Gramian of the Eastern U.S.\ power grid. (A) Exact Gramian $\bm{W}_{\text{c}}^{\infty}$. (B) Approximate Gramian $\widetilde{\bm{W}}_{\text{c}}^{\infty}$ obtained by our localized method with information neighborhood size $L=\lceil N/100 \rceil$. The network consists of $N = 3907$ generator nodes, each described by a phase $\delta_i$ and frequency $\omega_i$, and is constructed from data as described in \textcolor{blue}{\emph{SI Text 2}}. Eq.~\textbf{\ref{consys}} for this system is specified by Eq.~\textbf{\ref{linpower}} in \textcolor{blue}{\emph{Materials and Methods}}, in which the mechanical power input of every generator is directly controlled.}
      
         \label{WcWcest}
\end{figure}

Defining $\widetilde{\bm{W}}_{\text{c}i}^{\infty} := \bm{N}_i \bm{W}_{\text{c}i}^{\infty} \bm{N}_i^T$, we see that $\bm{W}_{\text{c}i}^{\infty}$ can be approximated well by $\bm{N}_i^T\widetilde{\bm{W}}_{\text{c}i}^{\infty}\bm{N}_i$, and this $\widetilde{\bm{W}}_{\text{c}i}^{\infty}$ can be directly obtained by solving the projected Lyapunov equation,
\begin{equation}\label{lyapeqatnode}
    \widetilde{\bm{C}}_i\widetilde{\bm{W}}_{\text{c}i}^{\infty} + \widetilde{\bm{W}}_{\text{c}i}^{\infty}\widetilde{\bm{C}}_i^T+ \bm{N}_i\bm{e}_{i}\bm{b}_{i}\bm{b}_{i}^T\bm{e}_{i}^T\bm{N}_i^T=\bm{0},
\end{equation}
where $\widetilde{\bm{C}}_i := \bm{N}_i \bm{C} \bm{N}_i^T$. This Lyapunov equation is generally of much lower dimension than Eq.~\textbf{\ref{lyacon}} and involves only the portions of the system  inside the information neighborhood $\mathcal{N}_i(\tau_i)$. Eq.~\textbf{\ref{lyapeqatnode}} can be solved at each node independently so that the computation can be distributed across all nodes. After obtaining all $\widetilde{\bm{W}}_{\text{c}i}^{\infty}$, the entire controllability Gramian can be approximated as $ \widetilde{\bm{W}}_{\text{c}}^{\infty} = \sum_{i=1}^N  \bm{N}_i^T\widetilde{\bm{W}}_{\text{c}i}^{\infty}\bm{N}_i$. Fig.~\ref{WcWcest}A and \ref{WcWcest}B show the exact $\bm{W}_{\text{c}}^{\infty}$ and the corresponding approximation $\widetilde{\bm{W}}_{\text{c}}^{\infty}$, respectively, for the Eastern U.S.\ power grid, showing that the localized method developed here can accurately capture the structure of the exact $\bm{W}_{\text{c}}^{\infty}$. Furthermore, 
our numerics confirm the high accuracy of the approximation across various model and empirical networks, including the Eastern U.S.\ power grid, the global air transportation network, and a human brain network (\textcolor{blue}{\emph{SI Table S1}}). Thus, our analysis establishes that, for localized networks, each component $\bm{{W}}_{\text{c}i}^{\infty}$ of the Gramian can be approximated accurately by solving Eq. \textbf{\ref{lyapeqatnode}} independently.

\subsection*{Driver Placement Algorithm Exploiting Locality}

We now consider the problem of optimally placing drivers on the network to maximize the smallest eigenvalue of the controllability Gramian given an allowed number of drivers $d_\text{max}$. The localized methods developed above to approximate the smallest eigenvalue (as validated in Fig.~\ref{localappeig}) and the Gramian itself (Fig.~\ref{WcWcest} and 
\textcolor{blue}{\emph{SI Table S1}}) can be combined to design a scalable algorithm for the driver placement problem. Here, we propose a gradient-based greedy algorithm (Algorithm 2 in \textcolor{blue}{\emph{Materials and Methods}}), which at each iteration seeks to add a driver node leading to the largest increase in $\widetilde{\lambda}_{\text{min}}(\widetilde{\bm{W}}_\text{c}^{\infty})$ and can be used to obtain a provably near-optimal solution. This is based on the fact that $\lambda_{\text{min}}(\bm{W}_{\text{c}}^{\infty})$ is a submodular function of the driver set \cite{summers2015submodularity}, meaning that the gain in $\lambda_{\text{min}}(\bm{W}_{\text{c}}^{\infty})$ from adding a driver is larger when the original driver set is smaller. Our algorithm far outperforms random placement while requiring computation time that scales only sub-quadratically with the network size, as demonstrated for model networks in Fig.~\ref{driverpalceperform}A--C. 
The advantage over random placement is substantial also for empirical networks, as observed in Fig.~\ref{driverpalceperform}D--F.

\section*{Localized Optimal Control}

\subsection*{Locality of the Optimal Responses}

We can now proceed to explore network locality in the optimal control problem in which we seek a control strategy that achieves the best trade-off between dynamical performance and control effort. The problem is mathematically formulated as
\begin{equation}\label{lqr}
    \begin{aligned}
      \underset{\bm{u} \in L_2[0,+\infty)}{\text{min}}&\ J=\int_0^\infty \bm{x}(\tau)^{T}\bm{Q}\bm{x}(\tau)+\bm{u}(\tau)^{T}\bm{R}\bm{u}(\tau) d\tau\\
      &  \text{s.t. } \dot{\bm{x}}=\bm{C}\bm{x}+\bm{B}\bm{u},\ \bm{x}(0)=\bm{x}_0,
    \end{aligned}
\end{equation}
whose objective $J$ is an integral quadratic functional with positive definite weighting matrices $\bm{Q}$ and $\bm{R}$ for the node states and control inputs, respectively. The control task in this formulation is to drive the system state towards the origin, which does not involve loss of generality since many practical problems with non-trivial target states, such as equilibrium stabilization, trajectory tracking, and command following, can be cast in this form (\textcolor{blue}{\emph{Materials and Methods}}). \reviseTen{The global optimal control strategy takes the form of a state feedback:}
\reviseTen{
\begin{equation}\label{optfeedback}
   \bm{u}(t)=\bm{K}\bm{x}(t)=-\bm{R}^{-1}\bm{B}^{T}\bm{P}\bm{x}(t),
\end{equation}
}
where $\bm{P}$ is the stabilizing solution of the Riccati equation.

We now show that, if the network system is localized, locality is preserved in the optimal responses and the computation needed to approximate the optimal feedback law can be performed locally and in parallel at different driver nodes. Our arguments are based on the theory of \emph{System-level synthesis} \cite{wang2016system,anderson2019system}. Based on this theory, the time-domain problem in Eq.~\textbf{\ref{lqr}} can be Laplace-transformed and decomposed into $N$ independent problems in the complex $s$-domain given by
\begin{equation} \label{slslqr2}
\begin{aligned}
&\underset{\bm{\phi}_j, \bm{h}_j \in \frac{1}{s}\mathcal{RH}_{\infty} }{\text{min}} \ 
\norm{\begin{bmatrix} \bm{Q}^{1/2} & \\ & \bm{R}^{1/2}\end{bmatrix} \begin{bmatrix}\bm{\phi}_j(s) \\\bm{h}_j(s)  \end{bmatrix}
}_{\mathcal{H}_2}^2\\
&\qquad \qquad  \text{s.t.}
\begin{array}{l}
\begin{bmatrix}
            s\bm{I}-\bm{C} & -\bm{B}
        \end{bmatrix}
       \begin{bmatrix}
            \bm{\phi}_j(s) \\ \bm{h}_j(s)
        \end{bmatrix}=\bm{e}_j.
\end{array}
\end{aligned}
\end{equation}
for $j=1,\ldots,N$.
For each $j$, this optimization problem seeks the optimal response of the system {for} {the initial condition $\bm{x}_0 = \bm{e}_j$, which is fully concentrated on node $j$}. In particular, $\bm{h}_j(s)$ and $\bm{\phi}_j(s)$ represent the transfer functions for the optimal control $\bm{u}(t)$ and the corresponding optimal state response $\bm{x}(t)$, respectively. The solution of the problem in Eq.~\textbf{\ref{slslqr2}} is given by $\bm{\phi}_j(s)=\bm{\Phi}(s)\bm{e}_j$ and $\bm{h}_j(s)=\bm{H}(s)\bm{e}_j$, where $\bm{\Phi}(s)=(s\bm{I}-\bm{C}+\bm{B}\bm{R}^{-1}\bm{B}^T\bm{P})^{-1}$, $\bm{H}(s)=-\bm{R}^{-1}\bm{B}^T\bm{P}(s\bm{I}-\bm{C}+\bm{B}\bm{R}^{-1}\bm{B}^T\bm{P})^{-1}$, $\bm{I}$ is the identity matrix, and $\bm{P}$ is the solution to the Riccati equation (see \textcolor{blue}{\emph{SI Text 6}} for details). It has been proved in \cite{curtain2011riccati} that, if the matrices $\bm{C}$, $\bm{B}\bm{R}^{-1}\bm{B}^T$, and $\bm{Q}$ all belong to the Banach algebra $\mathcal{L}_{v,\rho}$, then the solution $\bm{P}$ is also localized and belongs to $\mathcal{L}_{v,\rho}$. This implies that the optimal feedback matrix $\bm{K}=-\bm{R}^{-1}\bm{B}^{T}\bm{P}$ exhibits off-diagonal decay and is concentrated on small information neighborhoods of the driver nodes.
In addition, since $\mathcal{L}_{v,\rho}$ is closed under matrix addition, multiplication, and inversion, the coefficient matrices $\bm{\Phi}(s)$ and $\bm{B}\bm{H}(s)$ both belong to $\mathcal{L}_{v,\rho}$. Furthermore, $\bm{\phi}_j(s)$ and $\bm{B}\bm{h}_j(s)$ are the $j$th column blocks of $\bm{\Phi}(s)$ and $\bm{B}\bm{H}(s)$, respectively, and hence by the definition of $\mathcal{L}_{v,\rho}$ there exist constant $\kappa_1$ and $\kappa_2$ such that $\norm{[\bm{\phi}_j(s)]_i} \leq \kappa_1 \cdot v(\rho({i,j}))^{-1}$ and $\norm{[\bm{B}\bm{h}_j(s)]_i } \leq \kappa_2 \cdot v(\rho({i,j}))^{-1}$. That is, the magnitude of the $i$th elements of $\bm{\phi}_j(s)$ and $\bm{B}\bm{h}_j(s)$ decay at least at a near-exponential rate as the information distance between nodes $i$ and $j$ increases. This result has an explicit physical meaning: the optimal controller always seeks to confine the disturbance to the information neighborhood of the disturbance location, and the control action to achieve this is also concentrated within the information neighborhood. In other words, both the disturbance propagation and control intervention must be localized in order for the controller to be optimal.

\subsection*{Localized Control Design}

Considering the locality of system responses under optimal control, we approximate the problem in Eq.~\textbf{\ref{slslqr2}} by a reduced problem involving only the system data and decision variables within the information neighborhood $\mathcal{N}_j$ of the initial disturbance at node $j$, where we use $\mathcal{N}_j$ as a short for $\mathcal{N}_j(\tau)$. Let $\bm{T}_j$ be the projection matrix that maps the entire input space $\mathbb{R}^r$ to the input subspace $\mathbb{R}^{\abs{\mathcal{T}_j}}$ associated with the neighborhood $\mathcal{N}_j$, where $ \mathcal{T}_j =\{ 1\leq k \leq r\ |\ [\bm{B}]_{ik}\neq 0 \text{ for some }  i \in  \mathcal{N}_j  \}$. Using $\bm{T}_j$ along with the projection matrix $\bm{N}_j$ defined earlier for the state space, we let $\widetilde{\bm{C}}_j=\bm{N}_j \bm{C} \bm{N}_j^{T}$, $\widetilde{\bm{B}}_j = \bm{N}_j \bm{B} {\bm{T}}_j^{T}$, $\widetilde{\bm{Q}}_j=\bm{N}_j \bm{Q} \bm{N}_j^{T}$, $\widetilde{\bm{R}}_j={\bm{T}}_j \bm{R} {\bm{T}}_j^{T}$, and $\widetilde{\bm{e}}_j = \bm{N}_j {\bm{e}}_j$. Eq.~\textbf{\ref{slslqr2}} can then be rewritten in terms of $\widetilde{\bm{C}}$, $\widetilde{\bm{B}}$, $\widetilde{\bm{Q}}$, $\widetilde{\bm{R}}$, and $\widetilde{\bm{e}}_j$ to obtain a projected version of the problem, whose solution $(\widetilde{\bm{\phi}}_j(s),\widetilde{\bm{h}}_j(s))$ is given by $\widetilde{\bm{\phi}}_j(s)=(s\bm{I}-\widetilde{\bm{C}}_j+\widetilde{\bm{B}}_j\widetilde{\bm{R}}_j^{-1}\widetilde{\bm{B}}_j^{T}\widetilde{\bm{P}}_j)^{-1} \widetilde{\bm{e}}_j$ and $\widetilde{\bm{h}}_j(s)=-\widetilde{\bm{R}}_j^{-1}\widetilde{\bm{B}}_j^{T}\widetilde{\bm{P}}_j\widetilde{\bm{\phi}}_j(s)$, where $\widetilde{\bm{P}}_j$ is the solution of the projected Riccati equation $\widetilde{\bm{C}}_j^{T}\widetilde{\bm{P}}_j+\widetilde{\bm{P}}_j\widetilde{\bm{C}}_j-\widetilde{\bm{P}}_j\widetilde{\bm{B}}_j\widetilde{\bm{R}}_j^{-1}\widetilde{\bm{B}}_j^{T}\widetilde{\bm{P}}_j+\widetilde{\bm{Q}}_j=\bm{0}$. Once this is solved for all $j$, we can construct the full optimal control law as $\bm{u}(s) = \widetilde{\bm{K}}(s)\bm{x}(s)=\widetilde{\bm{H}}(s) \widetilde{\bm{\Phi}}(s)^{-1}\bm{x}(s)$, where $\widetilde{\bm{\Phi}}(s)$ and $\widetilde{\bm{H}}(s)$ are the concatenations of $\bm{N}_{j}^T \widetilde{\bm{\phi}}_j(s)$ and $\bm{T}_{j}^T \widetilde{\bm{h}}_j(s)$, respectively. We expect the solution $(\widetilde{\bm{\phi}}_j(s),\widetilde{\bm{h}}_j(s))$ of the reduced problem to approximate $({\bm{\phi}}_j(s),\bm{h}_j(s))$ well if the size of the information neighborhood $\mathcal{N}_j$ is not too small. We show that controllers designed using these projected models do enjoy stability and a near-optimality guarantee when implemented on the actual original system in Eq. \textbf{\ref{consys}}. We refer to this formulation as \emph{disturbance-oriented localization}, since it is based on the decomposition of the optimal control problem into $N$ independent problems given in Eq.~\textbf{\ref{slslqr2}}, each localized around the perturbed node {(see \textcolor{blue}{\emph{SI Text 7}} for details).}

To respond optimally to disturbances, a driver at node $i$ must react to perturbations at all nodes belonging to its \emph{control neighborhood} $\mathcal{C}_i := \{ 1\leq j\leq N | \  i\in \mathcal{N}_j\}$, i.e., the set of nodes whose information neighborhoods contain node $i$. Although the optimal control law from each sub-problem is a static state feedback $\widetilde{\bm{K}}_j$ (Eq.~\textbf{S36} in \textcolor{blue}{\emph{SI Text 7}}), the aggregate controller $\widetilde{\bm{K}}(s)$ may not be static because two different sub-problems may ask for different feedback gains from the same state-driver pair (Eq.~\textbf{S42} in \textcolor{blue}{\emph{SI Text 7}}). To resolve this issue, we take a driver-centric viewpoint and design a control law for the driver at node $i$ by projecting the original problem onto the information neighborhood of $\mathcal{C}_i$, which we call the \emph{controller-oriented localization} (see \textcolor{blue}{\emph{SI Text 8}} for details). This approach leads to a fully decentralized method to design localized near-optimal static controllers (Algorithm 3 in \textcolor{blue}{\emph{Materials and Methods}}), in which each driver only needs feedback signals from its control neighborhood.

\begin{figure}[!t]
\centering
     \begin{overpic}[width=0.33\textwidth,tics=5]{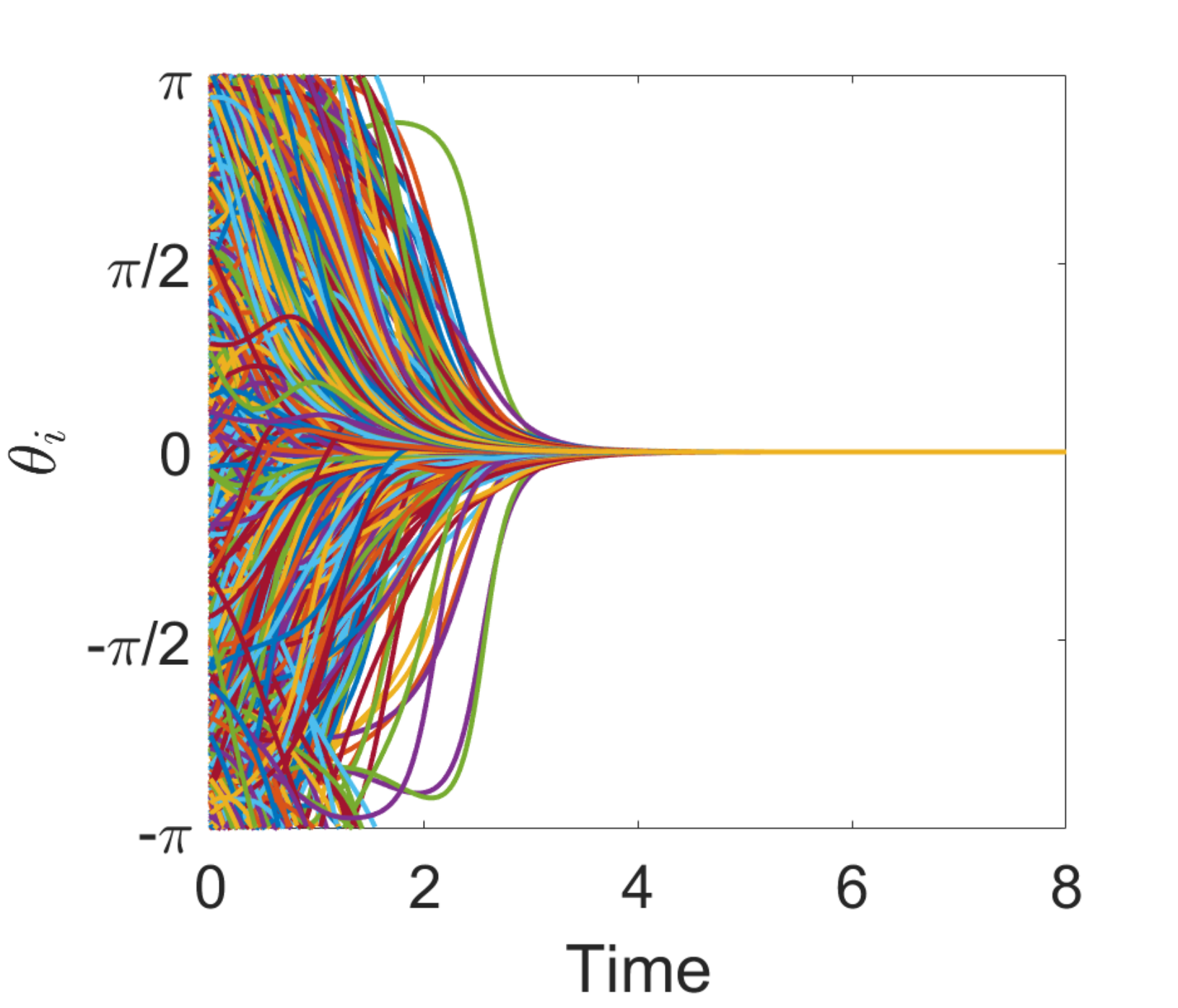}
\put (0,78) {\textbf{A}}
\end{overpic}
      \begin{overpic}[width=0.33\textwidth,tics=5]{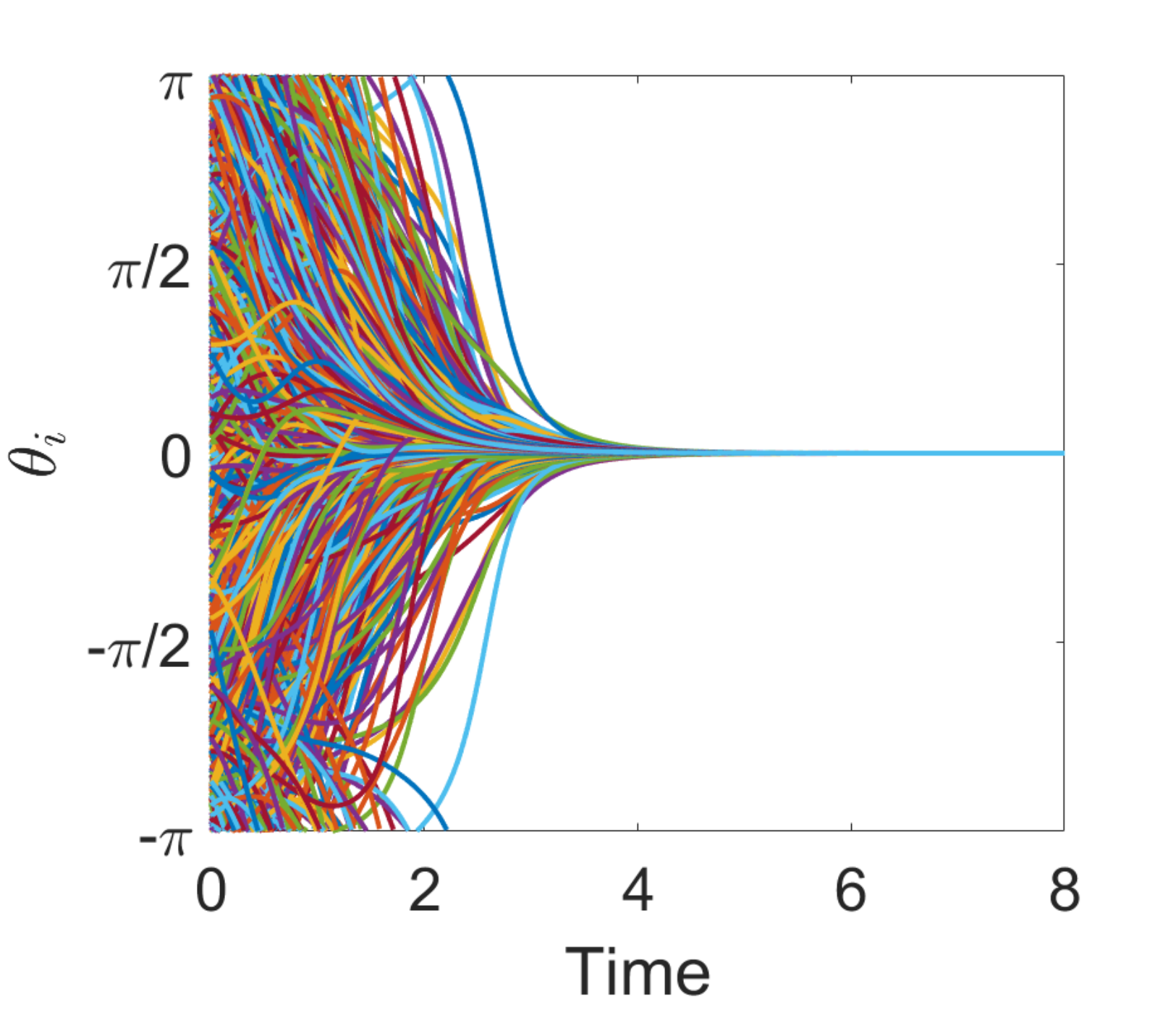}
\put (0,78) {\textbf{B}}
\end{overpic}
      \begin{overpic}[width=0.33\textwidth,tics=5]{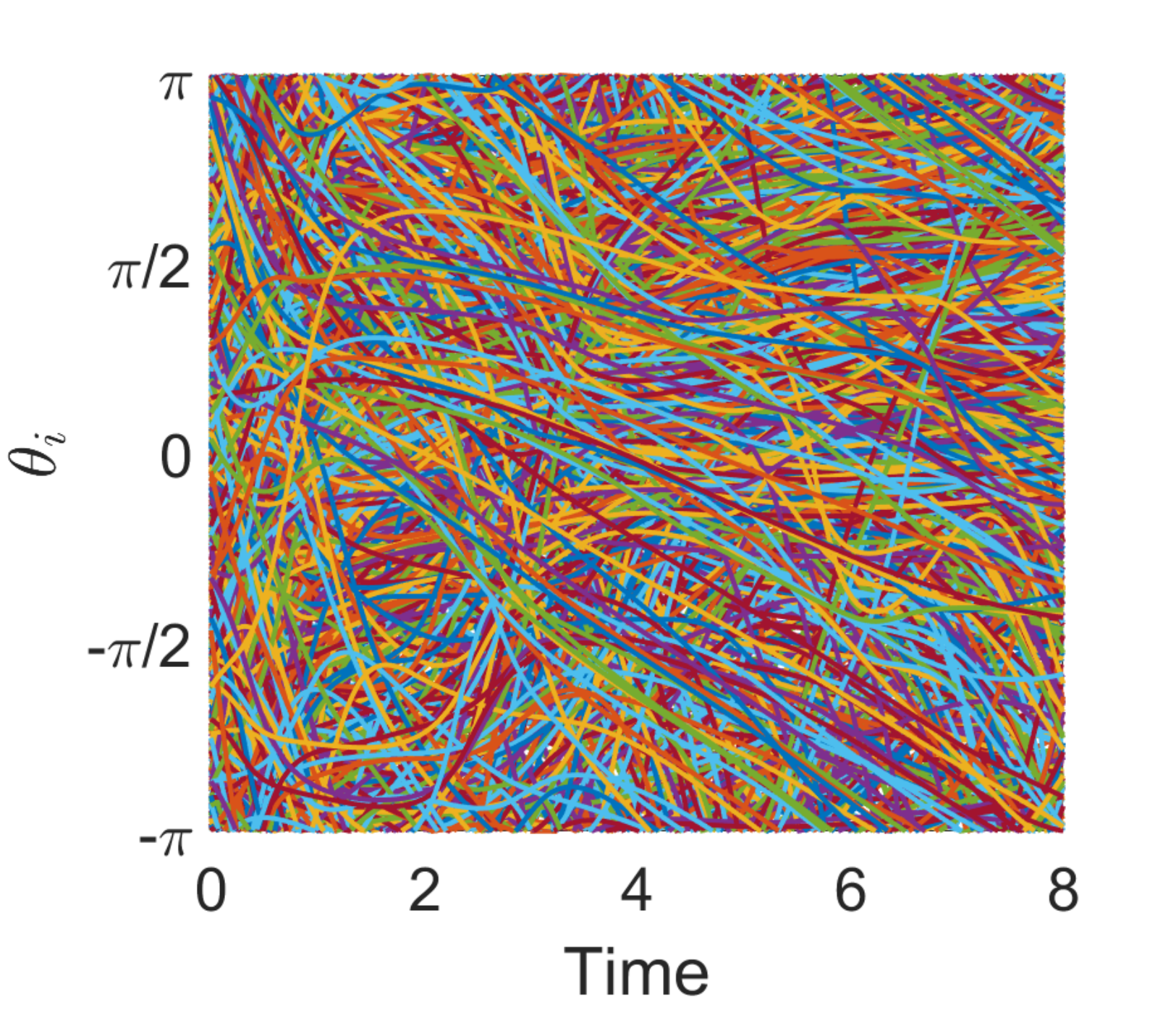}
\put (0,78) {\textbf{C}}
\end{overpic}
      \begin{overpic}[width=0.33\textwidth,tics=5]{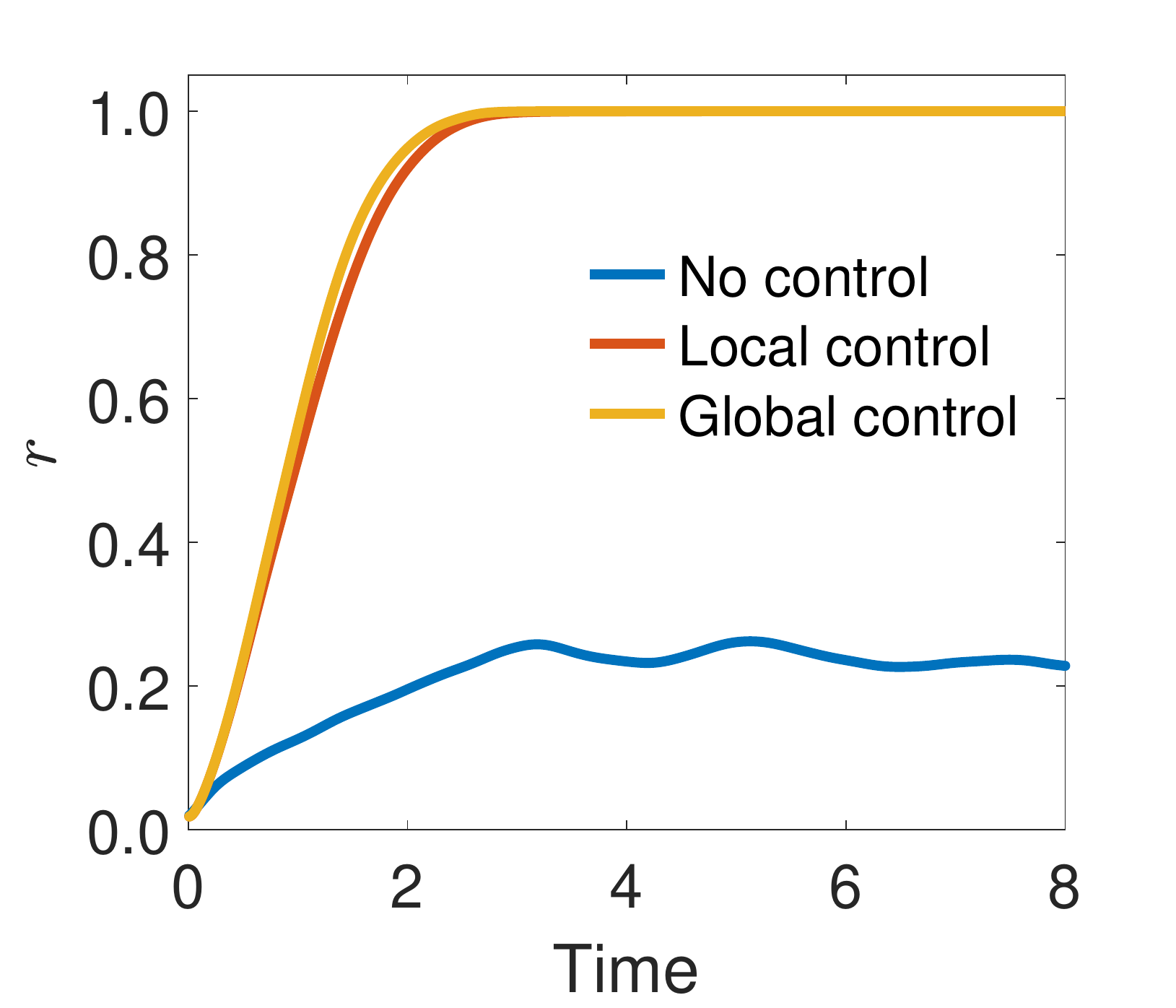}
\put (0,78) {\textbf{D}}
\end{overpic}
\\
\begin{overpic}[width=0.22\textwidth,tics=5]{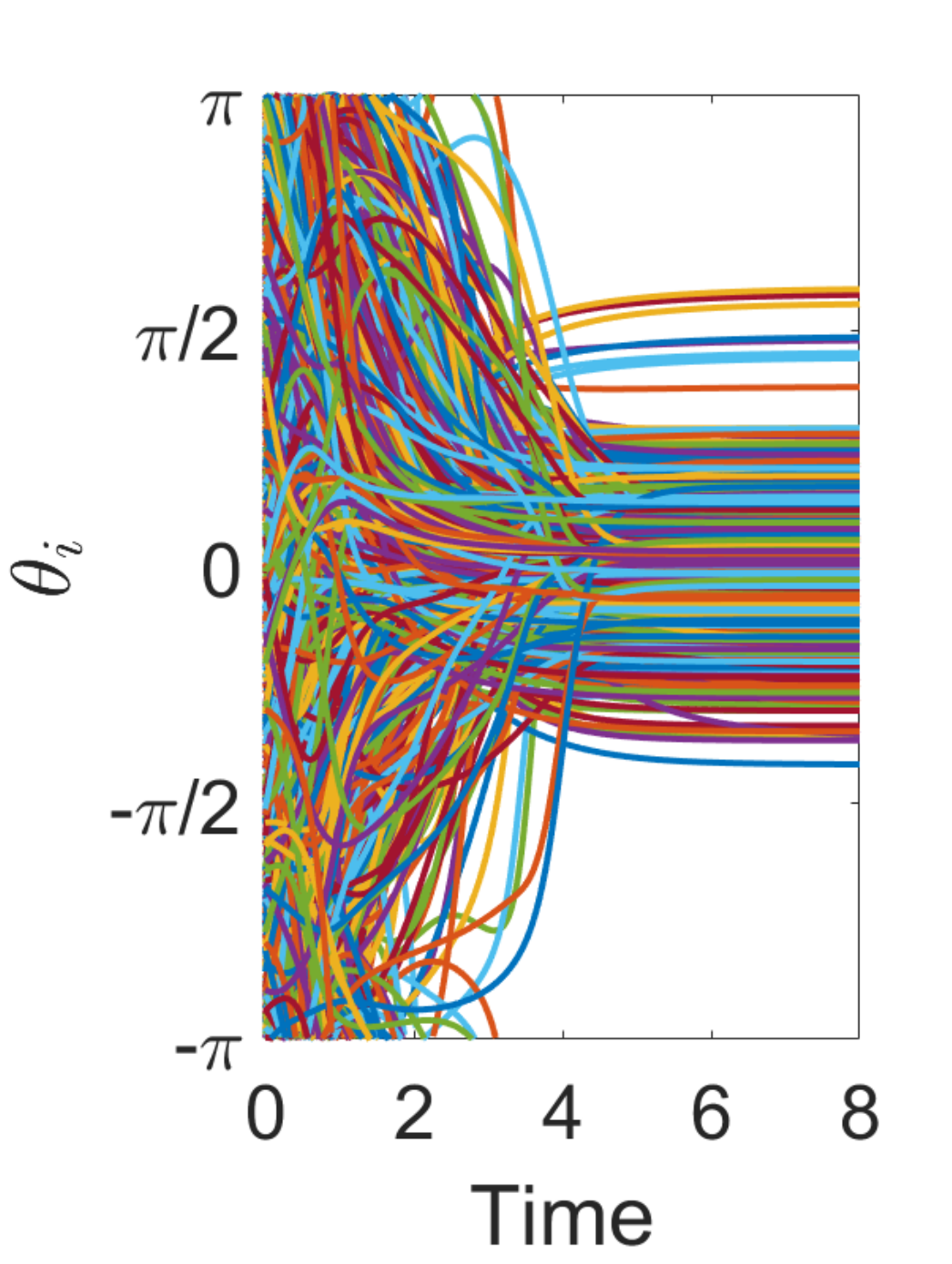}
\put (0,90) {\textbf{E}}
\end{overpic}
      \begin{overpic}[width=0.22\textwidth,tics=5]{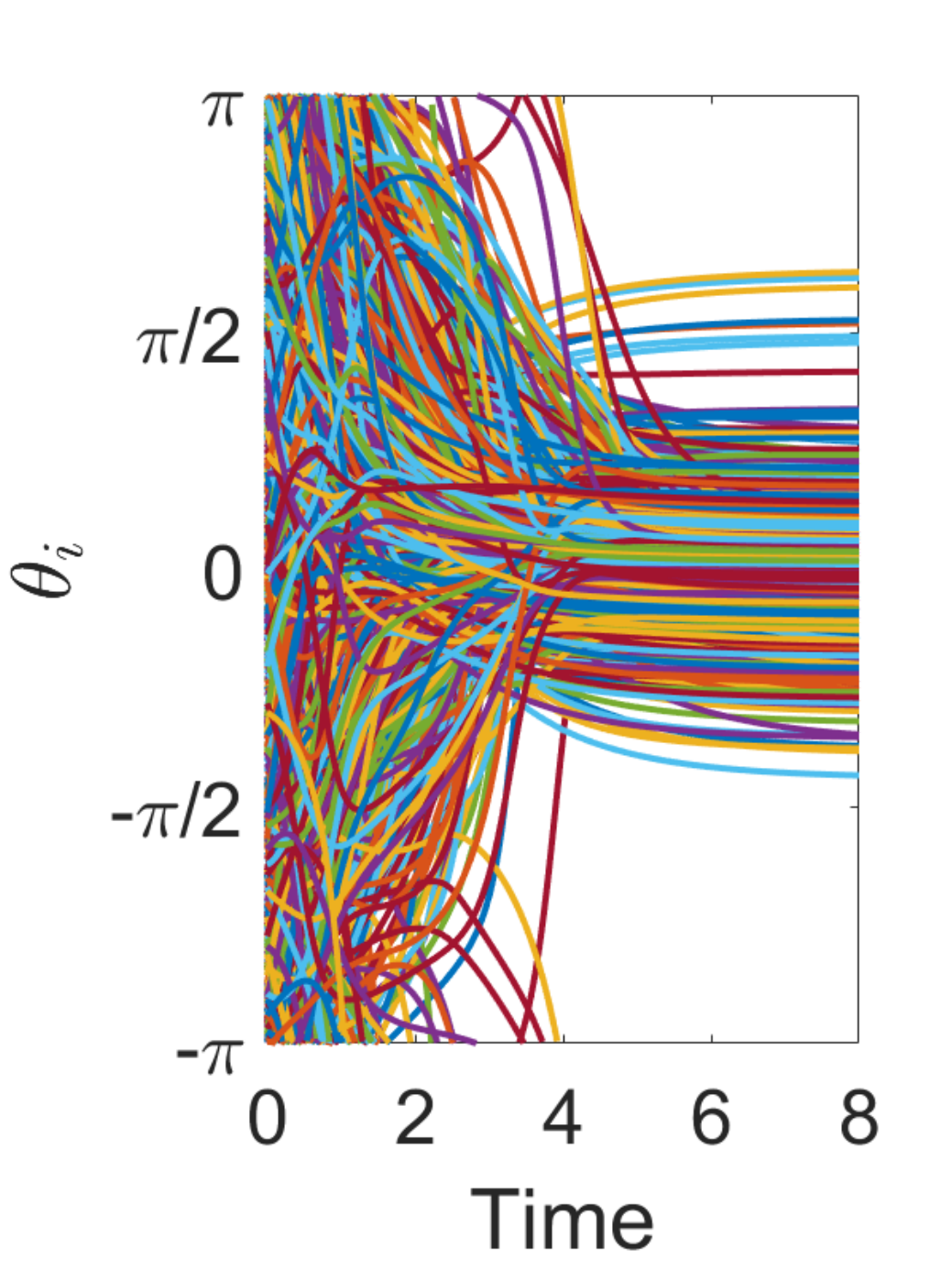}
\put (0,90) {\textbf{F}}
\end{overpic}
      \begin{overpic}[width=0.22\textwidth,tics=5]{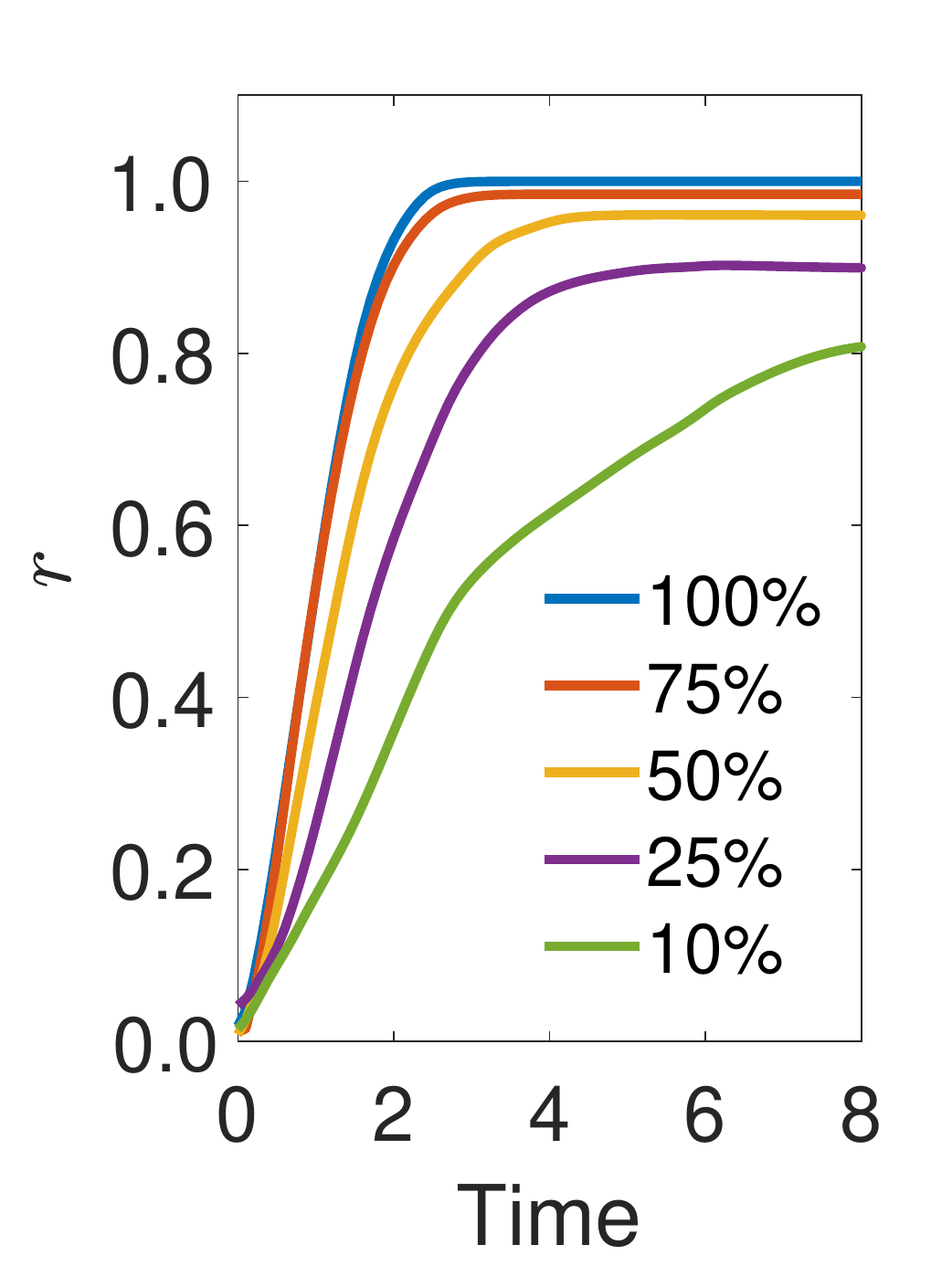}
\put (0,90) {\textbf{G}}
\end{overpic}
     \caption{Synchronization control of coupled Kuramoto oscillators. (A)--(C) Simulations of the oscillator network under global control (A), local control (B), and no control (C), with $100\%$ of the nodes controlled directly. The oscillators are coupled by a (weighted) WS network with $N=1000$, $\bar{d}=20$, and $p=0.1$. The natural frequencies and the initial phases of the oscillators are both sampled uniformly from the interval $(-\pi,\pi)$. The control problem is to track a frequency-synchronized trajectory whose common frequency $\omega^*$ is the average of the natural frequencies over all nodes. (D) Evolution of the order parameter $r=\frac{1}{N}\sum_{i=1}^N e^{{\text{j}}\theta_i}$ in (A)--(C). (E), (F) Phase trajectories under global (E) and local (F) control with only $50\%$ of nodes randomly selected and directly controlled. When only part of the nodes are directly controlled, $\omega^*$ is set to be the average of the natural frequencies over the uncontrolled nodes. (G)~Evolution of the order parameter when different fractions of nodes are directly controlled using the localized method, where each curve corresponds to $100$ realizations for randomly selected nodes. The information neighborhood size used is $L = 10$.}
     \label{Kurasim}
    
\end{figure}

\begin{figure*}[t]
\centering
 \begin{overpic}[width=0.98\textwidth,tics=2]{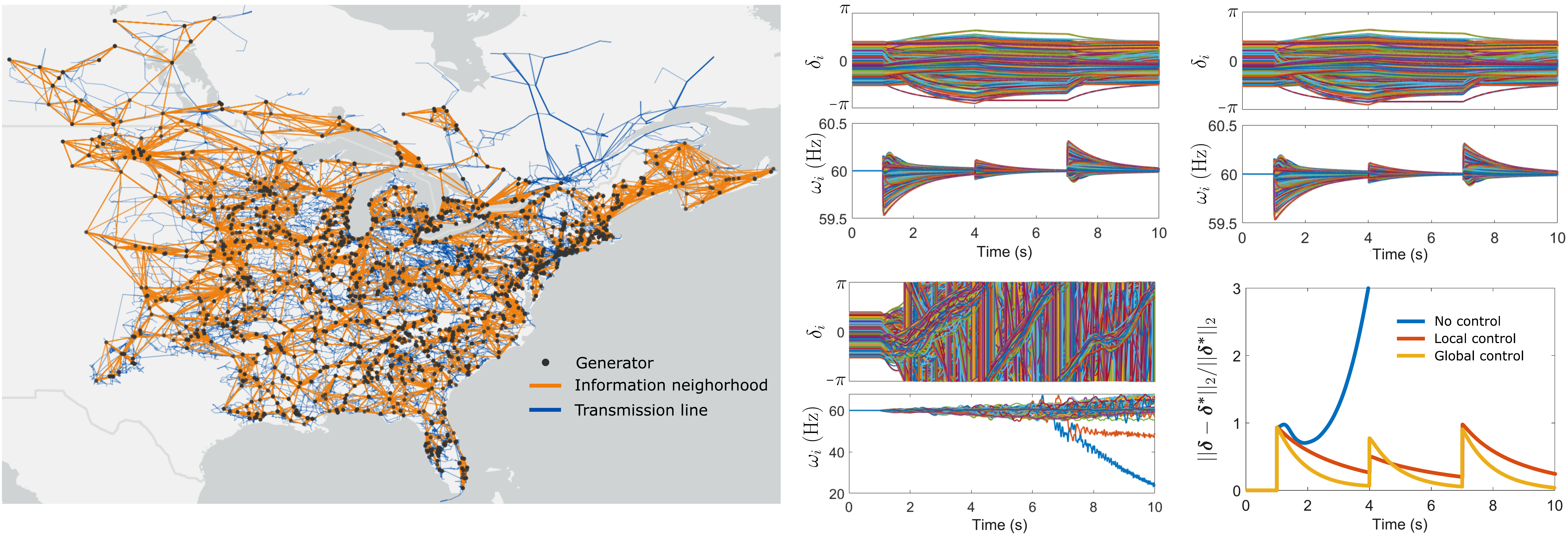}
\put (2,34.5) {\textbf{A}}
\put (51.5,34.5) {\textbf{B}}
\put (76.5,34.5) {\textbf{C}}
\put (51.5,17) {\textbf{D}}
\put (76.5,17) {\textbf{E}}
\end{overpic}

\caption{Stability control of the Eastern U.S.\ power grid. (A) Physical topology of the power grid along with its information neighborhood network for $L=10$, constructed by creating a link between nodes $i$ and $j$ if node $j$ is among the first $L$ information neighbors of node $i$. A black dot represents one generator or a set of co-located generators. To assess the control performance, we simulate a scenario in which the system suddenly loses $80\%$ of the renewable generation (accounting for $30\%$ of the total load) at $t=1$~s, recovers to $50\%$ of the original level of renewables at $t=4$ s, and then fully recovers at $t=7$ s. (B)--(D) Transient responses under global control (B), local control (C), and no control (D). (E) Relative distances to the target power angles (obtained from the post-contingency power flow solution) under the three control scenarios in (B)--(D). In (C) and (E), the local control is for the same information neighborhood network as in (A). For comparison purpose, the frequency and angle deviations under no control are shown beyond what power system operation allows without actually triggering protection actions.}
\label{powergrid}

\end{figure*}

\subsection*{Applications to Nonlinear Dynamical Networks}

\reviseTen{Our local control approach is applicable to nonlinear networks in general, which follows by employing suitable linearization methods in the control design. This significantly extends the scope of our theory since most real networks are nonlinear.}
The performance in nonlinear networks will depend on the control task and linearization method used. We present the general solutions to three control tasks---equilibrium stabilization, trajectory tracking, and command following---using two linearization methods, specifically the Jacobian linearization and the extended linearization (see \textcolor{blue}{\emph{SI Text 9}} for details). 
We demonstrate the effectiveness of the proposed localized control design through four concrete applications, namely the synchronization control of Kuramoto oscillators, the stability control of the Eastern U.S.\ power grid, the mitigation of epidemic spreading through the global air transportation network, and the control of pathological brain network dynamics for managing Alzheimer's disease. 
\reviseTen{
The formulation of the problems and control methods are presented in \textcolor{blue}{\emph{Materials and Methods}}, with the data sources given in \textcolor{blue}{\emph{SI Text~2}}.
}

In the synchronization control of coupled Kuramoto oscillators, complete phase synchronization is achieved when all nodes are controlled by the global or the local method, while the synchrony is lost when they are not controlled (Fig.~\ref{Kurasim}A--D). When $50\%$ of the nodes are directly controlled, only frequency synchronization can be achieved by both global (Fig.~\ref{Kurasim}E) and local (Fig.~\ref{Kurasim}F) control. The higher the fraction of nodes directly controlled, the higher is the phase coherence that can be achieved in the frequency-synchronized orbit (Fig.~\ref{Kurasim}G). However, regardless of the fraction of driver nodes, the local control performs similarly to the global control. For the stability control of the Eastern U.S.\ power grid, we visualize in Fig.~\ref{powergrid}A the information neighborhood network among generators for $L=10$ on top of the physical network topology. The figure shows a stark contrast between the information and physical topology of the network. When the system is disturbed by intermittent renewable generation, both global (Fig.~\ref{powergrid}B) and local (Fig.~\ref{powergrid}C) methods are effective to control the system toward the target equilibrium points, while the system would lose stability in the absence of control (Fig.~\ref{powergrid}D). 

\begin{figure}[h]
    \centering
    \begin{overpic}[width=0.66\textwidth,tics=5]{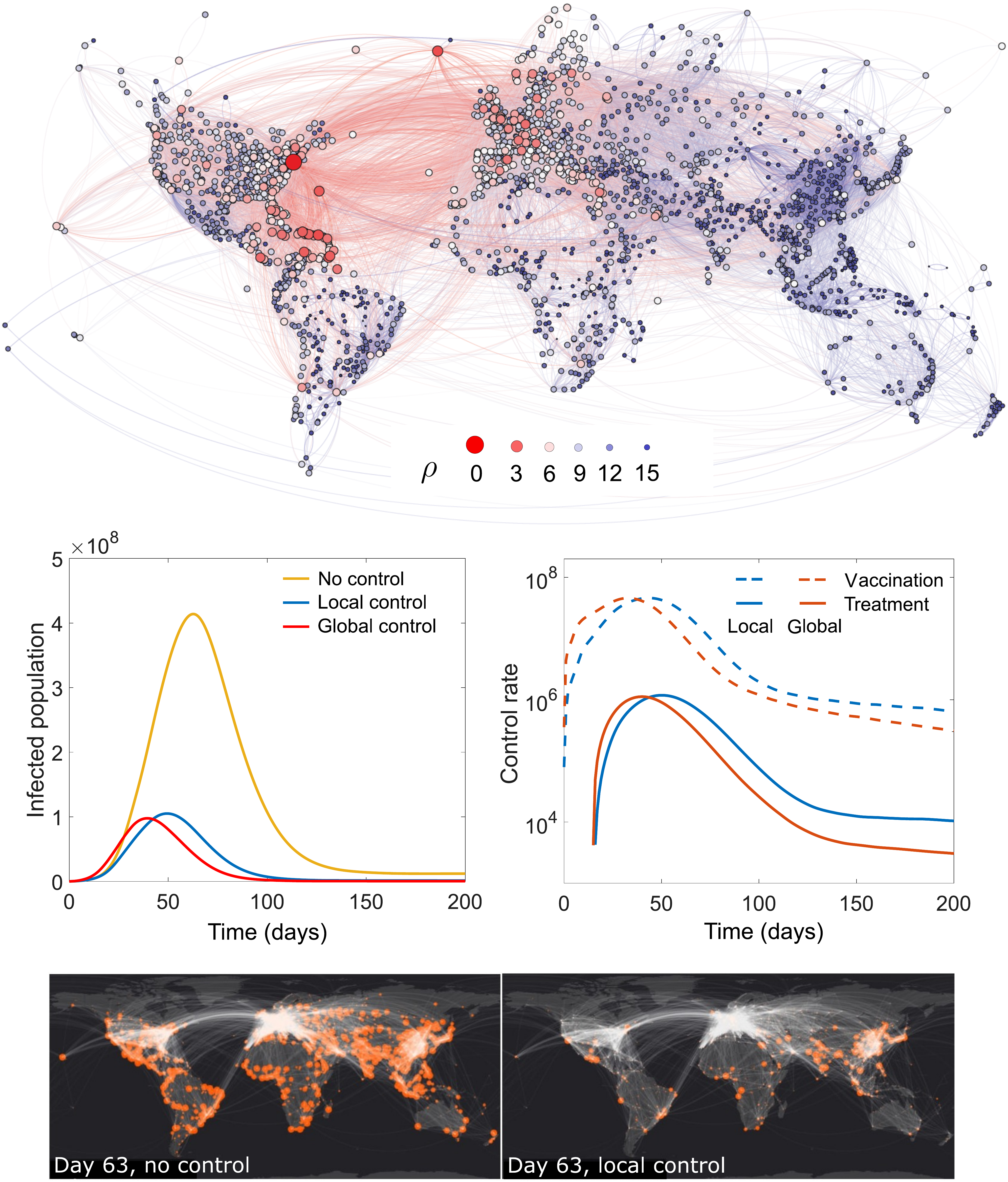}
\put (1,95) {\textbf{A}}
\put (0,53) {\textbf{B}}
\put (42,53) {\textbf{C}}
\put (0,19) {\textbf{D}}
\end{overpic}
         \caption{Controlling epidemic spreading through the global air transportation network. (A)~Network of $2219$ major cities connected by $59151$ edges representing the flow of passengers between the cities. The warmer color and larger size of the circles represent shorter information distance $\rho$ to New York (largest red circle). (B) Total infected population worldwide as a function of time from the onset of the spreading under no control, local control, and global control. (C) Computed control actions (the number of people treated and vaccinated per day) under local control (blue curves) and global control (red curves). (D) Distribution of the infected population around the world (indicated by the sizes of the orange dots) on day $63$ of the outbreak under no control (left) and local control (right). The information neighborhood size used is $L = 20$ for (B), (C), and (D).}
         \label{epidemic}
\end{figure}

\begin{figure}[h]
\centering
    \begin{overpic}[width=0.66\textwidth,tics=5]{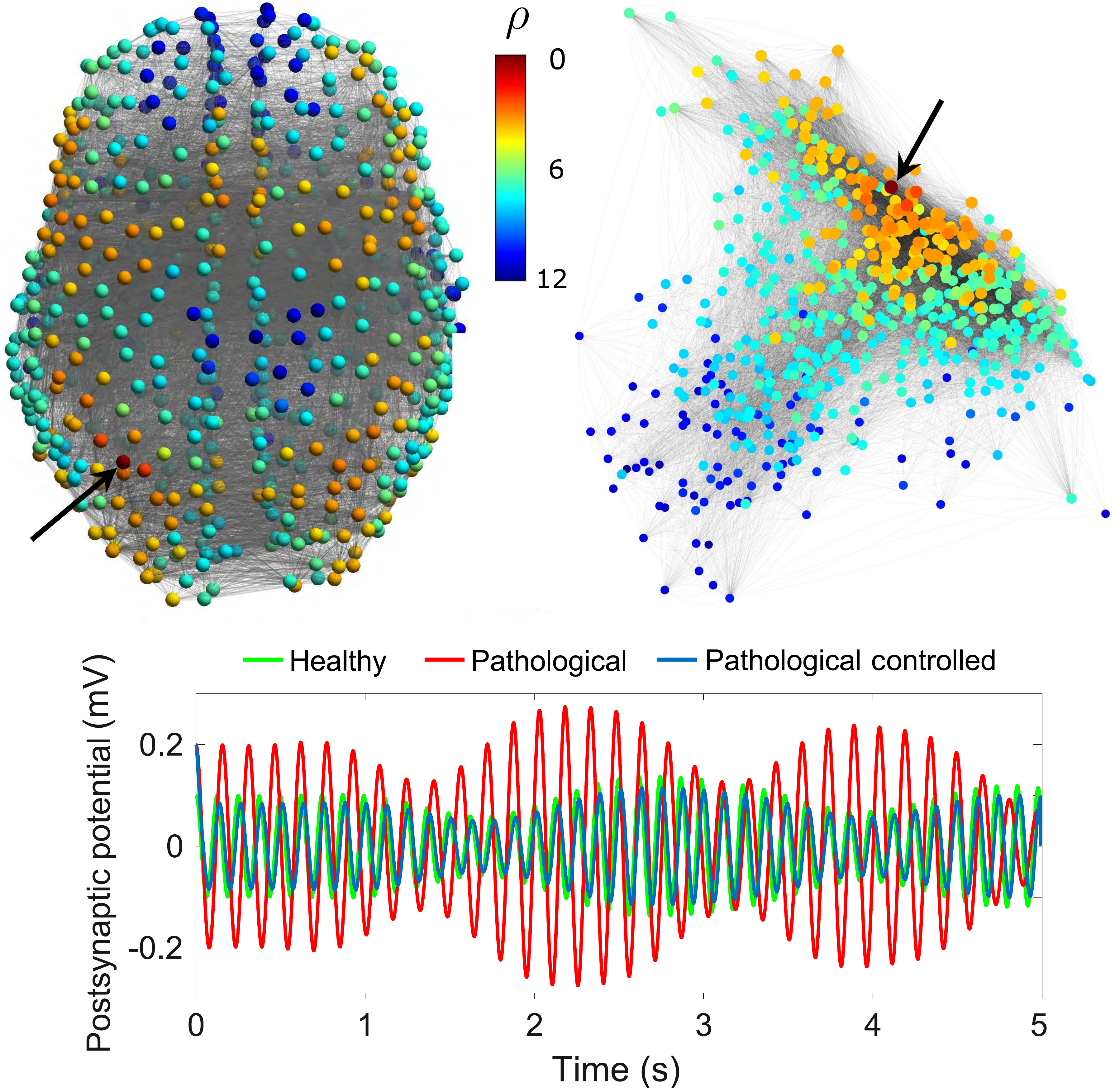}
\put (2,87) {\textbf{A}}
\put (76,87) {\textbf{B}}
\put (2,36) {\textbf{C}}
\end{overpic}
\caption{Controlling a whole-brain network by electrical stimulation. (A) Physical layout of the network, with $638$ nodes representing predefined regions of a human brain and $18625$ edges representing co-activations of pairs of different areas. The node colors represent the information distances $\rho$ to the reference node indicated by the arrow. Warmer colors and larger circles indicate shorter information distances. (B)~Flattened layout of the same network
for the same node color scale. 
(C) Trajectory of the reference node's state for the network under a healthy condition and a pathological condition with and without the local control implemented. The trajectories for the other nodes are similar. The information neighborhood size used is $L=10$.}
\label{brainnet}

\end{figure}

In our application to epidemic control, we visualize in Fig.~\ref{epidemic}A the information distances between New York City and all other major cities on top of the global air transportation network. The local and global methods generate vaccination and treatment strategies that result in similar curves of infected population and they are comparably effective in suppressing the outbreak, as shown in Fig.~\ref{epidemic}B--D. In the application to brain network control, we also visualize the information distances between one particular node and all other nodes of the brain co-activation network (Fig.~\ref{brainnet}A--B). As shown in Fig.~\ref{brainnet}C, by applying the brain stimulation strategy generated by the local control method, the electrical activity in a brain under a pathological condition is led to closely follow the activity observed under healthy conditions. Thus, within this model, local interventions are predicted to 
alleviate the symptoms of Alzheimer's disease.

\begin{figure}[h]
    \centering
\begin{overpic}[width=0.9\textwidth,tics=5]{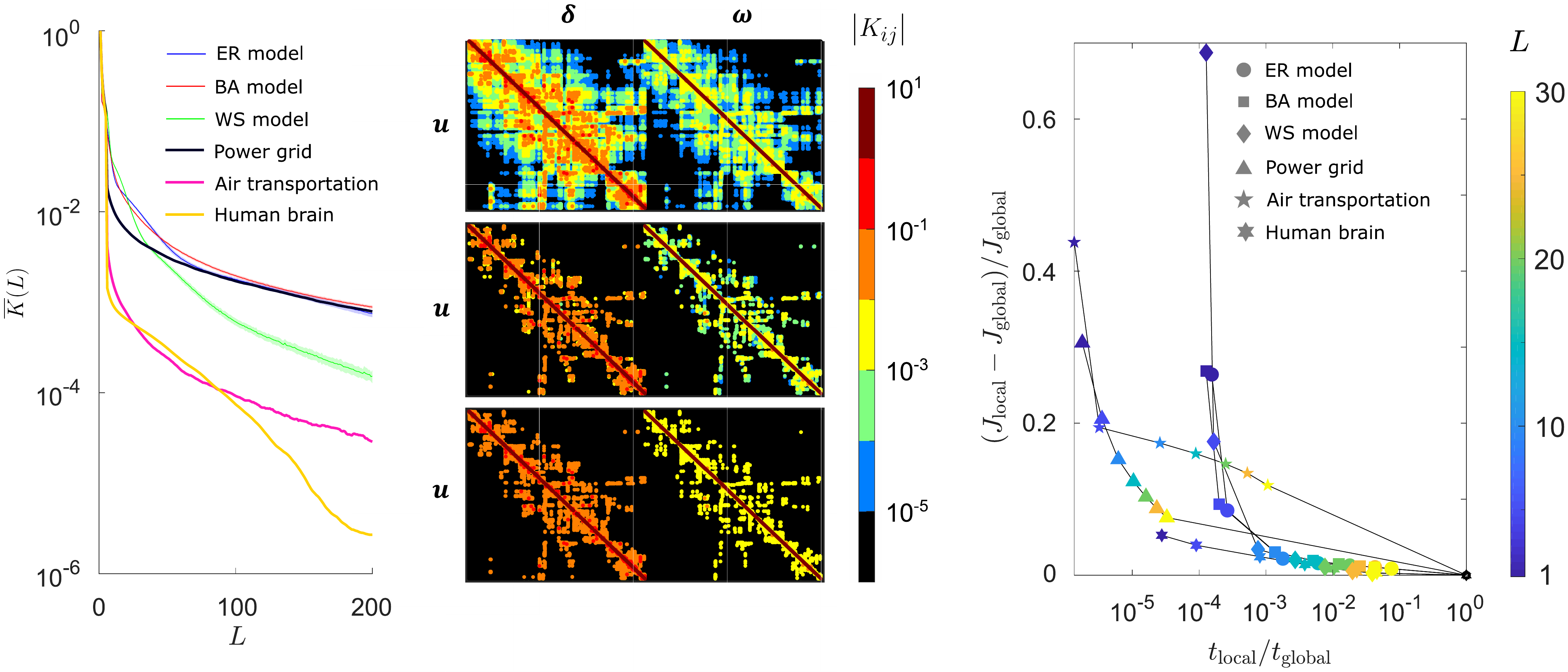}
\put (0,40) {\textbf{A}}
\put (27,40) {\textbf{B}}
 \put (64,40) {\textbf{C}}
\end{overpic}

\caption{Locality of optimal responses and performance of localized control design. (A) Off-diagonal decay of the globally designed optimal feedback matrix $\bm{K}$ for the model networks in Fig.~\ref{localappeig} and empirical networks in Fig.~\ref{driverpalceperform}D--F. Here, $\overline{K}(L)=\frac{1}{N}\sum_{i=1}^N\sum_{j\in \widehat{\mathcal{N}}_i(L)/\widehat{\mathcal{N}}_i(L-1)}\abs{K_{ij}}/\abs{K_{ii}}$ is the average relative magnitude of the matrix elements corresponding to the $L$th information neighbor.
For the model networks (ER, BA, and WS), the magnitude is further averaged over $100$ network realizations, and the shadings around the curves indicate one standard deviation. (B) Exact optimal feedback matrix (top row), optimal feedback matrix designed by our localized method (middle row), and thresholded version of the exact optimal feedback matrix (bottom row) for the Eastern U.S.\ power grid, showing a close match between the middle and bottom rows. 
The thresholding removes all elements with magnitude $<10^{-2}$ for the feedback from $\bm{\delta}$ to $\bm{u}$ and $<10^{-3}$ for the feedback from $\bm{\omega}$ to $\bm{u}$. (C) Control performance vs.\ computational time for the networks in (A) of Algorithm 3 for the local optimal control color coded by the size $L$ of the information neighborhood used, where the subscripts distinguish the localized and global design quantities. For the localized design, the computational time is the average time it takes for one driver to determine its localized optimal control law. For model networks, the quantities on the vertical and horizontal axes are averaged over $100$ network realizations. For the model networks in (A) and (C), we used the coupled Kuramoto oscillator dynamics as described in \textcolor{blue}{\emph{Materials and Methods}}. For all networks, a driver is placed at each node.}
         \label{kdecay}
       
\end{figure}

As evidenced in Figs.~\ref{powergrid}A, \ref{epidemic}A, and \ref{brainnet}A, proximity in the network-topological and geographical/physical distance does not necessarily imply proximity in information distance. As already noted in Fig.~\ref{smallworldfig}, this indicates that the information distance captures quantitative features of direct and indirect interactions beyond what is captured by commonly used network representations. \reviseTen{Fig.~\ref{kdecay}A verifies the off-diagonal decay in $\bm{K}$ for all application examples. Fig.~\ref{kdecay}B visually shows for the Eastern U.S.\ power grid that the localized feedback matrix obtained with Algorithm 3 closely matches the exact optimal feedback matrix. We find that the localized controllers can achieve performance levels close to those of global controllers with relatively small information neighborhoods and orders-of-magnitude less computational time, as illustrated in Fig.~\ref{kdecay}C for both model and empirical networks.}
Our application results show that, despite the diversity of systems and tasks, the local control drives the system towards the same state (albeit with slightly different transients) as the global optimal control and achieves the control objective with near-optimal dynamical performance.

\section*{Discussion}
In many real-world applications, the ability to implement a control method locally is not just an additional benefit---it is a necessity, for two reasons. First, it would be costly, if not impossible, to build the communication infrastructure that allows real-time all-to-all information exchange, as needed for global control. Second, the nonlinearity of the system would require the feedback strategy to be updated in real time, and hence the computation would have to be faster than the system dynamics being controlled. As a result, global control would be prohibitive in terms of both communication and computation requirements for large-scale nonlinear dynamical networks \reviseTen{such as those considered here. However, as our results show,} many empirical networks enjoy a high degree of locality, even when the network is densely connected. Crucially, for such networks, our results show that communication and computation limited to small information neighborhoods of the driver nodes are sufficient to generate near-globally optimal control actions. 

These results also suggest natural extensions to be explored in future research. In particular, based on the concept of target controllability \cite{morse1971output,gao2014target}, the analysis can be generalized to the control of a target subset of nodes (instead of the entire network) to establish that the target nodes can be controlled with small control effort only if they lie in a small information neighborhood of the set of driver nodes (\textcolor{blue}{\emph{SI Text 3}}). By the duality between controllability and observability \cite{zhou1996robust}, the analysis on full and target controllability Gramians also carry over to their observability counterparts. In all cases, an outstanding question for future research is: in addition to locality, are there other network properties that can further help control the system? For example, symmetries in the network may be inherited by the optimal control strategy and potentially simplify the analysis and design problems (regardless of the impact of network symmetries on controllability itself \cite{whalen2015observability}). More broadly, this study shows that it is promising to pursue \emph{structure-exploiting network control}, capitalizing on common network-specific properties (beyond purely topological ones) to develop improved control approaches that are effective, efficient, and broadly applicable to complex systems across diverse domains.

\newpage

\clearpage

\section*{Materials and Methods}

\subsection*{Algorithms}
The pseudocode for the three algorithms introduced above are shown in Algorithms 1--3: the UCS algorithm to construct information distances and information neighborhoods (Algorithm 1), the gradient-based greedy algorithm to solve the driver placement problem (Algorithm 2), and the local control design algorithm for optimal controllers (Algorithm 3). Our MATLAB implementation of these algorithms and four example applications are available at our GitHub repository \cite{githubrepo}. 
\begin{algorithm}[h]
{\small
        \caption{Uniform Cost Search for $\rho(i,\cdot)$}
        \begin{algorithmic}[1]
            \STATE Initialize $\mathcal{F}=\{i\}$, $\mathcal{H}=\emptyset$, $\rho(i,i)=0$, $\rho(i,j)=+\infty$ for all $j\neq i$.
            \WHILE{$\mathcal{F}\neq \emptyset $}
            \STATE $k =\text{arg }\underset{j\in \mathcal{F}}{\text{min}}\  \rho(i,j)$.
                \FOR{each node $p$ adjacent to node $k$ in $\widetilde{G}$}
                  \STATE $\rho(i,p):=\text{min}\{\rho(i,p),\rho(i,k)+\widetilde{\rho}_{kp}\}$.
                  \STATE If $p \not\in \mathcal{F}$ and $p \not\in \mathcal{H}$, $\mathcal{F}:=\mathcal{F} \cup \{p\}$.
                  \ENDFOR
                  \STATE $\mathcal{F}:=\mathcal{F} \setminus \{k\}$, $\mathcal{H}:=\mathcal{H} \cup \{k\}$.
            \ENDWHILE
             \STATE Output: $\rho(i,\cdot)$ and an ordered set $\mathcal{H}$ of information neighbors.
        \end{algorithmic}
        }
\end{algorithm}

\begin{algorithm}[!h]
{\small
        \caption{Gradient-Based Greedy Driver Placement}
        \begin{algorithmic}[1]
        \STATE Input the target for controllability measure $\lambda_{\text{min}}^*$.
        \STATE Input the maximum number of drivers $d_{\max}$.
            \STATE Initialize $\mathcal{X}=\mathcal{N}$, $\mathcal{D}=\emptyset$, $\widetilde{\bm{W}}_{\text{c}}=\bm{0}$, $ \widetilde{\lambda}_{\text{min}}=0$, $j=1$, $\bm{v}_i=\bm{N}_i\bm{e}_i$ for $i=1,2,\ldots,N$.
            \WHILE{$\widetilde{\lambda}_{\text{min}}<\lambda_{\text{min}}^*$ and $\abs{\mathcal{D}}<d_{\text{max}}$}
                  \FOR{$k\in \mathcal{X}$ such that $\mathcal{N}_k \cap \mathcal{N}_j \neq \emptyset$}
                  \STATE $g_k = \bm{v}_j^T \bm{N}_j \bm{N}_k^T\widetilde{\bm{W}}_{\text{c}k}\bm{N}_k \bm{N}_j^T \bm{v}_j$.
                  \ENDFOR
                  \STATE $i=\text{arg max}_{k\in \mathcal{N}} \  g_k$.
                 \STATE $\widetilde{\bm{W}}_{\text{c}}:=\widetilde{\bm{W}}_{\text{c}}+\bm{N}_i^T\widetilde{\bm{W}}_{\text{c}i}\bm{N}_i$.
                 \STATE $\mathcal{X}:=\mathcal{X}\setminus \{ i \}$, $\mathcal{D}:=\mathcal{D}\cup \{ i \}$.
                 \FOR{$k\in \mathcal{N}$}
                 \STATE $(\lambda_k,\bm{v}_k)$ is the smallest eigenvalue pair of $\bm{N}_k\widetilde{\bm{W}}_{\text{c}}\bm{N}_k^T$.
                 \ENDFOR
                 \STATE $\widetilde{\lambda}_{\text{min}}=\min_{k\in \mathcal{N}}\ \lambda_k$, $j=\text{arg min}_{k\in \mathcal{N}}\ \lambda_k$.
            \ENDWHILE
            \STATE Output: the driver node set $\mathcal{D}$.
        \end{algorithmic}
        }
\end{algorithm}

\begin{algorithm}[!h]
{\small
        \caption{Local Control Design}
        \begin{algorithmic}[1]
            \STATE Obtain the system data $\bm{C}$, $\bm{B}$, $\bm{Q}$, and $\bm{R}$.
            \STATE Choose an information neighborhood size $L$.
            \STATE Construct the neighborhood $\mathcal{N}_j$ of size $L$ for each node $j$.
            \FOR{$i=1,2,\ldots,N$} in parallel
            \STATE Construct control neighborhood $  \mathcal{C}_i = \{ 1\leq j\leq n \  | \  i\in \mathcal{N}_j\}$.
            \STATE Construct $  \mathcal{\widehat{N}}_i = \bigcup\limits_{j\in \mathcal{C}_i} \mathcal{N}_j$ and projection matrix $\widehat{\bm{{N}}}_i$.
            \STATE Construct $ \mathcal{\widehat{T}}_i=\{ 1\leq k \leq r\ |\ [\bm{B}]_{jk}\neq 0, \text{ for some }  j \in \mathcal{\widehat{N}}_i  \}$ and  projection matrix $\widehat{\bm{{T}}}_i$.
            \STATE Set $\widehat{\bm{C}}_i=\widehat{\bm{N}}_i \bm{C} \widehat{\bm{N}}_i^{T} $, $\widehat{\bm{B}}_i = \widehat{\bm{N}}_i \bm{B} \widehat{\bm{T}}_i^{T} $, $\widehat{\bm{Q}}_i=\widehat{\bm{N}}_i \bm{Q} \widehat{\bm{N}}_i^{T}$, and $    \widehat{\bm{R}}_i=\widehat{\bm{T}}_i \bm{R} \widehat{\bm{T}}_i^{T} $.
            \STATE Solve equation $ \widehat{\bm{C}}_i^{T}\widehat{\bm{P}}_i+\widehat{\bm{P}}_i\widehat{\bm{C}}_i-\widehat{\bm{P}}_i\widehat{\bm{B}}_i\widehat{\bm{R}}_i^{-1}\widehat{\bm{B}}_i^{T}\widehat{\bm{P}}_i+\widehat{\bm{Q}}_i=\bm{0}$.
            \STATE Define $\widehat{\bm{K}}_i=-\widehat{\bm{R}}_i^{-1}\widehat{\bm{B}}_i^{T}\widehat{\bm{P}}_i$.
            \STATE Calculate the control law at node $i$: $\bm{k}_i=\bm{f}_i^{T} \widehat{\bm{{T}}}_i^T  \widehat{\bm{K}}_i \widehat{\bm{{N}}}_i$.
            \ENDFOR
            \STATE Output: full feedback control matrix $\bm{K}=[\bm{k}_1^T,\bm{k}_2^T,\cdots,\bm{k}_N^T]^T$.
        \end{algorithmic}
        }
    \end{algorithm}

\setlength{\abovedisplayskip}{0.7pt}
\setlength{\belowdisplayskip}{0.7pt}

\subsection*{Control of Synchronization in Kuramoto Oscillator Networks}

Consider $N$ phase oscillators coupled through a weighted directed network:
\begin{equation}\label{kuramoto}
    \dot{\theta}_i = \omega_i + \sum_{j=1}^{N}A_{ij} \text{sin}(\theta_j - \theta_i) +b_i u_i,
\end{equation}
where $\omega_i$ and $\theta_i$ are respectively the natural frequency and the phase of the $i$th oscillator, $A_{ij}$ denotes the elements of the (generally weighted and asymmetric) adjacency matrix of the network, and $u_i$ is the control input to the $i$th oscillator with the coefficient $b_i = 1$ if the oscillator is directly controlled and $b_i=0$ otherwise \cite{arenas2008synchronization}.

We consider a trajectory tracking task in which the target trajectory is a solution of the system synchronized to a common frequency $\omega^*$. Such a target trajectory can be expressed as $\theta_i(t) = \theta_i^* + \omega^*t$, where the constants $\theta_i^*$ are obtained by solving the nonlinear equation 
$
    \omega^* = \omega_i + \sum_{j=1}^{N}A_{ij} \text{sin}(\theta_j^* - \theta_i^*) +b_i u_i^*,\ 1 \leq i\leq N.
$
Here, $\omega^*$ can be chosen to be any value for which this equation has a solution. By further setting $u_i^* = \omega^* - \omega_i$, a sufficient condition for the solvability of the steady-state equation is given by $\norm{\bm{L}^{\dagger} \widetilde{\bm{\omega}}}_{\infty}<1$, where $\bm{L}^{\dagger}$ denotes the Moore–Penrose inverse of the corresponding Laplacian matrix $\bm{L}$ and the vector $\widetilde{\bm{\omega}}\in \mathbb{R}^N$ is such that $\widetilde{\omega}_i = \omega^*-\omega_i$ if the $i$th oscillator is uncontrolled and $\widetilde{\omega}_i=0$ otherwise \cite{dorfler2013synchronization}. The equation is then neither under- or over-determined and can be solved using the Newton--Raphson method. In the extreme case of directly controlling all nodes (i.e., $b_i=1,\ \forall i$), the equation always admits the phase-synchronized solution $\theta_1^*=\theta_2^*=\cdots=\theta_N^*=0$ for any given $\omega^*$. In the typical case of directly controlling a subset of nodes, there will be nonzero phase differences among the oscillators in the target trajectory. 

Defining $\Delta \theta_i = \theta_i-\theta_i^*-\omega^*t$ and $\Delta u_i = u_i-u_i^*$, we have 
$
 \Delta   \dot{\theta}_i = \sum_{j=1}^{N}A_{ij} \text{sin}(\Delta \theta_j - \Delta \theta_i) +b_i \Delta u_i,\ 1 \leq i\leq N.
$
Given the sinusoidal form of the coupling terms, we consider feedback laws of the form $\Delta u_i = \sum_{j=1}^{N}K_{ij} \text{sin} (\Delta \theta_j)$, which are generalizations of the control law in \cite{skardal2015control}. Employing the Jacobian linearization around the equilibrium $\Delta \theta_i=0,\ \forall i$, we obtain
$
    \Delta \dot{\bm{\theta}} = -\bm{L} \Delta \bm{\theta}+\bm{B}\bm{K}\Delta \bm{\theta}.
$
\reviseTen{The feedback matrix $\bm{K}$ can then be designed using Eq. \textbf{\ref{optfeedback}} (global control) and Algorithm 3 (local control) in which the weighting matrices in the control objective function are set to $\bm{Q}=5\bm{I}_N$ and $\bm{R}=\bm{I}$.}


\subsection*{Control of Stability in Power-Grid Networks}

We consider the classical model for the electro-mechanical dynamics of a power grid~\cite{machowski2011power}:
\begin{equation}
    \begin{aligned}
  &  \dot{{\delta}}_i   = {\omega}_i - {\omega}_\text{s}, \ \ 
  \dot{{\omega}}_i = \frac{1}{\Gamma_i} \big({P}_{\text{m}i} - {P}_{\text{e}i}-D_i ({\omega}_i - {\omega}_\text{s})\big),
    \end{aligned}
\end{equation}
where $\omega_\text{s}$ is the nominal frequency of the system, $\delta_i$ and $\omega_i$ are respectively the rotor angle and frequency of the $i$th generator, $\Gamma_i$ and $D_i$ are generator's inertia and damping constants, and ${P}_{\text{m}i}$ is the generator's mechanical power input. Here, ${P}_{\text{e}i}$ is the generator's electrical power output given by
$
    P_{\text{e}i} = \sum_{j=1}^N E_i E_j [\text{Im}(Y_{ij})\text{sin}(\delta_i-\delta_j)+\text{Re}(Y_{ij})\text{cos}(\delta_i-\delta_j)],
$
where $E_i$ is its internal voltage and $\bm{Y} = (Y_{ij})$ is the effective admittance matrix of the network.

In a steady state, a power system operates at an equilibrium in which generation matches consumption. However, since such balance is continuously challenged by the time-varying power generation and consumption, the system has to be operated in a series of quasi-steady states. Thus, the control problem is an equilibrium stabilization problem, where the target equilibrium at each time is determined by the power flow equation \cite{machowski2011power}. The desired equilibrium is specified by the target power flow solution $(\bm{E}^*,\bm{\delta}^*)$ and target frequency $\bm{\omega}^* = \omega_\text{s} \bm{1}$, where $\bm{1}$ denotes the vector of all ones. Assuming that the internal voltages are directly set to target values by the excitation systems, we seek a strategy to control the mechanical power input of the generators to drive the system towards the target and then stabilize it there. Noting that the target $(\bm{E}^*,\bm{\delta}^*)$ satisfies
$
     P_{\text{m}i}^* = P_{\text{e}i}^* = \sum_{j=1}^N E_i^* E_j^* [\text{Im}(Y_{ij})\, \text{sin}(\delta_i^*-\delta_j^*)+\text{Re}(Y_{ij})\, \text{cos}(\delta_i^*-\delta_j^*)]
$
and applying the Jacobian linearization, we obtain
\begin{equation}\label{linpower}
    \begin{bmatrix}
    \Delta \dot{\bm{\delta}} \\
   \Delta \dot{\bm{\omega}}
    \end{bmatrix}
    =
        \begin{bmatrix}
    \bm{0}  &  \bm{I} \\
   -\bm{P}  & - \bm{\Gamma}^{-1}\bm{D}
    \end{bmatrix}
        \begin{bmatrix}
    \Delta {\bm{\delta}} \\
   \Delta {\bm{\omega}}
    \end{bmatrix}
    +    \begin{bmatrix}
    \bm{0} \\
   \bm{\Gamma}^{-1}
    \end{bmatrix}
    \Delta \bm{P}_\text{m},
\end{equation}
where $\Delta \bm{\delta}=\bm{\delta}-\bm{\delta}^*$, $\Delta \bm{\omega}=\bm{\omega}-\bm{\omega}^*$, and $\Delta \bm{P}_\text{m}=\bm{P}_\text{m}-\bm{P}_\text{m}^*$. Here, $\bm{\Gamma}$ and $\bm{D}$ denote the diagonal matrices with $\Gamma_i$ and $D_i$ on their diagonals, respectively, and $\bm{P}$ denotes the equilibrium-dependent matrix whose elements are given by $P_{ij}=\frac{E_i^*E_j^*}{\Gamma_i} [\text{Re}(Y_{ij})\, \text{sin}(\delta_i^*-\delta_j^*)-\text{Im}(Y_{ij})\, \text{cos}(\delta_i^*-\delta_j^*)]$ for $i\neq j$, and $P_{ij}=-\sum_{k\neq i}P_{ik}$ for $i=j$. Then, Eq.~\textbf{\ref{linpower}} leads to the feedback law of the form
$
    \bm{P}_\text{m} = \bm{P}_\text{m}^* + \bm{K}_{\delta}(\bm{\delta}-\bm{\delta}^*)+\bm{K}_{\omega}(\bm{\omega}-\bm{\omega}^*).
$ 
The feedback matrix $\bm{K}=[\bm{K}_{\delta} \ \bm{K}_{\omega}]$ can be designed \reviseTen{using Eq. \textbf{\ref{optfeedback}} and Algorithm 3 for global and local control, respectively,} in which the weighting matrices are set to $\bm{Q}=10\bm{I}_{2N}$ and $\bm{R}=\bm{I}$.

\subsection*{Control of Epidemic Spreading through the Air Transportation Network}

We consider a human infectious disease whose spread is mediated by the global air transportation network \cite{colizza2006role}. To suppress the spreading, we implement a control intervention through treatment and vaccination. We consider the epidemic dynamics governed by the network-coupled susceptible-infectious model:
\begin{equation}\label{epdem}
\left\{
    \begin{aligned}
       & \dot{s}_k = -\beta s_k \ell_k- \sum_{j\neq k} a_{jk}\frac{s_k}{\nu_k} + \sum_{j\neq k}a_{kj} \frac{s_j}{\nu_j} -v_{k},\\
       & \dot{\ell}_k = \beta s_k \ell_k-\alpha \ell_k-\sum_{j\neq k} a_{jk}\frac{\ell_k}{\nu_k} + \sum_{j\neq k}a_{kj} \frac{\ell_j}{\nu_j} -w_{k},
    \end{aligned}\right.
\end{equation}
where each node $k$ represents a population of size $\nu_k$. Here, $s_k$ and $\ell_k$ are respectively the sizes of the susceptible and infected populations at node $k$, $\beta$ is the infection rate, $\alpha$ is the recovery rate, and $a_{jk}$ represents the rate at which people travel from node $k$ to node $j$. For the air transportation network considered, $a_{jk}$ is the number of travellers per day and each node represents an airport and the main city served by that airport. The control variable $v_{k}$ represents reduction of the susceptible population at node $k$ through vaccination, while $w_{k}$ represents reduction of the infected population through treatment. The goal is to design a vaccination/treatment strategy that can suppress the epidemic spreading using minimal medical resources. We formulate this as an optimal control problem in which we seek a control strategy $\bm{u}(t)=[\bm{v}(t)^T \ \bm{w}(t)^T]^T$ to minimize the quadratic cost function $\int_0^{\infty}\bm{\ell}^T(t)\bm{Q}\bm{\ell}(t)+\bm{u}(t)^T\bm{R}\bm{u}(t) dt$, where the first term measures the severity of the epidemics and the second quantifies the cost of the control strategy. This is a trajectory tracking problem in which the feasible trajectory is the manifold of disease-free solutions ${\ell}_k=0, \ \forall k$. To solve this problem, we first write the system in Eq. \textbf{\ref{epdem}} as $\dot{\bm{s}}= \bm{
L}'\bm{s}  -\beta\, \text{diag}(\bm{s})\bm{\ell}-\bm{v}$, $\dot{\bm{\ell}}=\beta \, \text{diag}(\bm{\ell})\bm{s} + (\bm{L}'-\alpha \bm{I})\bm{\ell}-\bm{w}$, where $\bm{L}'$ is a Laplacian-like matrix defined by $ L'_{kj}= \frac{a_{kj}}{\nu_j},\ k\neq j$, and $L'_{kk}=-\sum_{j\neq k}  \frac{a_{jk}}{\nu_k}$. Applying the extended linearization to this system, we obtain the state-dependent system matrix 
$
    \bm{C}(\bm{s},\bm{\ell}) = 
    \begin{bmatrix}
    \bm{L}' & -\beta \, \text{diag}(\bm{s}) \\
    \beta \, \text{diag}(\bm{\ell}) & \bm{L}'-\alpha \bm{I}
\end{bmatrix}.
$ 
The state-dependent feedback law
$
    \begin{bmatrix}
\bm{v}^T &
\bm{w}^T
\end{bmatrix}^T
=\bm{K}(\bm{s},\bm{\ell})
\begin{bmatrix}
\bm{s}^T  &
\bm{\ell}^T
\end{bmatrix}^T
$
can then be designed \reviseTen{using Eq. \textbf{\ref{optfeedback}} and Algorithm 3 for global and local control, respectively. In this case,
the weighting matrices are set to $\bm{Q}=\text{diag}(\bm{0}_{N},\bm{I}_N)$ and $\bm{R}=\text{diag}(5\bm{I}_{N},500\bm{I}_N)$.}

\subsection*{Control of Alzheimer's Disease Dynamics in Brain Networks}

Brain stimulation has been an active area of research in neuroscience for its potential to treat various neurological disorders, such as Alzheimer's disease, epilepsy, and Parkinson's disease \cite{taylor2015optimal,scheid2021time,schiff2010towards}. 
It has been widely reported that neurological disorders often manifest themselves as distinctive patterns of electrical activity detectable by electroencephalogram (EEG). For example, abnormal activity may be characterized by high-amplitude regular spike-wave oscillations \cite{taylor2015optimal} and can be modeled by a network of nonlinear oscillators. 
For Alzheimer's disease, coupled Duffing oscillators given by
\begin{equation}
      \dot{x}_i = {y}_i, \ \
       \dot{y}_i = -\alpha x_i -\gamma x_i^3+\beta \sum_{j=1}^N W_{ij}x_j+u_i
\end{equation}
have been used to describe the electrical activity of connected 
regions of the brain \cite{sanchez2018design}. Here, the state variables $\bm{x}=(x_i)$ and $\bm{y}=(y_i)$ describe excitatory postsynaptic potentials and their derivatives, respectively; the coupling matrix $\bm{W}=(W_{ij})$ reflects the relative connection strengths among brain regions; and the parameters $\beta$ and $\gamma$ capture the overall coupling strength and oscillator nonlinearity, respectively. In addition, EEG activities under different conditions are modeled by different values of parameter $\alpha$: a higher value $\alpha = \alpha_\text{h}$ produces low-amplitude high-frequency oscillations representing those observed under healthy conditions, whereas a lower value $\alpha = \alpha_\text{p}$ yields high-amplitude low-frequency oscillations representing those observed under pathological conditions. The control problem is then to generate an electrical stimuli $\bm{u}=(u_i)$ that steers the pathological system (with $\alpha = \alpha_\text{p}$) toward a trajectory of the healthy system (with $\alpha = \alpha_\text{h}$). This can be regarded as a command following task in which the healthy system generates a command signal for the pathological system to follow.

Using the procedure for command following presented above, we augment the system by introducing an integral state. That is, we write the controlled pathological system as $\dot{\bm{x}}_\text{p} = \bm{y}_\text{p}$, $\dot{\bm{y}}_\text{p} = \big( -\alpha \bm{I}-\gamma \, \text{diag}(\bm{x}_\text{p}^2)+\beta \bm{W} \big)\bm{x}_\text{p} + \bm{u}$, and $\dot{\bm{z}}_\text{p} = \bm{x}_\text{p} - \bm{x}_\text{h}$, where $\bm{x}_\text{h}$ is the state of the healthy system that serves as the command signal. This system is already in a form suitable for extended linearization, with the linearized \reviseTen{equation} defined by matrices
\begin{equation}\nonumber
  {\bm{C}}(\bm{x}_\text{p}) =
    \begin{bmatrix}
    \bm{0} & \bm{I} & \bm{0} \\
    -\alpha \bm{I}-\gamma \, \text{diag}(\bm{x}_\text{p}^2)+\beta \bm{W}  & \bm{0} & \bm{0} \\
    \bm{I} & \bm{0} & \bm{0}
    \end{bmatrix},
        \ \ \ 
    {\bm{B}} =  \begin{bmatrix}
     \bm{0} \\
     \bm{I} \\
     \bm{0}
    \end{bmatrix}.
\end{equation}
The state-dependent feedback law
$
    \bm{u} = \bm{K}_1(\bm{x}_\text{p})(\bm{x}_\text{p}-\bm{x}_\text{h})+\bm{K_2}(\bm{x}_\text{p})\bm{y}_\text{p}+\bm{K}_3(\bm{x}_\text{p})\bm{z}_\text{p}
$
\reviseTen{can once again be designed using Eq. \textbf{\ref{optfeedback}} and Algorithm 3 for global and local control, respectively,} in which the weighting matrices in the control objective function are set to $\bm{Q}=\text{diag}(100\bm{I}_{N},\bm{0}_N,\bm{I}_N)$ and $\bm{R}=10^{-6}\bm{I}_N$.


\bigskip\noindent{\bf\large Acknowledgements}

{This work was supported by ARO Grant No.\ W911NF-19-1-0383, ARPA-E Award No.\ DE-AR0000702, and the Institute for Sustainability and Energy at Northwestern.}

\bigskip\noindent{\bf\large Author contributions}

C.D., T.N., and A.E.M. designed
research; C.D. performed research; C.D. developed the
theory and performed numerical simulations; C.D., T.N.,
and A.E.M. analyzed data; and C.D., T.N., and A.E.M. wrote
the paper.

\bigskip\noindent{\bf\large Competing interests}

\noindent
The authors declare no competing interests.

\bigskip\noindent{\bf\large Materials \& Correspondence}

\noindent
Correspondence and material requests should be addressed to A.E.M.\\
(E-mail: motter@northwestern.edu).

\clearpage


\clearpage


\baselineskip18pt

\setcounter{page}{1}

\setcounter{equation}{0}
\renewcommand{\theequation}{S\arabic{equation}}
\setcounter{figure}{0}
\renewcommand{\thefigure}{S\arabic{figure}}

\renewcommand{\bibnumfmt}[1]{[A#1]} 
\renewcommand{\citenumfont}[1]{A#1} 

\begin{center}
{\Large Supplementary Materials for}

\bigskip\noindent{\bf\Large Prevalence and scalable control of localized networks}
\vspace{0.5cm}

\noindent{Chao Duan, Takashi Nishikawa, and Adilson E. Motter*}
\vspace{0.2cm}

*Corresponding author. Email: motter@northwestern.edu

\end{center}

\setlength{\cftsecindent}{1.5cm}
\setlength{\cftsubsubsecindent}{4cm}
\renewcommand\contentsname{}

\vspace{1cm}

\bigskip\noindent{\bf\large List of supplementary materials}

\vspace{1cm}

\vspace{0.3cm}

SI Text
\vspace{-1.8cm}
{\footnotesize
\vspace{1cm} \tableofcontents
\thispagestyle{empty}
}

SI References

Fig. S1

Table S1



\renewcommand{\thesection}{S\arabic{section}}
\renewcommand{\thesubsection}{S\arabic{section}.\arabic{subsection}}

\bigskip\noindent

\renewcommand{\theequation}{S\arabic{equation}}

\renewcommand{\thesection}{S\arabic{section}}

\renewcommand{\thesubsection}{S\arabic{section}.\arabic{subsection}}

\newpage
\section{Basic Implications of Locality}

\subsection*{Locality of the Solutions for Linear Equations}

Here we show that, if $\bm{M}$ is localized, then the locality of $\bm{b}$ implies the locality of the solution of $\bm{M}\bm{x}=\bm{b}$.  To see this, suppose that an invertible $\bm{M}$ belongs to $\mathcal{L}_{v,\rho}$ (i.e., $\bm{M}$ is localized) and that $\norm{\bm{b}_j}\leq \kappa \cdot v(\rho(i,j))^{-1}$ for all $j$ (i.e., $\bm{b}$ is localized around a given node $i$). Since $\mathcal{L}_{v,\rho}$ is closed under the inverse operation, we have $\bm{M}^{-1} \in \mathcal{L}_{v,\rho}$. It then follows that the solution $\bm{x}^{*}=\bm{M}^{-1}\bm{b}$ is also localized around node $i$, i.e., there is a constant $\kappa'$ such that $\norm{\bm{x}^*_j}\leq \kappa'\cdot v(\rho(i,j))^{-1}$. In the special case of $\bm{b}=\bm{e}_i\bm{b}'$ (recalling that $\bm{e}_i \in \mathbb{R}^{m\times n_i}$ is a matrix mapping the state space of node $i$ to that of the entire network), this implies that the solution of $\bm{M}\bm{x}=\bm{e}_i\bm{b}'$ is localized around node $i$ for any $n_i$-dimensional vector $\bm{b}'$.

The locality of the solution $\bm{x}^*$ has further implications. Consider the projection of the equation $\bm{M}\bm{x}=\bm{b}$ from the state space of the entire network to the subspace corresponding to the information neighborhood $\mathcal{N}_i(\tau)$:
\begin{equation}\label{projeq}
    \bm{N}_i \bm{M} \bm{N}_i^T\bm{z}= \bm{N}_i \bm{b},
\end{equation}
where $\bm{N}_i$ is the corresponding projection matrix. The solution $\bm{z}^{*}(\tau)$ of this equation satisfies
\begin{equation}
    \bm{N}_i \bm{M} \bm{N}_i^T (\bm{z}^{*}(\tau)-\bm{N}_i \bm{x}^*)  = -\bm{N}_i\bm{M}(\bm{N}_i^T\bm{N}_i-\bm{I})\bm{x}^*.
\end{equation}
The locality of $\bm{x}^*$ shown above ensures that the right hand side of this equation quickly decreases to zero as $\tau$ is increased. More precisely, we have
\begin{equation}
    \norm{\bm{N}_i\bm{M}(\bm{N}_i^T\bm{N}_i-\bm{I})\bm{x}^*}_{\infty} = \mathcal{O}\big(v(\tau)^{-1}\big),
\end{equation}
which implies
\begin{equation}
    \norm{\bm{z}^{*}(\tau)-\bm{N}_i \bm{x}^*}_{\infty} = \mathcal{O}\big(v(\tau)^{-1}\big)
\end{equation}
and
\begin{equation}
    \norm{\bm{N}_i^T \bm{z}^{*}(\tau)-\bm{x}^*}_{\infty} = \mathcal{O}\big(v(\tau)^{-1}\big).
\end{equation}
Thus, the locality of $\bm{M}$ and $\bm{b}$ guarantees that the solution of $\bm{M}\bm{x}=\bm{b}$ can be well approximated by solving its projection, Eq.~\textbf{\ref{projeq}}. Indeed, by decomposing any given vector $\bm{b}$ into individual nodes as $\bm{b} = \sum_{i} \bm{e}_i \bm{b}_i$ with $\bm{b}_i\in \mathbb{R}^{n_i\times 1}$ and solving $\bm{N}_i \bm{M} \bm{N}_i^T\bm{z}_i = \bm{N}_i \bm{e}_i$ to obtain the $m\times n_i$ solution matrix $\bm{z}_i$ for each $i$, we have an approximate solution of the original equation as $\bm{x} = \sum_{i} \bm{N}_i^T \bm{z}_i \bm{b}_i$. If the sizes of the information neighbourhoods are chosen to be independent of the system size, this leads to a linear-time algorithm for computing approximate solutions of large-scale localized linear equations. In addition, the quadratic form $\bm{b}'^T\bm{M}^{-1}\bm{b}$ for an arbitary vector $\bm{b}'$ can be well approximated by $(\bm{N}_i \bm{b}')^T(\bm{N}_i \bm{M} \bm{N}_i^T)^{-1}(\bm{N}_i \bm{b})$, which can be seen by noting that
\begin{equation}\label{quadapprox}
\begin{aligned}
\abs{(\bm{N}_i \bm{b}')^T(\bm{N}_i \bm{M} \bm{N}_i^T)^{-1}(\bm{N}_i \bm{b})-\bm{b}'^T\bm{M}^{-1}\bm{b}}
=&
    \abs{(\bm{N}_i \bm{b}')^T\bm{z}^{*}(\tau)-\bm{b}'^T\bm{x}^*}\\
    =&
    \abs{\bm{b}'^T(\bm{N}_i^T\bm{z}^{*}(\tau)-\bm{x}^*)} \\
    \leq & \norm{\bm{b}'}_1 \norm{\bm{N}_i^T \bm{z}^{*}(\tau)-\bm{x}^*}_{\infty}= \mathcal{O}\big(v(\tau)^{-1}\big).
    \end{aligned}
\end{equation}
Here, we have assumed that the projected matrix $\bm{N}_i \bm{M} \bm{N}_i^T$ is invertible. However, even when this matrix is not invertible, this analysis still applies to the minimum norm solution of Eq. ~\textbf{\ref{projeq}} obtained by using Moore–Penrose inverse of matrix $\bm{N}_i \bm{M} \bm{N}_i^T$.

\subsection*{Locality of the Matrix Exponential, Riccati Equation, and Lyapunov Equation}

{An implication of $\mathcal{L}_{v,\rho}$ being a Banach algebra is that it is closed under matrix exponential,} i.e., $e^{\bm{C}}\in \mathcal{L}_{v,\rho}$ if $\bm{C}\in \mathcal{L}_{v,\rho}$. This is easy to see from the polynomial expansion of matrix exponential $e^{\bm{C}} = \sum_{k=0}^{\infty}\frac{1}{k!}\bm{C}^{k}$ and the fact that a Banach algebra is complete and closed under addition and multiplication.


In control theory, the Riccati equation,
\begin{equation}\label{riccati}
    \bm{C}^{T}\bm{P}+\bm{P}\bm{C}-\bm{P}\bm{B}\bm{R}^{-1}\bm{B}^{T}\bm{P}+\bm{Q}=\bm{0},
\end{equation}
and the Lyapunov equation,
\begin{equation}\label{lyaeq}
    \bm{C}^{T}\bm{P}'+\bm{P}'\bm{C}+\bm{Q}'=\bm{0},
\end{equation}
are of central importance for system analysis and synthesis. The Lyapunov equation can be seen as a special case of the Riccati equation in the limit of $\bm{R}^{-1}$ approaching zero. The fact that $\mathcal{L}_{v,\rho}$ is an inverse-closed Banach algebra has significant implications for the solutions of both equations. It has been shown in ref.~\cite{curtain2011riccati_SI} that, if the matrices $\bm{C}$, $\bm{B}\bm{R}^{-1}\bm{B}^{T}$, and $\bm{Q}$ belong to an inverse-closed Banach algebra, then the unique stabilizing solution of the Riccati equation,
Eq.~\textbf{\ref{riccati}}, also belongs to the same algebra, as shown in ref.~\cite{zhou1996robust_SI}. Applying this result to $\mathcal{L}_{v,\rho}$ with the characteristic function $v(z)=e^{\alpha {z}^{\beta}} (1+z)^q$, we have the following result.
\begin{lemma}
Assume that the characteristic function is $v(z)=e^{\alpha {z}^{\beta}} (1+z)^q$ with $\alpha>0,\ 0<\beta<1$, and $q>1$. If i) the matrices $\bm{C}$, $\bm{B}\bm{R}^{-1}\bm{B}^{T}$, and $\bm{Q}$ belong to $\mathcal{L}_{v,\rho}$, ii) the matrix pair $(\bm{C},\bm{B})$ is stabilizable, and iii) the matrix pair $(\bm{C},\bm{Q}^{1/2})$ is detectable, then the stabilizing solution $\bm{P}$ of the Riccati equation (Eq.~\textbf{\ref{riccati}}) belongs to $\mathcal{L}_{v,\rho}$. As a special case, if $\bm{Q}'\in \mathcal{L}_{v,\rho}$ and $(\bm{C},\bm{Q}'^{1/2})$ is detectable, then the solution $\bm{P}'$ of the Lyapunov equation (Eq.~\textbf{\ref{lyaeq}}) also belongs to $\mathcal{L}_{v,\rho}$.
\end{lemma}

In this Lemma, $(\bm{C},\bm{B})$ is said to be stabilizable if the matrix $\begin{bmatrix}\bm{C}-\lambda \bm{I} & \bm{B}\end{bmatrix}$ has full row rank for all $\text{Re}\lambda \geq 0$, and $(\bm{C},\bm{Q}^{1/2})$ is said to be detectable if the matrix $
\begin{bmatrix}
\bm{C}-\lambda \bm{I} \\ \bm{Q}^{1/2}
\end{bmatrix}
$ has full column rank for all $\text{Re}\lambda \geq 0$.

There are other notions of network locality developed in the literature, such as those in refs. \cite{bartolucci2020emerging_SI, favero2021locality_SI}. However, those notions of locality are not developed for applications in network control and do not lead to the basic implications described in this section.

\section{Network Data}

The network data used in this paper is summarized in the following table.
\renewcommand\footnoterule{}
\setlength{\footnotesep}{0pc}
\begin{table}[!h]
\begin{minipage}{\textwidth}
{\footnotesize
\begin{tabular}{@{}m{0.03\textwidth}m{0.35\textwidth}|m{0.03\textwidth}|m{0.03\textwidth}@{}}
\toprule
\multicolumn{2}{c|}{\textbf{Networks}} &  \multicolumn{1}{c|}{\textbf{Figures}}   &          \multicolumn{1}{c}{\textbf{References}}                                                                                             \\ \toprule
\multicolumn{1}{l}{ KONECT dataset\footnote{Network names reproduced as in the KONECT dataset.}} & 
{

Advogato, 
Air traffic control, 
arXiv astro-ph, 
arXiv cond-mat, 
arXiv hep-ph (citation), 
arXiv hep-ph (coauthor), 
arXiv hep-th (citation), 
arXiv hep-th (coauthor), 
Blogs, 
Brightkite, 
CAIDA, 
Chicago, 
Cora citation, 
DBLP, 
Digg, 
DNC co-recipient, 
DNC emails co-recipients, 
Enron, 
Epinions, 
Euroroad, 
Facebook (NIPS), 
Facebook friendships, 
Facebook wall posts, 
FOLDOC, 
Gnutella, 
Google.com internal, 
Google+, 
Hamsterster friendships, 
Hamsterster full, 
Human protein (Figeys), 
Human protein (Stelzl), 
Human protein (Vidal), 
Internet topology, 
Linux kernel mailing list replies, 
OpenFlights, 
Pretty Good Privacy, 
Protein, 
Reactome, 
Route views, 
Slashdot threads, 
Slashdot Zoo, 
Twitter lists, 
U. Rovira i Virgili, 
UC Irvine messages, 
US airports, 
US power grid, 
Wikibooks (fr), 
Wikinews (fr), 
Wikipedia elections, 
Wikipedia threads (de)
}
&  \multicolumn{1}{c|}{Fig. 2}   & \multicolumn{1}{c}{\cite{konect_url}}                                        \\ \midrule
\multicolumn{2}{l|}{Eastern U.S.\ power grid}&  \multicolumn{1}{c|}{Fig. 4, Fig. 5D, Fig. 6A-C, Fig. 8} & \multicolumn{1}{c}{\cite{FERC}} \\
\midrule
\multicolumn{2}{l|}{Global air transportation network}&  \multicolumn{1}{c|}{Fig. 5E, Fig. 6A, Fig. 6C, Fig. 9} & \multicolumn{1}{c}{\cite{OpenFlights}}                                                        \\ \midrule
\multicolumn{2}{l|}{Whole brain network}&  \multicolumn{1}{c|}{Fig. 5F, Fig. 6A, Fig. 6C, Fig. 10}  & \multicolumn{1}{c}{\cite{Crossley2013_SI}}                                                  \\ \bottomrule
\end{tabular}
    }
      \end{minipage}
\end{table}

For the KONECT dataset \cite{konect_url}, we downloaded the edge list for each {network that has $10^3$--$10^5$ nodes and is in one of the following categories: communication, social, online contact, infrastructure, computer, hyperlink, authorship, citation \& coauthorship, and metabolic.}
From each list, we constructed the adjacency and Laplacian matrices of the (possibly directed and/or weighted) network. For these networks, unavailable edge weights are set to one, self-links are ignored, and the weights of parallel edges are combined.
For the Eastern U.S.\ power grid, the network considered represents a snapshot of the summer of 2017 obtained from Federal Energy Regulatory Commission (FERC) \cite{FERC}. This network was analyzed using MATPOWER \cite{zimmerman2010matpower_SI}, a MATLAB-based power system analysis toolbox. 
The effective admittance matrix representing the coupling among generators was obtained from the full admittance matrix $\bm{Y}_{\text{full}}$ through the Kron reduction \cite{machowski2011power_SI} given by $\bm{Y} = \bm{Y}_{\text{gg}}-\bm{Y}_{\text{gl}}\bm{Y}_{\text{ll}}^{-1}\bm{Y}_{\text{lg}}$, where $\bm{Y}_{\text{gg}}$ and $\bm{Y}_{\text{ll}}$ are the principal submatrices of $\bm{Y}_{\text{full}}$ induced by the set of generator nodes and load nodes, respectively; $\bm{Y}_{\text{gl}} $ ($\bm{Y}_{\text{lg}} $) is the submatrix of $\bm{Y}_{\text{full}}$ whose rows {correspond to} generator (load) nodes and whose columns {correspond to} load (generator) nodes. 
The steady state $(\bm{E}^*,\bm{\delta}^*)$ of the system was obtained by solving the power flow equation with the MATPOWER function \texttt{runpf}. Since the FERC data does not contain generator dynamic parameters, we sampled the values for the inertia constants $\Gamma_i$ uniformly from the interval $[4,8]$ seconds and set $D_i=0.1$ p.u.\ for each generator, {which are typical values for these parameters} \cite{1994power_SI} in power systems. To create testing scenarios, we also assume that generators with output power between 0 and 200MW are renewable generation units. For the global air transportation network, the OpenFlights data \cite{OpenFlights} provides information on $67663$ airline routes, including the source airport, the destination airport, and the aircraft type designator. We identified the city served by each airport using either the International Air Transport Association (IATA) code or the International Civil Aviation Organization (ICAO) code provided in the data set, along with the mapping between the airport codes and city names available at \href{https://www.world-airport-codes.com}{https://www.world-airport-codes.com}. We also obtained the population and geographical location of the cities from the World Cities Database available at \href{https://simplemaps.com/data/world-cities}{https://simplemaps.com/data/world-cities}. We consider only the $2219$ cities with a population larger than $10000$, corresponding to a total population of $2.42$ billion in this model. We obtained the seating capacity of each airplane type from the maker's website (as in the case of Airbus and Boeing) or other Internet sources; the seating capacity was then used to estimate the number of passengers in each flight (considering full occupancy for simplicity, which is an assumption that does not impact the qualitative results). The passenger flow network constructed connects all $2219$ cities considered, where the entry $C_{jk}$ of the adjacency matrix $\bm{C}$ in the dynamical model represents the fraction of the population in city $j$ that travel to city $k$ on average per day. 
For the whole brain network, the coupling matrix $\bm{W}$ {in ref.~\cite{Crossley2013_SI}} is readily available from the Brain Connectivity Toolbox website \href{https://sites.google.com/site/bctnet/datasets}{https://sites.google.com/site/bctnet/datasets} (as \texttt{Coactivation\_matrix.mat}). It represents the functional coactivation strengths among 638 similarly sized regions of a human brain. The parameters $\alpha_\text{h}$, $\alpha_\text{p}$, $\beta$, and $\gamma$ of the associated dynamical model were obtained from ref. \cite{sanchez2018design_SI}.

\section{Target Controllability of Localized Networks}

In many practical problems of controlling dynamical networks, it is not necessary to control the state of all nodes; instead, the goal is to steer {a target subset of nodes $\mathcal{S}\subseteq \mathcal{N}$ to desired states}, regardless of the states of the other nodes. This was the motivation for the concept of \emph{target controllability} \cite{gao2014target,zhao2016non_SI}. Let $\bm{S}$ be the projection matrix from the entire state space to that associated with the subset $\mathcal{S}$. The target subset $\mathcal{S}$ is said to be controllable if, for any initial state $\bm{x}_0$ at $t=t_0$, final target-set state $\bm{y}_1$, and finite time $t_1>t_0$, there exists an input $\bm{u}$ that drives the system from $\bm{x}(t_0)=\bm{x}_0$ to $\bm{x}(t_1)=\bm{x}_1$ for some $\bm{x}_1$ such that $\bm{S}\bm{x}_1=\bm{y}_1$; that is, the nodes in $\mathcal{S}$ can be steered to the desired states. When $\mathcal{S}=\mathcal{N}$, target controllability reduces to the usual notion of controllability. It is shown in ref.~\cite{zhao2016non_SI} that $\mathcal{S}$ is controllable if and only if the principal minor of the controllability Gramian indexed by $\mathcal{S}$, i.e., the matrix $\bm{S}\bm{W}_{\text{c}}^t\bm{S}^T$, is positive definite for any $t>0$ (from this point on we assume that $t_0=0$). It is straightforward to verify that, for any $\bm{y}\in \text{Null}(\bm{W}_{\text{c}}^{t_1})$, the control input
\begin{equation}\label{partialcontrol}
    \bm{u}(t) = \bm{B}^T e^{\bm{C}^T(t_1-t)}\left( \bm{S}^T(\bm{S}\bm{W}_{\text{c}} ^{t_1}\bm{S}^T)^{-1}\bm{S}\bm{x}_1+\bm{y}\right),\ 0<t<t_1,
\end{equation}
steers the system from $\bm{x}_0=\bm{0}$ to a state $\bm{x}(t_1)$ such that $\bm{S}\bm{x}(t_1)=\bm{y}_1$. In addition, the control input given by Eq.~\textbf{\ref{partialcontrol}} has the minimum possible control energy:
\begin{equation}
    \int_0^{t_1}\norm{\bm{u}(\tau)}_2^{2}d\tau = (\bm{S}\bm{x}_1)^{T}(\bm{S}\bm{W}_{\text{c}} ^{t_1}\bm{S}^T)^{-1}(\bm{S}\bm{x}_1).
\end{equation}
\reviseTen{In the special case of full controllability, $\mathcal{S}=\mathcal{N}$ and $\bm{S}=\bm{I}$, and hence $\int_0^{t_1}\norm{\bm{u}(\tau)}_2^{2}d\tau = \bm{x}_1^T (\bm{W}_{\text{c}} ^{t_1})^{-1} \bm{x}_1 \leq \lambda_{\text{min}}^{-1} (\bm{W}_{\text{c}}^{t_1})\cdot \norm{\bm{x}_1}_2^2 $, where the equality is attained when $\bm{x}_1$ becomes parallel with the eigenvector of $\bm{W}_{\text{c}}^{t_1}$ corresponding to its smallest eigenvalue. This implies that the worst-case control energy is inversely proportional to $\lambda_{\text{min}} (\bm{W}_{\text{c}}^{t_1})$. Therefore, the smallest eigenvalue of controllability Gramian can be considered a controllability measure: the larger the value of $\lambda_{\text{min}} (\bm{W}_{\text{c}}^{t_1})$, the more controllable the system is.}

Therefore, analogously to the case of full controllability, the worst-case minimum control energy needed to steer the target subset of nodes to the desired states is inversely proportional to the smallest eigenvalue of the projected Gramian $\bm{S}\bm{W}_{\text{c}}^{t_1}\bm{S}^T$. Since the smallest eigenvalue of $\bm{S}\bm{W}_{\text{c}}^{t_1}\bm{S}^T$ is upper-bounded by the smallest diagonal element of the matrix, we have
\begin{equation}\label{lamWopartial}
\begin{aligned}
     \lambda_{\text{min}} (\bm{S}\bm{W}_{\text{c}}^{t_1}\bm{S}^T) &\leq  {\breve{\kappa}_{t_1}} \eta^2 \cdot \min_{j\in \mathcal{S}} \sum_{i\in \mathcal{D}} v\big( \rho(i,j) \big)^{-2}\\
     &\leq  {\breve{\kappa}_{t_1}} \eta^2\cdot \abs{\mathcal{D}} \cdot \min_{j\in \mathcal{S}} v\big({\rho_{\text{H}}}(\mathcal{D},\{j\})\big)^{-2}\\
     &= {\breve{\kappa}_{t_1}} \eta^2\cdot \abs{\mathcal{D}} \cdot  v\big({\rho_{\text{H}}} (\mathcal{D},\mathcal{S})\big)^{-2},
     \end{aligned}
\end{equation}
where ${\rho_{\text{H}}} (\mathcal{D},\mathcal{S})=\max_{j\in \mathcal{S}}\,\min_{i\in \mathcal{D}}\,\rho(i,j)$. This inequality also establishes a crucial implication of locality for target controllability: the target subset of nodes $\mathcal{S}$ can be controlled using a smaller amount of energy if $\mathcal{S}$ lies closer to the driver set $\mathcal{D}$ in terms of the information distance ${\rho_{\text{H}}}(\cdot,\cdot)$. This is a more general form of the result in Eq.~\blue{\textbf{5}} of the main text.

\section{Local Approximability of Controllability Measure}


{For notational convenience, we use $\bm{W}_{\text{c}}$ to denote $\bm{W}_{\text{c}}^t$ for any $t\in (0,+\infty]$ since the analysis below applies to both finite- and infinite-time Gramian matrices.} Let $\bm{W}_{\text{c}}(\mathcal{N},\mathcal{M})$ be the submatrix of the controllability Gramian $\bm{W}_{\text{c}}$ whose rows and columns are induced by node sets $\mathcal{N}$ and $\mathcal{M}$, respectively. {If $\mathcal{M}$ is a singleton, i.e., $\mathcal{M}=\{i\}$,} we simply write $\bm{W}_{\text{c}}(\mathcal{N},i)$ (likewise, when $\mathcal{N}$ is a singleton).  Here, we show that, when a network is localized, the smallest eigenvalue $\lambda_{\text{min}} (\bm{W}_{\text{c}})$ of the entire Gramian $\bm{W}_{\text{c}}$ can be well approximated by $\lambda_{\text{min}} (\bm{W}_{\text{c}}(\mathcal{N}_i,\mathcal{N}_i))$, where $\bm{W}_{\text{c}}(\mathcal{N}_i,\mathcal{N}_i)$ is the principal submatrix of $\bm{W}_{\text{c}}$ induced by an information neighborhood $\mathcal{N}_i$ of some node $i$. Indeed, we show that there exist a node $i$, such that ${\lambda_{\text{min}} (\bm{W}_{\text{c}}(\mathcal{N}_i,\mathcal{N}_i))-\lambda_{\text{min}} (\bm{W}_{\text{c}})}=\mathcal{O}\big(v(\tau)^{-1}\big)$.

Suppose that the controllability Gramian $\bm{W}_{\text{c}}$ is localized (in addition to being symmetric and positive semi-definite by definition). Without loss of generality, we can assume that $\lambda_{\text{min}} (\bm{W}_{\text{c}})=0$; otherwise, we can instead consider $\bm{W}_{\text{c}}-\lambda_{\text{min}}(\bm{W}_{\text{c}})\bm{I}$ since locality, symmetry, and semi-definiteness are all preserved under the subtraction of $\lambda_{\text{min}}(\bm{W}_{\text{c}})\bm{I}$. If there exists a diagonal block $\bm{W}_{\text{c}}(i,i)$ of $\bm{W}_{\text{c}}$ for which $\lambda_{\text{min}}(\bm{W}_{\text{c}}(i,i))=0$, the desired information neighborhood is trivially $\mathcal{N}_i=\{i\}$. If not, we can show that the desired information neighborhood is that of a node $i$ satisfying the following condition.
\begin{condition}\label{condnb}
There exist a subset of nodes $\mathcal{M}$ and a information radius $\hat{\tau}<\infty$ such that $i\in \mathcal{M}\subseteq \mathcal{N}_i(\hat{\tau})$, $\bm{W}_{\text{c}}( \mathcal{M},\mathcal{M})$ is singular, and $\bm{W}_{\text{c}}\big( \mathcal{M}\setminus \{i\},\mathcal{M}\setminus \{i\}\big)$ is non-singular.
\end{condition}
We now show how this condition leads to the desirable information neighbourhood when $\lambda_{\text{min}}(\bm{W}_{\text{c}}(i,i))>0$ for all $i$.
For any given $\tau \leq \hat{\tau}$, we define the set $\mathring{\mathcal{N}}_i(\tau) = {\mathcal{N}_i}(\tau)\cap \mathcal{M} \setminus \{i\}$ and consider the following function of the radius $\tau$:
\begin{equation}\label{eqn:defftau}
    f(\tau) = \lambda_{\text{min}}\left(  \bm{W}_{\text{c}}(i,i)-\bm{W}_{\text{c}}(i,\mathring{\mathcal{N}}_i(\tau))  \bm{W}_{\text{c}}( \mathring{\mathcal{N}}_i(\tau),\mathring{\mathcal{N}}_i(\tau))^{-1} \bm{W}_{\text{c}}(\mathring{\mathcal{N}}_i(\tau),i) \right),
\end{equation}
i.e., the smallest eigenvalue of the Schur complement of $\bm{W}_{\text{c}}( \mathcal{M}\setminus \{i\},\mathcal{M}\setminus \{i\})$ in $\bm{W}_{\text{c}}( \mathcal{M},\mathcal{M})$. This function is well defined because $\bm{W}_{\text{c}}\big( \mathcal{M}\setminus \{i\},\mathcal{M}\setminus \{i\}\big)$ and its principal submatrices are non-singular. When $\tau =0$, the set $\mathring{\mathcal{N}}_i(\tau)$ is empty, so we have $f(0)=\lambda_{\text{min}}(\bm{W}_{\text{c}}(i,i))>0$. From the Schur determinant formula,
\begin{equation}
\begin{aligned}
 &\text{det}\big(\bm{W}_{\text{c}}( \mathcal{M},\mathcal{M})\big) \\=    &\text{det}\big(\bm{W}_{\text{c}}(i,i)\big)~ \text{det}\big(\bm{W}_{\text{c}}(i,i)-\bm{W}_{\text{c}}(i,\mathring{\mathcal{N}}_i(\tau))  \bm{W}_{\text{c}}( \mathring{\mathcal{N}}_i(\tau),\mathring{\mathcal{N}}_i(\tau))^{-1} \bm{W}_{\text{c}}(\mathring{\mathcal{N}}_i(\tau),i)\big),
 \end{aligned}
\end{equation}
and the assumptions that $\bm{W}_{\text{c}}( \mathcal{M},\mathcal{M})$ is singular and $\lambda_{\text{min}}(\bm{W}_{\text{c}}(i,i))>0$, we conclude that $f(\hat{\tau})=0$. By replacing $\bm{M}$ in Eq.~\textbf{\ref{quadapprox}} with $\bm{W}_{\text{c}}$ and taking $\bm{b}$ and $\bm{b}'$ to be any two (possibly identical) columns of $\bm{W}_{\text{c}}(\mathcal{N},i)$, we see that the locality of $\bm{W}_{\text{c}}$ leads to 
\begin{equation}
    \norm{
  \bm{W}_{\text{c}}(i,\mathcal{N})  \bm{W}_{\text{c}}^{-1}\bm{W}_{\text{c}}(\mathcal{N},i)-\bm{W}_{\text{c}}(i,\mathring{\mathcal{N}}_i(\tau))  \bm{W}_{\text{c}}( \mathring{\mathcal{N}}_i(\tau),\mathring{\mathcal{N}}_i(\tau))^{-1} \bm{W}_{\text{c}}(\mathring{\mathcal{N}}_i(\tau),i)
    } =  \mathcal{O}\big(v(\tau)^{-1}\big),
\end{equation}
where $\norm{\cdot}$ can be any matrix norm but, for convenience, we use the matrix norm induced by the vector $2$-norm. Therefore, 
{\small
\begin{equation}\label{eqn:ftauconverge}
\begin{aligned}
&{f(\tau)} \\  & \leq \lambda_{\text{min}} \big(\bm{W}_{\text{c}}(i,i)-\bm{W}_{\text{c}}(i,\mathcal{N})  \bm{W}_{\text{c}}^{-1}\bm{W}_{\text{c}}(\mathcal{N},i)\big) \\
    &  \ \ \ \ \ \ \ \ \ \ \ \    + \lambda_{\text{max}}\big(
  \bm{W}_{\text{c}}(i,\mathcal{N})  \bm{W}_{\text{c}}^{-1}\bm{W}_{\text{c}}(\mathcal{N},i)-\bm{W}_{\text{c}}(i,\mathring{\mathcal{N}}_i(\tau))  \bm{W}_{\text{c}}( \mathring{\mathcal{N}}_i(\tau),\mathring{\mathcal{N}}_i(\tau))^{-1} \bm{W}_{\text{c}}(\mathring{\mathcal{N}}_i(\tau),i)
    \big)\\
    & \leq 0 + \norm{
  \bm{W}_{\text{c}}(i,\mathcal{N})  \bm{W}_{\text{c}}^{-1}\bm{W}_{\text{c}}(\mathcal{N},i)-\bm{W}_{\text{c}}(i,\mathring{\mathcal{N}}_i(\tau))  \bm{W}_{\text{c}}( \mathring{\mathcal{N}}_i(\tau),\mathring{\mathcal{N}}_i(\tau))^{-1} \bm{W}_{\text{c}}(\mathring{\mathcal{N}}_i(\tau),i)
    } \\
    &= \mathcal{O}\big(v(\tau)^{-1}\big),
\end{aligned}
\end{equation}
}
where the first inequality is from the Weyl theorem and the second inequality is due to the assumption that $\bm{W}_{\text{c}}$ is singular and the fact that the spectral radius is a lower bound of any matrix norm \cite{horn2012matrix_SI}.

We further note that
{\small
\begin{equation}\label{eqn:upperboudnftau}
\begin{aligned}
   &\lambda_{\text{min}}\big(\bm{W}_{\text{c}}({\mathcal{N}_i}(\tau)\cap \mathcal{M} ,{\mathcal{N}_i}(\tau)\cap \mathcal{M} )\big)\\
   =& \min_{\bm{x},\bm{y}} \ \frac{\bm{x}^T\bm{W}_{\text{c}}(i,i)\bm{x}+2\bm{x}^T\bm{W}_{\text{c}}(i,\mathring{\mathcal{N}}_i(\tau)) \bm{y}+\bm{y}^T\bm{W}_{\text{c}}(\mathring{\mathcal{N}}_i(\tau),\mathring{\mathcal{N}}_i(\tau)) \bm{y}}{\bm{x}^T\bm{x}+\bm{y}^T
   \bm{y}}\\
   \leq & \min_{\bm{x}} \ \frac{\bm{x}^T\left( 
    \bm{W}_{\text{c}}(i,i)-\bm{W}_{\text{c}}(i,\mathring{\mathcal{N}}_i(\tau))  \bm{W}_{\text{c}}( \mathring{\mathcal{N}}_i(\tau),\mathring{\mathcal{N}}_i(\tau))^{-1} \bm{W}_{\text{c}}(\mathring{\mathcal{N}}_i(\tau),i) 
   \right)\bm{x}}{\bm{x}^T\bm{x}+\bm{x}^T
   \left( 
   \bm{W}_{\text{c}}(i,\mathring{\mathcal{N}}_i(\tau))  \bm{W}_{\text{c}}( \mathring{\mathcal{N}}_i(\tau),\mathring{\mathcal{N}}_i(\tau))^{-2} \bm{W}_{\text{c}}(\mathring{\mathcal{N}}_i(\tau),i) 
   \right)
   \bm{x}}\\
   \leq & \min_{\bm{x}} \ \frac{\bm{x}^T\left( 
    \bm{W}_{\text{c}}(i,i)-\bm{W}_{\text{c}}(i,\mathring{\mathcal{N}}_i(\tau))  \bm{W}_{\text{c}}( \mathring{\mathcal{N}}_i(\tau),\mathring{\mathcal{N}}_i(\tau))^{-1} \bm{W}_{\text{c}}(\mathring{\mathcal{N}}_i(\tau),i) 
   \right)\bm{x}}{\bm{x}^T\bm{x}}\\
   =& f(\tau),
\end{aligned}
\end{equation}
}
where the first inequality is obtained by taking
$
    \bm{y}=-\bm{W}_{\text{c}}( \mathring{\mathcal{N}}_i(\tau),\mathring{\mathcal{N}}_i(\tau))^{-1} \bm{W}_{\text{c}}(\mathring{\mathcal{N}}_i(\tau),i)\bm{x}.
$
Hence, we have
\begin{equation}
\lambda_{\text{min}}\big(\bm{W}_{\text{c}}({\mathcal{N}_i}(\tau) ,{\mathcal{N}_i}(\tau) )\big)
\leq
  \lambda_{\text{min}}\big(\bm{W}_{\text{c}}({\mathcal{N}_i}(\tau)\cap \mathcal{M} ,{\mathcal{N}_i}(\tau)\cap \mathcal{M} )\big)  \leq f(\tau),
\end{equation}
where the first inequality follows from the smallest eigenvalue of a principal submatrix being an upper bound of the smallest eigenvalue of the entire matrix. Combined with Eq.~\textbf{\ref{eqn:ftauconverge}}, this implies $\lambda_{\text{min}}\big(\bm{W}_{\text{c}}({\mathcal{N}_i}(\tau) ,{\mathcal{N}_i}(\tau) )\big)=\mathcal{O}\big(v(\tau)^{-1}\big)$, i.e., the smallest eigenvalue of the entire Gramian can be well approximated by the smallest eigenvalue of the projected Gramian in an information neighborhood of node $i$.

The analysis above relies on the existence of a node satisfying the three properties in Condition~\textbf{\ref{condnb}}. We now show that such a node indeed exists by providing an iterative procedure to find it. We note that this procedure only serves to prove the validity of the assumption and is not intended as an efficient algorithm for practical use. The procedure is as follows. First, initialize $\mathcal{M}$ as $\mathcal{M}=\mathcal{N}$. Then:
\begin{enumerate}[label={\arabic*)}]
    \item Pick any $i\in \mathcal{M}$.
    \item If $\bm{W}_{\text{c}}\big( \mathcal{M}\setminus \{i\},\mathcal{M}\setminus \{i\}\big)$ is non-singular, stop and output $i$ and $\mathcal{M}$; otherwise, set $\mathcal{M}:=\mathcal{M}\setminus \{i\}$ and go back to step 1).
\end{enumerate}
The procedure always terminates in a finite number of steps, since $\abs{\mathcal{M}}$ decreases by $1$ at each iteration and $\bm{W}_{\text{c}}\big( \mathcal{M}\setminus \{i\},\mathcal{M}\setminus \{i\}\big)$ is guaranteed to be non-singular when $\mathcal{M}\setminus \{i\}$ contains only one node. From the assumption that $\bm{W}_{\text{c}}(\mathcal{N},\mathcal{N})$ is singular, it follows that the matrix $\bm{W}_{\text{c}}( \mathcal{M},\mathcal{M})$ is always singular. As a result, when the procedure terminates, it is guaranteed to output the desirable $i$ and $\mathcal{M}$ that satisfy Condition~\textbf{\ref{condnb}}. The parameter $\hat{\tau}$ can then be determined as $\hat{\tau} = \max_{j\in \mathcal{M}} \rho(i,j)$.

Summarizing all of the above, given a localized network, we have shown that there exists a node $i$ for which the smallest eigenvalue of the controllability Gramian can be well approximated by $\lambda_{\text{min}} (\bm{W}_{\text{c}}(\mathcal{N}_i,\mathcal{N}_i))$. In practice, it is generally difficult to determine the desirable node $i$ \textit{a priori}. We thus consider instead the minimum among the smallest eigenvalues of all locally projected Gramians,
\begin{equation}\label{minlam_SI}
    \widetilde{\lambda}_{\text{min}} (\tau)= \min_{i}\,\lambda_{\text{min}} \big(\bm{W}_{\text{c}}(\mathcal{N}_i(\tau),\mathcal{N}_i(\tau))\big),
\end{equation}
which enjoys the same convergence $\mathcal{O}\big(v(\tau)^{-1}\big)$ due to the (guaranteed) existence of a special node $i$ satisfying Condition \textbf{\ref{condnb}}.

\section{Controllability Gramian of Diffusively Coupled Networks}

%
A diffusively coupled network always has a zero eigenvalue, which causes the divergence of the integral defining the infinite-horizon controllability Gramian (i.e., $\bm{W}_{\text{c}}^{t}$ for $t \rightarrow \infty$). This prevents the use of the Lyapunov equation to compute the infinite-horizon Gramian and its smallest eigenvalue, even though such a divergence does not affect the smallest eigenvalue itself. To overcome this issue, we perturb the system matrix as $\bm{C}-\epsilon \bm{I}$ with a small $\epsilon >0$ to remove the zero eigenvalue from the system and consider the Lyapunov equation for the perturbed system,
\begin{equation}
    (\bm{C}-\epsilon \bm{I}) \bm{W}_{\text{c}}^{(\epsilon)} + \bm{W}_{\text{c}}^{(\epsilon)}(\bm{C}-\epsilon \bm{I})^T + \bm{B}\bm{B}^T =\bm{0},
\end{equation}
which has a unique stablizing solution and thus is solvable by the Bartels--Stewart algorithm \cite{bartels1972solution_SI}. Once the solution $\bm{W}_{\text{c}}^{(\epsilon)}$ is obtained, we can project it to the orthogonal complement of the eigenvector $\bm{v}$ associated with the zero eigenvalue of $\bm{C}$. Denoting by $\bm{\Pi}$ an orthogonal basis for the orthogonal complement, we compute 
\begin{equation}\label{projectgram}
    \bm{W}_{\text{c}}^{\infty} = \bm{\Pi}~\bm{\Pi}^{T} \bm{W}_{\text{c}}^{(\epsilon)}  \bm{\Pi}~\bm{\Pi}^{T}
\end{equation}
as a projected version of the controllability Gramian. This $\bm{W}_{\text{c}}^{\infty}$ is insensitive to the choice of parameter $\epsilon$ when $\epsilon \ll \abs{\text{Re}\lambda_1}$, where $\lambda_1$ is the rightmost nonzero eigenvalue of $\bm{C}$. {The matrix $\bm{W}_{\text{c}}^{\infty}$ calculated using Eq.~\textbf{\ref{projectgram}} approaches the projected infinite-horizon Gramian of the original system (Eq.~\blue{\textbf{3}} in the main text) as $\epsilon$ approaches zero.}

\section{System-Level Synthesis}

Before presenting the main result of the System-Level Synthesis (SLS) theory, we briefly introduce the control-theoretic set-up for signals and systems. The time-domain signals are assumed to be in $L_2[0,+\infty)$, i.e., the space of square integrable functions supported on $t\geq 0$ with inner product
\begin{equation}
    \langle \bm{F}(t),\bm{E}(t) \rangle =\int_{-\infty}^{+\infty}\text{Trace}[\bm{F}^{\dagger}(\tau)\bm{E}(\tau)] d\tau
\end{equation}
and norm
\begin{equation}
    \norm{\bm{F}(t)}_{{L}_2}=\sqrt{ \langle \bm{F}(t),\bm{F}(t)\rangle },
\end{equation}
where $\bm{F}^{\dagger}$ denotes the conjugate transpose of matrix $\bm{F}$.
Through the Laplace transform
\begin{equation}
    \bm{F}(s) = \int_{0}^{+\infty} \bm{F}(t) e^{-st} dt,
\end{equation}
one can see that the time-domain signal space $L_2[0,+\infty)$ is isomorphic to $\mathcal{H}_2$. Here, $\mathcal{H}_2$ denotes the space of complex functions that are analytic in the open right-half plane $\text{Re}(s)>0$, which is endowed with inner product
\begin{equation}
    \langle \bm{F}(s),\bm{E}(s) \rangle=\frac{1}{2\pi}\int_{-\infty}^{+\infty}\text{Trace}[\bm{F}^{\dagger}(\text{j}w)\bm{E}(\text{j}w)] dw
\end{equation}
and norm
\begin{equation}
    \norm{\bm{F}(s)}_{\mathcal{H}_2}=\sqrt{ \langle \bm{F}(s),\bm{F}(s)\rangle},
\end{equation}
where $\text{j}$ denotes the imaginary unit.
The Plancherel theorem \cite{zhou1996robust_SI} states that $ \norm{\bm{F}(t)}_{{L}_2}= \norm{\bm{F}(s)}_{\mathcal{H}_2}$, where $\bm{F}(s) \in \mathcal{H}_2$ is the Laplace transform of $\bm{F}(t) \in L_2[0,+\infty)$. Linear dynamical systems can then be regarded as linear maps $\bm{G}$ from $L_2[0,+\infty)$ to $L_2[0,+\infty)$ or, equivalently, from $\mathcal{H}_2$ to $\mathcal{H}_2$. The induced norm is denoted by 
\begin{equation}
    \norm{\bm{G}}_{\mathcal{H}_{\infty}} = \text{ess }\underset{w\in \mathbb{R}}{\text{sup}} \ {\sigma}\big(\bm{G}(\text{j}w)\big),
\end{equation}
where ${\sigma}(\cdot)$ is the largest singular value and $\mathcal{H}_{\infty}$ denotes the associated normed space of all linear maps \cite{zhou1996robust_SI}. Furthermore, the subspaces of $\mathcal{H}_2$ and $\mathcal{H}_{\infty}$ consisting of proper rational matrix functions (i.e., matrices whose elements are rational functions in the complex variable $s$) are denoted by $\mathcal{RH}_2$ and $\mathcal{RH}_{\infty}$, respectively. In addition, the strictly proper rational function subspaces of $\mathcal{RH}_{2}$ and $\mathcal{RH}_{\infty}$ are denoted by $\frac{1}{s}\mathcal{RH}_{2}$ and $\frac{1}{s}\mathcal{RH}_{\infty}$, respectively.

By the Laplace transform, the optimal control problem in Eq.~\blue{\textbf{9}} of the main text,
\begin{equation}\label{lqr_SI}
    \begin{aligned}
      &\underset{\bm{u} \in L_2[0,+\infty)}{\text{min}}\ J=\int_0^\infty \bm{x}(\tau)^{T}\bm{Q}\bm{x}(\tau)+\bm{u}(\tau)^{T}\bm{R}\bm{u}(\tau) d\tau\\
      & \text{s.t. } \dot{\bm{x}}=\bm{C}\bm{x}+\bm{B}\bm{u},\ \bm{x}(0)=\bm{x}_0,
    \end{aligned}
\end{equation}
can be equivalently written in the $s$-domain as
\begin{equation}
\begin{aligned}
&\underset{\bm{u} \in \mathcal{H}_2}{\text{min}} \ 
\norm{\begin{bmatrix} \bm{Q}^{1/2} & \\ & \bm{R}^{1/2}\end{bmatrix} \begin{bmatrix}\bm{x}(s) \\\bm{u}(s)  \end{bmatrix}
}_{\mathcal{H}_2}^2\\
&\text{s.t.}
\begin{array}{l}
\bm{x}(s)=(s\bm{I}-\bm{C})^{-1}(\bm{B}\bm{u}(s)+\bm{x}_0).
\end{array}
\end{aligned}
\end{equation}
Given a feedback controller $\bm{K}(s)$, the transfer function from the initial state $\bm{x}_0$ to the state response $\bm{x}(s)$ is given by $\bm{\Phi}(s)=\big(s\bm{I}-\bm{C}-\bm{B}\bm{K}(s)\big)^{-1}$, and the response of the controller output $\bm{u}(s)$ is given by the transfer function $\bm{H}(s)=\bm{K}(s)\big(s\bm{I}-\bm{C}-\bm{B}\bm{K}(s)\big)^{-1}$. The parameterization of all $\bm{\Phi}(s)$ and $\bm{H}(s)$ that are achievable by some stabilizing controller $\bm{K}(s)$ is given by the following theorem.
\begin{theorem}[System-Level Parameterization for State Feedback Systems \cite{wang2016system_SI}]\label{slsthm}
The following are true for any linear system $\dot{\bm{x}}=\bm{C}\bm{x}+\bm{B}\bm{u}$: 1) The affine space defined by 
    \begin{equation} \label{slspara}
        \begin{bmatrix}
            s\bm{I}-\bm{C} & -\bm{B}
        \end{bmatrix}
       \begin{bmatrix}
            \bm{\Phi}(s) \\ \bm{H}(s)
        \end{bmatrix}=\bm{I}, \ \ \bm{\Phi},   \bm{H} \in \frac{1}{s}\mathcal{RH}_{\infty}
    \end{equation}
    parameterizes all system responses from $\bm{x}_0$ to $\bm{x}(s)$ and control responses from $\bm{x}_0$ to $\bm{u}(s)$ that are achievable by an internally stabilizing state feedback controller; 2) For all transfer matrices $\{\bm{\Phi}(s), \bm{H}(s)\}$ satisfying Eq.~\textbf{\ref{slspara}}, the controller $\bm{K}(s)=\bm{H}(s)\bm{\Phi}(s)^{-1}$ is internally stabilizing and achieves the desired responses $\bm{x}(s)=\bm{\Phi}(s)\bm{x}_0$ and $\bm{u}(s)=\bm{H}(s)\bm{x}_0$.
\end{theorem}
This theorem shows that there is a bijection between the stabilizing controllers and the responses in the affine space defined by Eq.~\textbf{\ref{slspara}}. Therefore, instead of directly designing the feedback control law $\bm{K}(s)$, we can equivalently design the transfer matrices $\bm{\Phi}(s)$ and $\bm{H}(s)$ under the affine constraint given by Eq.~\textbf{\ref{slspara}}. With this parameterization, the optimal control problem can be rewritten as 
\begin{equation} \label{slslqr}
\begin{aligned}
&\underset{\bm{\Phi}, \bm{H}\in \frac{1}{s}\mathcal{RH}_{\infty}}{\text{min}} \ 
J = \norm{\begin{bmatrix} \bm{Q}^{1/2} & \\ & \bm{R}^{1/2}\end{bmatrix} \begin{bmatrix}\bm{\Phi}(s) \\\bm{H}(s)  \end{bmatrix}\bm{x}_0
}_{\mathcal{H}_2}^2\\
&\text{s.t.}
\begin{array}{l}
\begin{bmatrix}
            s\bm{I}-\bm{C} & -\bm{B}
        \end{bmatrix}
       \begin{bmatrix}
            \bm{\Phi}(s) \\ \bm{H}(s)
        \end{bmatrix}=\bm{I}.
\end{array}
\end{aligned}
\end{equation}
The solution of this problem {for any given $\bm{x}_0$} is given by
\begin{equation} \label{slslqr_phi_sol}
    \bm{\Phi}(s) = (s\bm{I}-\bm{C}+\bm{B}\bm{R}^{-1}\bm{B}^T\bm{P})^{-1},
\end{equation}
\begin{equation} \label{slslqr_M_sol}
    \bm{H}(s) = -\bm{R}^{-1}\bm{B}^T\bm{P}(s\bm{I}-\bm{C}+\bm{B}\bm{R}^{-1}\bm{B}^T\bm{P})^{-1},
\end{equation}
where $\bm{P}$ is the solution of the Riccati equation, Eq.~\textbf{\ref{riccati}}. {The optimal controller thus becomes a static feedback law and is given by
\begin{equation}\label{fullopt}
 \bm{K}(s)=\bm{H}(s)\bm{\Phi}(s)^{-1} = -\bm{R}^{-1}\bm{B}^T\bm{P}.
\end{equation}
We note that $\bm{\Phi}(s)$ and $ \bm{H}(s)$ in Eqs.~\textbf{\ref{slslqr_phi_sol}} and \textbf{\ref{slslqr_M_sol}}, respectively, are independent of $\bm{x}_0$ and hence serve as the ``fundamental solution'' of the optimal control problem in Eq.~\textbf{\ref{slslqr}} for arbitrary initial conditions. The columns of $\bm{\Phi}(s)$ and $\bm{H}(s)$ can be computed independently. To see this, we set $\bm{x}_0 = \bm{e}_j$ in Eq.~\textbf{\ref{slslqr}}, rewrite the problem as the minimization over $\bm{\phi}_j(s)=\bm{\Phi}(s)\bm{e}_j$ and $\bm{h}_j(s)=\bm{H}(s)\bm{e}_j$, and eliminate the unnecessary constraints corresponding to all but the $j$th columns of $\bm{\Phi}(s)$ and $\bm{H}(s)$. This leads to Eq.~\blue{\textbf{10}} in the main text.}

\section{Disturbance-Oriented Localization}


By projecting the original optimal control problem in Eq.~\blue{\textbf{10}} of the main text onto the information neighborhood $\mathcal{N}_j$ of node $j$, we have the following projected optimization problem:
\begin{equation} \label{slslqr3}
\begin{aligned}
&\underset{\widetilde{\bm{\phi}}_j, \widetilde{\bm{h}}_j\in \frac{1}{s}\mathcal{RH}_{\infty}}{\text{min}} \ 
\norm{\begin{bmatrix} \widetilde{\bm{Q}}^{1/2}_j & \\ & \widetilde{\bm{R}}^{1/2}_j \end{bmatrix} \begin{bmatrix}\widetilde{\bm{\phi}}_j(s) \\\widetilde{\bm{h}}_j(s)  \end{bmatrix} \widetilde{\bm{e}}_j^T\bm{x}_0
}_{\mathcal{H}_2}^2\\
&\text{s.t.} 
\begin{array}{l}
\begin{bmatrix}
            s\bm{I}-\widetilde{\bm{C}}_j & -\widetilde{\bm{B}}_j
        \end{bmatrix}
       \begin{bmatrix}
            \widetilde{\bm{\phi}}_j(s) \\ \widetilde{\bm{h}}_j(s)
        \end{bmatrix}=\widetilde{\bm{e}}_j.
\end{array}
\end{aligned}
\end{equation}
To solve this projected problem, we invoke the inverse Laplace transform to go back to the time domain, which transforms the problem in Eq.~\textbf{\ref{slslqr3}} into
\begin{equation} \label{slslqr5}
\begin{aligned}
&\underset{\widetilde{\bm{u}}_j\in L_2[0,+\infty)}{\text{min}} \ 
 \widetilde{J}=\int_0^\infty
 \big[
 \widetilde{\bm{x}}_j(\tau)^{T}\widetilde{\bm{Q}}_j\widetilde{\bm{x}}_j(\tau)+\widetilde{\bm{u}}_j(\tau)^{T}\widetilde{\bm{R}}_j\widetilde{\bm{u}}_j(\tau)
 \big]
 d\tau   \\
&\text{s.t.}
\begin{array}{l}
\dot{\widetilde{\bm{x}}}_j=\widetilde{\bm{C}}\widetilde{\bm{x}}_j+\widetilde{\bm{B}}\widetilde{\bm{u}}_j, \ \widetilde{\bm{x}}_j(0)= \widetilde{\bm{e}}_j^T\bm{x}_0.
\end{array}
\end{aligned}
\end{equation}
This is simply a projected version of the optimal control problem defined by Eq.~\textbf{\ref{lqr_SI}}. The optimal control law is given by the static feedback matrix
\begin{equation}\label{feedbackK}
\widetilde{\bm{K}}_j=-\widetilde{\bm{R}}_j^{-1}\widetilde{\bm{B}}_j^{T}\widetilde{\bm{P}}_j,
\end{equation}
where $\widetilde{\bm{P}}_j$ is the solution to the Riccati equation
\begin{equation}\label{riccati2}
    \widetilde{\bm{C}}_j^{T}\widetilde{\bm{P}}_j+\widetilde{\bm{P}}_j\widetilde{\bm{C}}_j-\widetilde{\bm{P}}_j\widetilde{\bm{B}}_j\widetilde{\bm{R}}_j^{-1}\widetilde{\bm{B}}_j^{T}\widetilde{\bm{P}}_j+\widetilde{\bm{Q}}_j=\bm{0}.
\end{equation}
The $s$-domain solution corresponding to the problem in Eq.~\textbf{\ref{slslqr3}} is then given by
\begin{equation}\label{rjred}
   \widetilde{\bm{\phi}}_j(s)=(s\bm{I}-\widetilde{\bm{C}}_j-\widetilde{\bm{B}}_j\widetilde{\bm{K}}_j)^{-1} \widetilde{\bm{e}}_j,
\end{equation}
\begin{equation}\label{mjred}
    \widetilde{\bm{h}}_j(s)=\widetilde{\bm{K}}_j\widetilde{\bm{\phi}}_j(s).
\end{equation}
After solving the projected problem for all $1\leq j\leq N$, we can construct
\begin{equation}     \label{hatPhi}
    \widetilde{\bm{\Phi}}(s) =
    \begin{bmatrix}
    {\bm{N}}_{1}^T \widetilde{\bm{\phi}}_1(s) & {\bm{N}}_{2}^T \widetilde{\bm{\phi}}_2(s) & \cdots & {\bm{N}}_{n}^T \widetilde{\bm{\phi}}_n(s) 
    \end{bmatrix},
\end{equation}
\begin{equation} \label{hatM}
    \widetilde{\bm{H}}(s) =
    \begin{bmatrix}
    {\bm{T}}_{1}^T \widetilde{\bm{h}}_1(s) & {\bm{T}}_{2}^T \widetilde{\bm{h}}_2(s) & \cdots & {\bm{T}}_{n}^T \widetilde{\bm{h}}_n(s) 
    \end{bmatrix}.
\end{equation}
From Theorem~\ref{slsthm}, it then follows that the overall optimal control law takes the form
\begin{equation}\label{conLaw}
\bm{u}(s) = \widetilde{\bm{K}}(s)\bm{x}(s)=\widetilde{\bm{H}}(s) \widetilde{\bm{\Phi}}(s)^{-1}\bm{x}(s).
\end{equation}

Crucially, we now show that the controller $\widetilde{K}(s)$ is guaranteed to be stabilizing and its associated control objective value approaches that of the global controller when the system is sufficiently localized. That is, the controller designed based on the projected model, $\widetilde{\bm{K}}(s)$, will enjoy certain stability and near-optimal performance guarantees when implemented on the original system. Due to the locality of the system, in view of Eq.~\textbf{\ref{slspara}}, the following equality shows that the projected problem can be 
regarded as a perturbed version of the original one:
\begin{equation}\label{mismatch}
    \begin{aligned}
      & (s\bm{I}-\bm{C}) \bm{N}_j^T \widetilde{\bm{\phi}}_j(s)-\bm{B}\widetilde{\bm{T}}_j^T\widetilde{\bm{h}}_j(s)-\bm{e}_j = (\bm{I}-\bm{N}_j^T\bm{N}_j) (-\bm{C})\bm{N}_j^T\widetilde{\bm{\phi}}_j(s) := \bm{\epsilon}_j(s).
    \end{aligned}
\end{equation}
The perturbation term $\bm{\epsilon}_j(s)$ is expected to be very small in magnitude when $\mathcal{N}_j$ is sufficiently large. This is the case because: {1) the non-zeros elements of vector $\bm{N}_j^T\widetilde{\bm{\phi}}_j(s)$ are limited to those corresponding to the neighbourhood $\mathcal{N}_j$, and their magnitude decays as $\mathcal{O}\big(v(\tau)^{-1}\big)$ with the information distance $\tau$; 2) the multiplication by $-\bm{C}$ preserves the decay pattern due to the locality of the system; and 3) the further multiplication by $(\bm{I}-\bm{N}_j^T{\bm{N}}_j)$ turns the elements inside the information neighborhood ${\mathcal{N}}_j$ into zeros and leaves only negligible elements outside ${\mathcal{N}}_j$.} That is, the solution of the projected problem (Eq.~\textbf{\ref{slslqr3}} with mismatch $\bm{\epsilon}_j(s)$) when lifted to the original space is an approximate solution of the original problem in Eq.~\textbf{\ref{slslqr}}. Concatenating Eq.~\textbf{\ref{mismatch}} for all $1\leq j \leq n$ and accounting for Eq.~\textbf{\ref{hatPhi}} and Eq.~\textbf{\ref{hatM}}, we have
\begin{equation}
     (s\bm{I}-\bm{C}) \widetilde{\bm{\Phi}}(s)-\bm{B} \widetilde{\bm{H}}(s)= \bm{I} + \bm{\Sigma}(s),
\end{equation}
where $\bm{\Sigma}(s) =[\bm{\epsilon}_1(s), \bm{\epsilon}_2(s), \cdots, \bm{\epsilon}_n(s)]$. If the perturbation $\bm{\Sigma}(s)$ is sufficiently small such that 
\begin{equation}\label{sigmacon1}
( \bm{I} + \bm{\Sigma}(s))^{-1} \in \frac{1}{s}\mathcal{RH}_{\infty},
\end{equation}
then the matrices
\begin{equation}\label{sysred1}
  \widetilde{\bm{\Phi}}'(s) =  \widetilde{\bm{\Phi}}(s)\big( \bm{I} + \bm{\Sigma}(s)\big)^{-1},
\end{equation}
\begin{equation}\label{sysred2}
\widetilde{\bm{H}}'(s) = \widetilde{\bm{H}}(s)\big( I + \bm{\Sigma}(s)\big)^{-1}
\end{equation}
would satisfy exactly the condition given by Eq.~\textbf{\ref{slspara}} in Theorem~\ref{slsthm}. Hence, Eqs.~\textbf{\ref{sysred1}} and \textbf{\ref{sysred2}} constitute achievable system and control responses with the controller $\widetilde{\bm{K}}(s)=\widetilde{\bm{H}}'(s) \widetilde{\bm{\Phi}}'(s) ^{-1} =\widetilde{\bm{H}}(s) \widetilde{\bm{\Phi}}(s)^{-1}$, which stabilizes the original system even though this controller is designed based on the projected model.

A sufficient condition for Eq.~\textbf{\ref{sigmacon1}}, according to the small-gain theorem \cite{dullerud2013course_SI,anderson2019system_SI}, is given by any of the following:
\begin{equation}\label{Hcond1}
    \norm{\bm{\Sigma}}_{\mathcal{H}_{\infty}} < 1,
\end{equation}
\begin{equation}\label{Hcond2}
    \norm{\bm{\Sigma}}_{\mathcal{L}_{1}} < 1,
\end{equation}
\begin{equation}\label{L1smallgain}
    \norm{\bm{\Sigma}^T}_{\mathcal{L}_{1}} < 1,
\end{equation}
where $\norm{\cdot}_{\mathcal{L}_{1}}$ denotes the $\mathcal{L}_{1}$ system norm. The $\mathcal{L}_{1}$ system norm is defined through the impulse response in the time domain \cite{zhu2015stability_SI} as \begin{equation}\label{L1def}
    \norm{\bm{\Sigma}^T(t)}_{\mathcal{L}_{1}} = \underset{1\leq j \leq n}{ \text{max}} \sum_{i=1}^{n} \int_{0}^{\infty} \abs{\epsilon_{ij}(t)} dt,
\end{equation}
where $\epsilon_{ij}(t)$ is {the $i$th component of the perturbation (column) vector} $\bm{\epsilon}_{j}(s)$ in Eq.~\textbf{\ref{mismatch}} under inverse Laplace transform. The computation required to verify the small-gain condition in Eq.~\textbf{\ref{L1smallgain}} can be {distributed over} the columns of the impulse response matrix $\bm{\Sigma}(t)$, since Eq.~\textbf{\ref{L1smallgain}} is equivalent to 
\begin{equation}\label{L1small2}
    \sum_{i=1}^{n} \int_{0}^{\infty} \abs{\epsilon_{ij}(t)} dt <1, \ \ \forall \ 1\leq j \leq n.
\end{equation}


Once the controller defined by Eq.~\textbf{\ref{conLaw}} stabilizes the system, the next question is to determine
the extent to which the dynamical performance of this controller compares with that of the theoretical global optimal control.  Let $J^{*}$ be the optimal objective value for the original optimal control problem in Eq.~\textbf{\ref{slslqr}}, and define 
\begin{equation}
    \widetilde{J}^{*}=\norm{ \bm{Q}^{1/2}\widetilde{\bm{\Phi}}(s)}_{\mathcal{H}_2} + \norm{ \bm{R}^{1/2}\widetilde{\bm{H}}(s)}_{\mathcal{H}_2},
\end{equation}
i.e., the optimal objective value obtained from the solutions of the projected problems in Eq.~\textbf{\ref{slslqr3}}. Since $\widetilde{\bm{\Phi}}'(s)$ and $\widetilde{\bm{H}}'(s)$ form a feasible solution of the optimal control problem in Eq.~\textbf{\ref{slslqr}}, we have an upper bound for $J^{*}$:
\begin{equation}\label{est1}
\begin{aligned}
    J^{*} &\leq \norm{ \bm{Q}^{1/2}\widetilde{\bm{\Phi}}'(s)}_{\mathcal{H}_2} + \norm{ \bm{R}^{1/2}\widetilde{\bm{H}}'(s)}_{\mathcal{H}_2} \\
    &\leq \norm{ \bm{Q}^{1/2}\widetilde{\bm{\Phi}}(s)}_{\mathcal{H}_2}\norm{\bm{I} + \bm{\Sigma}(s)}_{\mathcal{H}_{\infty}} + \norm{ \bm{R}^{1/2}\widetilde{\bm{H}}(s)}_{\mathcal{H}_2} \norm{\bm{I} + \bm{\Sigma}(s)}_{\mathcal{H}_{\infty}}\\
    &\leq \widetilde{J}^* \norm{\bm{I} + \bm{\Sigma}(s)}_{\mathcal{H}_{\infty}}\\
    &\leq \frac{\widetilde{J}^*}{1-\norm{\bm{\Sigma}(s)}_{\mathcal{H}_{\infty}}}.
\end{aligned}
\end{equation}

To derive a lower bound for $J^{*}$, we interchange the roles of the original and projected systems in Eq.~\textbf{\ref{mismatch}} and obtain
\begin{equation}\label{mismatch2}
    \begin{aligned}
       (s\bm{I}-{\bm{C}}) \bm{N}_j {\bm{\phi}}_j(s)-{\bm{B}}{\bm{T}}_j{\bm{h}}_j(s)-\widetilde{\bm{e}}_j 
       =  - \bm{N}_j\bm{C}(\bm{N}_j^T\bm{N}_j-\bm{I}){\bm{\phi}}_j(s) := \bm{\xi}_j(s),
    \end{aligned}
\end{equation}
which implies
\begin{equation}
     (s\bm{I}-{\bm{C}}) {\bm{\Phi}}(s)-{\bm{B}} {\bm{H}}(s)= \bm{I} + \bm{\Xi}(s),
\end{equation}
where we define $\Xi (s) =[\xi_1(s), \xi_2(s), \cdots, \xi_N(s)]$. Due to the locality, the perturbation term can be small enough such that $( \bm{I} + \bm{\Xi}(s))^{-1} \in \frac{1}{s}\mathcal{RH}_{\infty}$, which is guaranteed by any of the inequalities in Eqs.~\textbf{\ref{Hcond1}--\ref{L1smallgain}}. Under this condition, ${\bm{\Phi}}(s)( \bm{I} + \bm{\Xi}(s))^{-1}$ and ${\bm{H}}(s)( \bm{I} + \bm{\Xi}(s))^{-1}$ form a feasible solution of the projected problem in Eq.~\textbf{\ref{slslqr3}}. This implies
\begin{equation}\label{est2}
\begin{aligned}
    \widetilde{J}^{*} & \leq \norm{\bm{Q}^{1/2}{\bm{\Phi}}(s)( \bm{I} + \bm{\Xi}(s))^{-1}}_{\mathcal{H}_2} + \norm{\bm{R}^{1/2}{\bm{H}}(s)( \bm{I} + \bm{\Xi}(s))^{-1}}_{\mathcal{H}_2}\\
    & \leq \big( \norm{\bm{Q}^{1/2}{\bm{\Phi}}(s)}_{\mathcal{H}_2} + \norm{\bm{R}^{1/2}{\bm{H}}(s)}_{\mathcal{H}_2} \big)\norm{( \bm{I} + \bm{\Xi}(s))^{-1}}_{\mathcal{H}_{\infty}}\\
    & = J^{*}\norm{( \bm{I} + \bm{\Xi}(s))^{-1}}_{\mathcal{H}_{\infty}}\\
    &\leq \frac{{J}^{*}}{1-\norm{\bm{\Xi}(s)}_{\mathcal{H}_{\infty}}}.
\end{aligned}
\end{equation}

Combining the inequalities in Eqs.~\textbf{\ref{est1}} and \textbf{\ref{est2}}, we have
\begin{equation}
\widetilde{J}^{*}(1-\norm{\bm{\Xi}(s)}_{\mathcal{H}_{\infty}})
\leq
  {J}^{*}\leq \frac{\widetilde{J}^*}{1-\norm{\bm{\Sigma}(s)}_{\mathcal{H}_{\infty}}}.
\end{equation}
This shows that, when the system is more localized, $\widetilde{J}^*\rightarrow J^*$ as $\bm{\Sigma}(s)\rightarrow 0$ and $\bm{\Xi}(s) \rightarrow 0$. Thus, when the residual terms $\bm{\Sigma}(s)$ and $\bm{\Xi}(s)$ are small in magnitude, the optimal objective value for the projected problem is close to that of the original problem, meaning that solving the projected problem provides a near-optimal solution to the original problem.

\section{Controller-Oriented Localization}
Although the controller in Eq.~\textbf{\ref{conLaw}} obtained by the disturbance-oriented localization is guaranteed to be stabilizing and near-optimal, this controller is a dynamic feedback controller, whereas the global optimal controller given by Eq.~\textbf{\ref{fullopt}} is static. As a consequence, the controller given by Eq.~\textbf{\ref{conLaw}} is a dynamical system for each driver node with the same state-space dimension as the entire network system, which makes the controller impractical for large networks. Here, we show how to convert the dynamical controller into static ones that are scalable to large networks.

To understand why the localized solutions in Eqs.~\textbf{\ref{hatPhi}} and \textbf{\ref{hatM}} lead to dynamical controllers and how they can be converted into static ones, we take the viewpoint of the controller installed at each node individually. Recall that the projected problem in Eq.~\textbf{\ref{slslqr3}} at node $j$ is obtained by projecting the original problem onto the information neighborhood $\mathcal{N}_j$. Also recall that, since the controller at node $i$ responds to disturbances at all node $j$ for which $i\in \mathcal{N}_j$, we define the set of all such nodes $j$ as the control neighborhood of $i$:
\begin{equation}
    \mathcal{C}_i = \{ 1\leq j\leq N \ | \  i\in \mathcal{N}_j\}.
\end{equation}
By using the projection matrix $\bm{f}_i^T$ defined in the main text, the controller installed at the $i$th node is $\bm{f}_i^T \widetilde{\bm{K}}(s)$, which is a linear map from the entire state space to the input of that node. The nonzero columns of $\bm{f}_i^T \widetilde{\bm{K}}(s)$ correspond to nodes belonging to $\mathcal{C}_i$. From Eq.~\textbf{\ref{conLaw}}, we have
\begin{equation}\label{Kequation}
    \bm{f}_i^T \widetilde{\bm{K}}(s) \widetilde{\bm{\Phi}}(s) = \bm{f}_i^T \widetilde{\bm{H}}(s),
\end{equation}
which is equivalent to
\begin{equation}\label{Fcondition}
    \bm{f}_i^T \widetilde{\bm{K}}(s) {\bm{N}}_{j}^T \widetilde{\bm{\phi}}_j(s) = \bm{f}_i^T  {\bm{T}}_{j}^T \widetilde{\bm{K}}_j\widetilde{\bm{\phi}}_j(s),\ \forall \ j\in \mathcal{C}_i,
\end{equation}
as verified using Eqs.~\textbf{\ref{mjred}}--\textbf{\ref{hatM}}. Suppose that the set of feedback matrices $\{\widetilde{\bm{K}}_j\}_{j=1}^{N}$ for the projected problem satisfies
\begin{equation}\label{eqcondition}
    \bm{f}_i^T  {\bm{T}}_{j}^T \widetilde{\bm{K}}_j  \bm{N}_j\bm{e}_p=  \bm{f}_i^T  {\bm{T}}_{k}^T \widetilde{\bm{K}}_k \bm{N}_k\bm{e}_p,\ \forall \, j,k\in \mathcal{C}_i, \ p\in \mathcal{N}_j\cap \mathcal{N}_k.
\end{equation}
That is, for any given node $p$, the feedback matrix from the state of node $p$ to the control input of node $i$ take the same value for all node pairs $(j,k)$ in $\mathcal{C}_i$ whose information neighborhoods $\mathcal{N}_j$ and $\mathcal{N}_k$ both contains node $p$. We denote this common matrix $ \bm{f}_i^T  {\bm{T}}_{j}^T \widetilde{\bm{K}}_j  \bm{N}_j\bm{e}_p$ in Eq.~\textbf{\ref{eqcondition}} as $\widetilde{\bm{K}}_{ip}$. It follows that
\begin{equation}\label{kicon}
  \bm{f}_i^T \widetilde{\bm{K}}(s) =  \sum_{j\in \mathcal{C}_i} \widetilde{\bm{K}}_{ij}\bm{e}_{j}^T
\end{equation}
satisfies Eq.~\textbf{\ref{Kequation}}. Thus, when the condition in Eq.~\textbf{\ref{eqcondition}} holds, the controller at node $i$ becomes a static feedback matrix.

However, since $\widetilde{\bm{K}}_j$ is determined independently for each $j\in \mathcal{C}_i$ by solving the projected Riccati equation (Eq.~\textbf{\ref{riccati2}}), the condition in Eq.~\textbf{\ref{eqcondition}} is not guaranteed to hold. To ensure that condition is satisfied by the set of feedback matrices $\{\widetilde{\bm{K}}_j\}_{j=1}^{N}$, we modify the way in which the projected problems defined by Eq.~\textbf{\ref{slslqr3}} are formulated. 
To do this, we merge $\mathcal{N}_j$ for all $j\in \mathcal{C}_i$ to form the set
\begin{equation}
    \mathcal{\widehat{N}}_i = \bigcup\limits_{j\in \mathcal{C}_i} \mathcal{N}_j,
\end{equation}
which is a superset of the information neighborhood of any $j\in \mathcal{C}_i$. Now, if we project the original problem onto a superset of the information neighborhood, the quality of the approximation of the projected problem would not be compromised. This gives an extra degree of freedom to construct the projected problems, so that the condition in Eq.~\textbf{\ref{eqcondition}} can be satisfied. Let $\widehat{\bm{N}}_i$ be the projection matrix from the entire state space to the state subspace corresponding to nodes in $\mathcal{\widehat{N}}_i$. In analogy with $\mathcal{T}_i$ and ${\bm{T}}_i$ in the main text, we introduce
\begin{equation}
     \mathcal{\widehat{T}}_i=\big\{ 1\leq k \leq r\ |\ [\bm{B}]_{jk}\neq 0 \text{ for some }  j \in \mathcal{\widehat{N}}_i  \big\}
\end{equation}
and define $\widehat{\bm{T}}_i$ as the projection matrix from the entire input space $\mathbb{R}^r$ to the input subspace $\mathbb{R}^{\widehat{L}}$ associated with $\mathcal{\widehat{T}}_i$, where $\widehat{L}=\abs{\mathcal{\widehat{T}}_i}$. We can then project the original problem onto $\mathcal{\widehat{N}}_i$ to obtain a new set of projected problems, one for each $j\in \mathcal{C}_i$:
\begin{equation} \label{slslqr4}
\begin{aligned}
&\underset{\widehat{\bm{\phi}}_j, \widehat{\bm{h}}_j \in \frac{1}{s}\mathcal{RH}_{\infty}}{\text{min}} \ 
\norm{\begin{bmatrix} \widehat{\bm{Q}}^{1/2}_j & \\ & \widehat{\bm{R}}^{1/2}_j \end{bmatrix} \begin{bmatrix}\widehat{\bm{\phi}}_j(s) \\\widehat{\bm{h}}_j(s)  \end{bmatrix}
}_{\mathcal{H}_2}^2\\
&\text{s.t.} 
\begin{array}{l}
\begin{bmatrix}
            s\bm{I}-\widehat{\bm{C}}_j & -\widehat{\bm{B}}_j
        \end{bmatrix}
       \begin{bmatrix}
            \widehat{\bm{\phi}}_j(s) \\ \widehat{\bm{h}}_j(s)
        \end{bmatrix}=\widehat{\bm{e}}_j,
\end{array}
\end{aligned}
\end{equation}
where $\widehat{\bm{C}}_i=\widehat{\bm{N}}_i \bm{C} \widehat{\bm{N}}_i^{T} $, $\widehat{\bm{B}}_i = \widehat{\bm{N}}_i \bm{B} \widehat{\bm{T}}_i^{T} $, $\widehat{\bm{Q}}_i=\widehat{\bm{N}}_i \bm{Q} \widehat{\bm{N}}_i^{T}$, $    \widehat{\bm{R}}_i=\widehat{\bm{T}}_i \bm{R} \widehat{\bm{T}}_i^{T} $, and $\widehat{\bm{e}}_j = \widehat{\bm{N}}_i \bm{e}_j$. We call this set of projected problems the \emph{controller-oriented localization} of the original problem, since each is a local version of the original problem around a driver node. The solution to Eq.~\textbf{\ref{slslqr4}} is obtained by solving the Riccati equation
\begin{equation}\label{projectriccati}
       \widehat{\bm{C}}_i^{T}\widehat{\bm{P}}_i+\widehat{\bm{P}}_i\widehat{\bm{C}}_i-\widehat{\bm{P}}_i\widehat{\bm{B}}_i\widehat{\bm{R}}_i^{-1}\widehat{\bm{B}}_i^{T}\widehat{\bm{P}}_i+\widehat{\bm{Q}}_i=\bm{0},
\end{equation}
which yields the optimal feedback $\widehat{\bm{K}}_i = -\widehat{\bm{R}}_i^{-1}\widehat{\bm{B}}_i^{T}\widehat{\bm{P}}_i$. Hence, for each $j\in \mathcal{C}_i$,
\begin{equation}\label{rjred2}
   \widehat{\bm{\phi}}_j(s)=(s\bm{I}-\widehat{\bm{C}}_i-\widehat{\bm{B}}_i\widehat{\bm{K}}_i)^{-1} \widehat{\bm{e}}_j,
\end{equation}
\begin{equation}\label{mjred2}
    \widehat{\bm{h}}_j(s)=\widehat{\bm{K}}_i\widehat{\bm{\phi}}_j(s).
\end{equation}
Compared to Eqs.~\textbf{\ref{rjred}} and \textbf{\ref{mjred}}, the optimal response in Eqs.~\textbf{\ref{rjred2}} and \textbf{\ref{mjred2}} is obtained by projecting the original problem onto a superset of the information neighborhood $\mathcal{N}_j$, and hence this solution provides at least the same accuracy as the solution in Eqs.~\textbf{\ref{rjred}} and \textbf{\ref{mjred}}. The resulting controller is static because
\begin{equation}
    \widetilde{\bm{K}}_j = \bm{T}_j\widehat{\bm{T}}_i^T \widehat{\bm{K}}_i  \widehat{\bm{N}}_i \bm{N}_j^T
\end{equation}
satisfies the condition in Eq.~\textbf{\ref{eqcondition}}. The corresponding responses in $\mathcal{N}_j$ achieved by $\widetilde{\bm{K}}_j $ are given by
\begin{equation}
    \widetilde{\bm{\phi}}_j(s) = \bm{N}_j\widehat{\bm{N}}_i^T\widehat{\bm{\phi}}_j(s),
\end{equation}
\begin{equation}
    \widetilde{\bm{h}}_j(s) = \bm{T}_j\widehat{\bm{T}}_i^T\widehat{\bm{h}}_j(s).
\end{equation}
It follows that
\begin{equation}
\begin{aligned}
 \big(\bm{f}_i^T \widehat{\bm{T}}_i^T\widehat{\bm{K}}_i \widehat{\bm{N}}_i \big) {\bm{N}}_{j}^T \widetilde{\bm{\phi}}_j(s) = \big(\bm{f}_i^T \bm{T}_j^T \bm{T}_j \widehat{\bm{T}}_i^T\widehat{\bm{K}}_i \widehat{\bm{N}}_i \big) {\bm{N}}_{j}^T \widetilde{\bm{\phi}}_j(s) = \bm{f}_i^T  \bm{T}_j^T \widetilde{\bm{K}}_j \widetilde{\bm{\phi}}_j(s),
\end{aligned}
\end{equation}
i.e., $\bm{f}_i^T \widetilde{\bm{K}}(s)=\bm{f}_i^T \widehat{\bm{T}}_i^T\widehat{\bm{K}}_i \widehat{\bm{N}}_i$ satisfies Eq.~\textbf{\ref{Fcondition}}. This implies that $\bm{f}_i^T \widehat{\bm{T}}_i^T\widehat{\bm{K}}_i \widehat{\bm{N}}_i$ is the appropriate feedback matrix for the controller at node $i$, and thus
\begin{equation}
    \bm{f}_i^T \bm{u}(t) = \bm{f}_i^T \widehat{\bm{T}}_i^T\widehat{\bm{K}}_i \widehat{\bm{N}}_i \bm{x}(t).
\end{equation}
Therefore, the design of the feedback law of each controller only requires solving the projected Riccati equation locally around that driver node and the computation can be performed in parallel for all drivers across the network. This provides a decentralized method for designing a near-optimal control strategy for the entire network.

\section{Basic Control Tasks and Linearization Methods}

In scientific and engineering applications, one is often required to actively control the
dynamics of complex networks so that they exhibit certain desirable behaviors and functionalities. Different applications require addressing different control tasks, and the most often encountered in practice are equilibrium stabilization, trajectory tracking, and command following. All these control tasks can be accomplished by proper design of feedback control laws, as described next. To proceed, we consider a general system
\begin{equation}\label{state1}
    \dot{\bm{x}} = \bm{f}(\bm{x})+\bm{B}\bm{u}
\end{equation}
in which the function $\bm{f}(\cdot)$ can be nonlinear.

\underline{\emph{Equilibrium stabilization}} refers to the control task in which the system is driven from a given initial condition to a desired equilibrium and held stably there. In power grids, for example, each power flow solution corresponds to an equilibrium of the system. A major task of power system controllers is to bring the system towards the most efficient and reliable power flow equilibrium and to maintain the system at the equilibrium in the presence of disturbances. Consider the system in Eq.~\textbf{\ref{state1}} with a desired equilibrium $(\bm{x}^*,\bm{u}^*)$, i.e.,
    $\bm{f}(\bm{x}^*)+\bm{B}\bm{u}^*=\bm{0}$.
The problem is to design a feedback law of the form $\bm{u}(t)=\bm{K}(\bm{x}(t)-\bm{x}^*)+\bm{u}^*$ that drives the system from the initial state $\bm{x}_0$ to the target equilibrium $\bm{x}^*$. In general, the map $\bm{K}(\cdot)$ can be nonlinear but is assumed to be homogeneous, i.e., $\bm{K}(\bm{0})=\bm{0}$. By defining $\Delta \bm{x} = \bm{x}-\bm{x}^*$ and $\Delta \bm{u} = \bm{u}-\bm{u}^*$, we have $\Delta \dot{\bm{x}} = \bm{f}(\bm{x}^*+\Delta \bm{x})-\bm{f}(\bm{x}^*)+\bm{B}\Delta \bm{u}$ (throughout the Article we use $\Delta \dot{\bm{x}}$ to denote the time derivative of variable $\Delta \bm{x}$ even in cases where $\bm{x}^*$ is time dependent). This reduces the original problem to the problem of designing a control $\Delta \bm{u}$ to stabilize the system at the origin. \reviseTen{To apply the linear-quadratic optimal control theory}, we replace the nonlinear term $\bm{f}(\bm{x}^*+\Delta \bm{x})-\bm{f}(\bm{x}^*)$ with $\bm{C(\bm{x},\bm{x}^*)}\Delta \bm{x}$, which is linear in $\Delta \bm{x}$. Linearization methods to obtain $\bm{C(\bm{x},\bm{x}^*)}$ are presented at the end of this section. Accordingly, the problem further reduces to $\Delta \dot{\bm{x}} = \bm{C(\bm{x},\bm{x}^*)}\Delta \bm{x}+\bm{B}\Delta \bm{u}$, which shares the same structure as the linear system in Eq.~\blue{\textbf{3}} of the main text. Thus, Algorithm 3 (in \emph{\textcolor{blue}{Materials and Methods}}) can be used to design $\bm{K}$ for this problem, which can be time independent or time varying, depending on the linearization method employed.

\underline{\emph{Trajectory tracking}} aims to drive the system to a given feasible trajectory and then force it to follow the trajectory stably. A feasible trajectory of the system is a pair of curves $(\bm{x}^*(t),\bm{u}^*(t))$ that satisfy Eq. \textbf{\ref{state1}}. For example, it can be the set of planned optimal trajectories for a group of autonomous spacecrafts in formation, or a selected common orbit for a group of periodic oscillators in a synchronous state. The task is to design a feedback control $\bm{u}(t)=\bm{K}\big(\bm{x}(t)-\bm{x}^*(t)\big)+\bm{u}^*(t)$ such that the system starting from an given initial condition converges to and then stays on the target trajectory. The design approach using Algorithm 3 is essentially the same as for equilibrium stabilization, except that $(\bm{x}^*(t),\bm{u}^*(t))$ is a time-varying target. 

\underline{\emph{Command following}} differs from the two tasks above in that a desired equilibrium or feasible trajectory is not known \textit{a priori}. The goal is to design a control law in which a subset $\bm{x}_1$ of the state vector elements will quickly follow any (possibly time-varying) control command $\bm{r}$ given in real time. Thus, $\bm{r}$ is unspecified at the design stage and the state vector can be written as $\bm{x}=[\bm{x}_1^T,\bm{x}_2^T]^T$, where $\bm{x}_2$ represents the other elements of the state vector. In order to completely track the command $\bm{r}$, there must exist an equilibrium $(\bm{x}^*,\bm{u}^*)$ such that $\bm{f}(\bm{x}^*)+\bm{B}\bm{u}^*=\bm{0}$, $\bm{x}_1^*-\bm{r}=\bm{0}$. Because $\bm{r}$ is unknown \textit{a priori}, $(\bm{x}^*,\bm{u}^*)$ is also not available in advance and therefore the controller cannot be designed against a known $(\bm{x}^*,\bm{u}^*)$ as done above. This problem is solved by augmenting the system with an internal state of the controller $\bm{z}$, representing the integral of the error in following $\bm{r}$. The augmented system reads $\dot{\bm{x}} = \bm{f}(\bm{x})+\bm{B}\bm{u}$, $\dot{\bm{z}} =\bm{x}_1-\bm{r}$. We seek to design a feedback control law in the form $\bm{u}=\bm{K}_1(\bm{x}_1-\bm{r})+\bm{K}_2\bm{x}_2+\bm{K}_3\bm{z}$, which can be seen as a high-dimension generalization of the proportional-integral control widely used in industrial applications \cite{aastrom1995pid_SI}. The closed-loop system is then given by $\dot{\bm{x}} = \bm{f}(\bm{x})+\bm{B}\bm{K}_1(\bm{x}_1-\bm{r})+\bm{B}\bm{K}_2\bm{x}_2+\bm{B}\bm{K}_3\bm{z}$, $\dot{\bm{z}} = \bm{x}_1-\bm{r}$, and the equilibrium equation is rewritten as $\bm{f}(\bm{x}^*)+\bm{B}\bm{K}_2\bm{x}_2^*+\bm{B}\bm{K}_3\bm{z}^*=\bm{0}$, $\bm{x}_1^*-\bm{r}=\bm{0}$. Defining $\Delta \bm{x}_1 = \bm{x}_1-\bm{r}$, $\Delta \bm{x}_2 = \bm{x}_2-\bm{x}_2^*$, and $\Delta \bm{z} = \bm{z}-\bm{z}^*$, we have $ \Delta\dot{\bm{x}} = \bm{f}(\bm{x})-\bm{f}(\bm{x}-\Delta \bm{x}) +\bm{B}\bm{K}_1 \Delta \bm{x}_1+\bm{B}\bm{K}_2 \Delta \bm{x}_2+\bm{B}\bm{K}_3 \Delta \bm{z}$, $\Delta \dot{\bm{z}} = \Delta \bm{x}_1$.
By linearizing $\bm{f}(\bm{x})-\bm{f}(\bm{x}-\Delta \bm{x})$ as $\bm{C}(\bm{x})\Delta \bm{x}$, the system becomes
\begin{equation}
    \begin{bmatrix}
     \Delta \dot{\bm{x}}_1 \\
     \Delta \dot{\bm{x}}_2 \\
     \Delta \dot{\bm{z}}
    \end{bmatrix}
    =
     \left[
\begin{array}{cc|c}
    \multicolumn{2}{c|}{\multirow{2}{*}{$\bm{C}(\bm{x})$}} & \multicolumn{1}{c}{\multirow{1}{*}{$\bm{0}$}} \\ 
    & & \multicolumn{1}{c}{\multirow{1}{*}{$\bm{0}$}} \\ \cline{1-3}
    \\
    \multicolumn{1}{c}{\multirow{1}{*}{$\bm{I}$}} & \multicolumn{1}{c|}{\multirow{1}{*}{$\bm{0}$}} & \multicolumn{1}{c}{\multirow{1}{*}{$\bm{0}$}}
\end{array} \right]
      \begin{bmatrix}
     \Delta {\bm{x}_1} \\
     \Delta {\bm{x}_2} \\
     \Delta {\bm{z}} 
    \end{bmatrix}
    +
      \begin{bmatrix}
     \bm{B} \\
     \bm{0} \\
     \bm{0}
    \end{bmatrix}
          \begin{bmatrix}
     \bm{K}_1 \ \ \bm{K}_2 \ \ \bm{K}_3
    \end{bmatrix}
     \begin{bmatrix}
     \Delta {\bm{x}_1} \\
     \Delta {\bm{x}_2} \\
     \Delta {\bm{z}} 
    \end{bmatrix},
\end{equation}
independently of the unknown equilibrium. The feedback law $[\bm{K}_1 \ \bm{K}_2 \ \bm{K}_3]$ can then be designed according to Algorithm 3.

All three control tasks just described rely on linearization of $\bm{f}(\bm{x}^*+\Delta \bm{x})-\bm{f}(\bm{x}^*)$ or, equivalently, of $\bm{f}(\bm{x})-\bm{f}(\bm{x}-\Delta \bm{x})$. The most basic linearization approach is the \emph{Jacobian linearization}, in which the first-order Taylor expansion around a known equilibrium or feasible trajectory is used as an approximation:
$
    \bm{f}(\bm{x}^*+\Delta \bm{x})-\bm{f}(\bm{x}^*) \approx \frac{\partial \bm{f}}{\partial \bm{x}}\bigr|_{\bm{x}^*} \Delta \bm{x}.
$
When the linearized system is stable, the approach guarantees that the original nonlinear system has a non-empty basin of attraction around $\bm{x}^*$, which can be enlarged by using high-gain feedback. Moreover, the feedback matrix does not change while the system is controlled (unless $\bm{x}^*$ varies in time), which is computationally appealing. Another common approach is the so-called \emph{extended linearization}, in which the nonlinear function is algebraically factorized as
$
     \bm{f}(\bm{x}^*+\Delta \bm{x})-\bm{f}(\bm{x}^*) = \bm{C}(\bm{x}^*,\Delta \bm{x})\Delta \bm{x}
$.
The specific form of $\bm{C}(\bm{x}^*,\Delta \bm{x})$ depends on the problem formulation and is in general not unique. An existence condition and a complete parameterization of all such factorizations can be found in \cite{liang2013analysis_SI,lin2015analytical_SI}. Due to the dependence of the linear coefficient matrix on the state, different Riccati equations are solved at different time steps to obtain a time-varying feedback law (owing to this, the method is also called the state-dependent Riccati equation approach). The appeal of this approach is that it not only guarantees local stability but also generates a system trajectory that satisfies the Hamilton–Jacobi–Bellman equation, which is a necessary condition for the optimality of the trajectory \cite{ccimen2008state_SI}. For the command following problem, which requires a linearization coefficient matrix that does not depend on the unknown equilibrium $\bm{x}^*$, a common practice is to first use the extended linearization to obtain $\bm{f}(\bm{x})=\bm{C}(\bm{x})\bm{x}$ and then approximate the nonlinear term as
$
    \bm{f}(\bm{x})-\bm{f}(\bm{x}-\Delta \bm{x}) \approx \bm{C}(\bm{x})\Delta \bm{x}.
$
Despite a certain lack of theoretical justification, this approximation has been successfully used in many practical problems \cite{cloutier2002capabilities_SI,ccimen2010systematic_SI}.

\renewcommand{\refname}{SI References}

\newpage

\begin{figure}[ht]
    \centering
      \begin{overpic}[width=0.5\textwidth,tics=5]{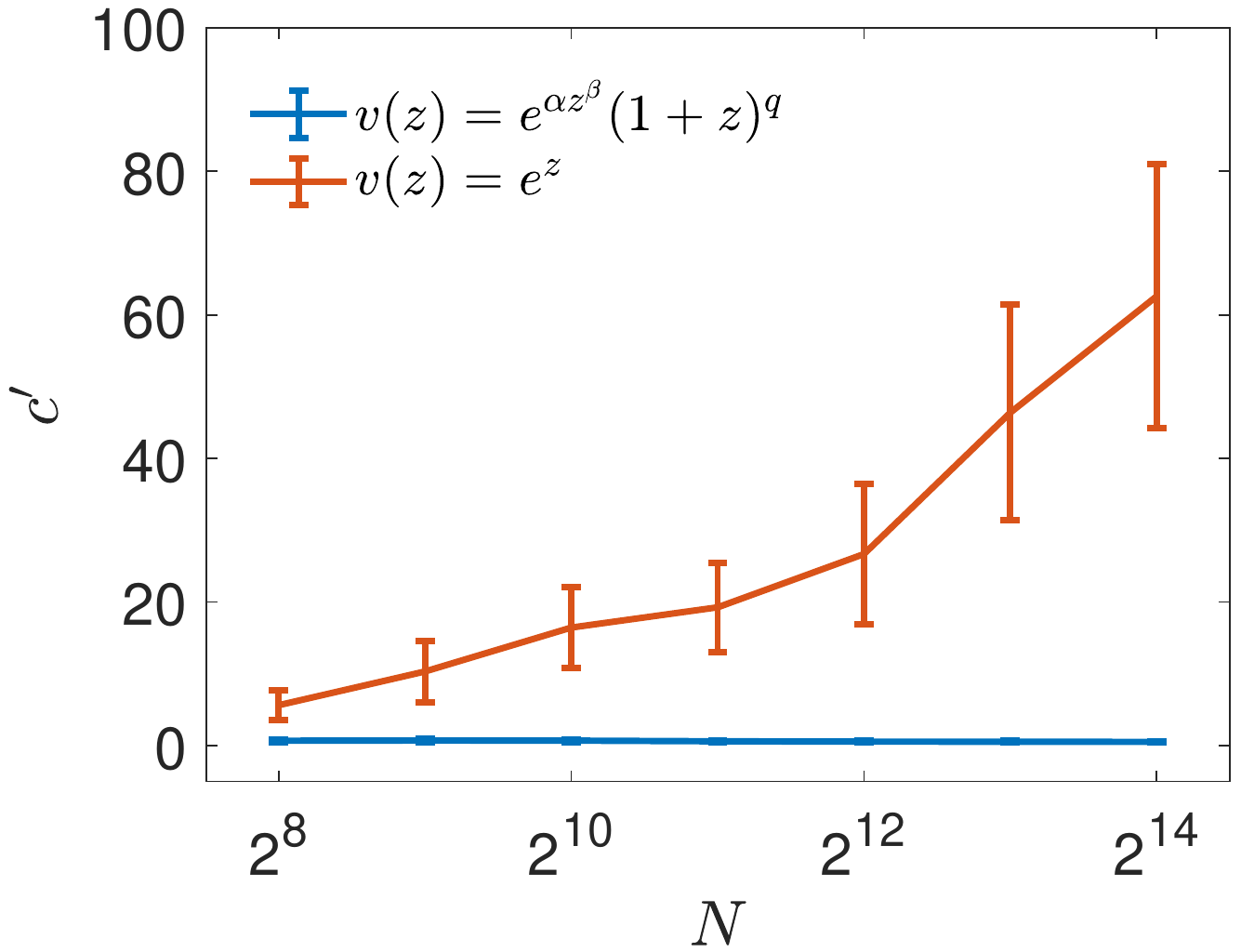}
\end{overpic}
         \caption{\reviseTen{The scaling of the off-diagonal decay constant for the inverse of matrix $\bm{C}$. Here, we consider $\bm{C} = -\bm{L}-\bm{I}$, where $\bm{L}$ is the Laplacian matrix of an ER random network with $\bar{d}=10$ and $\bm{I}$ is a identity matrix to ensure that matrix $\bm{C}$ is invertible. According to our construction of information distance, given a characteristic function $v(\cdot)$, we can identify a information distance $\rho(\cdot,\cdot)$ such that $\norm{\bm{C}_{ij}}\leq \kappa \cdot v\big(\rho(i,j)\big)^{-1}$ for all $i,j=1,2,\ldots,N$. The constant $\kappa$ can be chosen as the spectral radius $\lambda_r$ of matrix $\bm{C}$. We now consider the off-diagonal decay of $\bm{C}^{-1}$ and seek to identify a number $c'$ such that $\norm{(\bm{C}^{-1})_{ij}}\leq c' \lambda_r \cdot v\big(\rho(i,j)\big)^{-1}$ for all $i,j=1,2,\ldots,N$. When we use a characteristic function $v(\cdot)$ that satisfies the GRS condition (e.g., the one we use in this paper), the constant $c'$ can be chosen to be independent of the network size $N$ (blue curve). In contrast, if we use a characteristic function $v(\cdot)$ that violates the GRS condition (e.g., $v(z)=e^z$), the constant $c'$ may necessarily grow with $N$ (red curve). }
         }
         \label{localscale}
\end{figure}

\newpage

\renewcommand{\thetable}{S\arabic{table}}

\begin{table}[h]\centering
\begin{threeparttable}
{\small
\caption{Relative approximation error$^*$ for the smallest eigenvalue of the controllability Gramian.}
\label{tabeigWc}
\reviseTen{
\begin{tabular}{@{}l|ccc@{}}
\toprule
\multirow{2}{*}{\rotatebox[origin=c]{0}{\makecell{\ \ \ \ \ \ \ \ model  \,\,\,networks$^\dagger$ \ \ \ \ }}} & ER                  & BA                  & WS                  \\
                  & \ \ \ \ $(9.3\pm 7.7)\times 10^{-3}$ \ \ \ \ & \ \ \ \ $(2.2\pm 0.7)\times 10^{-2} $ \ \ \ \ & \ \ \ \ $(6.1\pm 5.0)\times 10^{-3}$ \ \ \ \ \\ \midrule
\multirow{2}{*}{\rotatebox[origin=c]{0}{\makecell{\ \ \ \ \ \ \ \ empirical  \,\,\,networks$^\ddagger$ \ \ \ \ }}} & power grid          & air transportation      & human brain               \\
                  & \ \ \ \ $2.13 \times 10^{-6}$   \ \ \ \         & \ \ \ \ $7.27 \times 10^{-4}$     \ \ \ \         & \ \ \ \ $1.21\times 10^{-2}$   \ \ \ \         \\ 
\bottomrule
\end{tabular}
}
 \begin{tablenotes}
      \footnotesize
      \item $^*$Computed as $\abs{\lambda_{\text{min}}(\widetilde{\bm{W}}_\text{c}^{\infty})-\lambda_{\text{min}}(\bm{W}_{\text{c}}^{\infty})}/\lambda_{\text{min}}(\bm{W}_{\text{c}}^{\infty})$, with each node chosen as a driver and assigned an information neighborhood of size $L=\lceil N/100 \rceil$.
      \item $^\dagger$Mean and standard deviation of the relative approximation error over $100$ network realizations for the same parameters as in Fig.~3.
      \item $^\ddagger$Relative approximation error for the empirical networks and dynamics used in Fig.~5D--F. 
    \end{tablenotes}
    }
  \end{threeparttable}
\end{table}

\newpage

\clearpage


\begin{thebibliography}{10}

\bibitem{newman2006structure}
M.E. Newman, A.L. Barab{\'a}si, D.E. Watts, {\em The Structure and Dynamics of Networks}
\newblock (Princeton University Press, 2006).

\bibitem{barrat2008dynamical}
A. Barrat, M. Barthelemy, A. Vespignani, {\em Dynamical Processes on Complex
  Networks}
\newblock (Cambridge University Press, 2008).

\bibitem{albert2002statistical}
R. Albert, A.L. Barab{\'a}si, Statistical mechanics of complex networks.
\newblock {\em\protect{Rev. Mod. Phys.}} \textbf{74}, 47--97
  (2002).

\bibitem{motter2013spontaneous}
A.E. Motter, S.A. Myers, M. Anghel, T. Nishikawa, Spontaneous synchrony in power-grid networks.
\newblock {\em\protect{Nat. Phys.}} \textbf{9}, 191--197
  (2013).

\bibitem{skardal2015control}
P.S. Skardal, A. Arenas, Control of coupled oscillator networks with application
  to microgrid technologies.
\newblock {\em\protect{Science Advances}} \textbf{1}, e1500339
  (2015).

\bibitem{bullo2009distributed}
F. Bullo, J. Cort{\'e}s, S. Mart{\'\i}nez, {\em Distributed Control of Robotic Networks: A
  Mathematical Approach to Motion Coordination Algorithms}
\newblock (Princeton University Press, 2009).


\bibitem{nagurney2006supply}
A. Nagurney, {\em Supply Chain Network Economics: Dynamics of Prices, Flows and Profits
}
\newblock (Edward Elgar Publishing, 2006).


\bibitem{feinberg2019foundations}
M. Feinberg, {\em Foundations of Chemical Reaction Network Theory}
\newblock (Springer, 2019).

\bibitem{wuchty2014controllability}
S. Wuchty, Controllability in protein interaction networks.
\newblock {\em\protect{Proc. Natl. Acad. Sci. U.S.A.}} \textbf{111}, 7156--7160 (2014).


\bibitem{lessard2005should}
R.B. Lessard, S.J. Martell, C.J. Walters, T.E. Essington, J.F. Kitchell, Should ecosystem management involve active control of species abundances?
\newblock {\em\protect{Ecology and Society}} \textbf{10}(2), 1 (2005).

\bibitem{sahasrabudhe2011rescuing}
S. Sahasrabudhe, A.E. Motter, Rescuing ecosystems from extinction cascades through compensatory perturbations.
\newblock {\em\protect{Nat. Commun.}} \textbf{2}, 170
  (2011).

\bibitem{galbiati2013power}
M. Galbiati, D. Delpini, S. Battiston, The power to control.
\newblock {\em\protect{Nat. Phys.}} \textbf{9}, 126--128 (2013).


\bibitem{cornelius2013realistic}
S.P. Cornelius, W.L. Kath, A.E. Motter, Realistic control of network dynamics.
\newblock {\em\protect{Nat. Commun.}} \textbf{4}, 1942
  (2013).

\bibitem{liu2016control}
Y.Y. Liu, A.L. Barab{\'a}si, Control principles of complex systems.
\newblock {\em\protect{Rev. Mod. Phys.}} \textbf{88},
  035006 (2016).





\bibitem{scholl2016control}
\reviseSix{
E. Sch{\"o}ll, S.H.L. Klapp, P. H{\"o}vel, {\em Control of Self-Organizing Nonlinear Systems}
\newblock (Springer, 2016).}


\bibitem{motter2015networkcontrology}
A.E. Motter, Networkcontrology.
\newblock {\em\protect{Chaos}} \textbf{25}, 097621 (2015).



  
  

  

\bibitem{lin1974structural}
C.T. Lin, Structural controllability.
\newblock {\em\protect{IEEE Trans. Automat. Contr.}}
  \textbf{19}, 201--208 (1974).
  
  
 
\bibitem{liu2011controllability}
Y.Y. Liu, J.J. Slotine, A.L. Barab{\'a}si, Controllability of complex networks.
\newblock {\em\protect{Nature}} \textbf{473}, 167--173 (2011).



  
  
\bibitem{zanudo2017structure}
J.G.T. Za{\~n}udo, G. Yang, R. Albert, Structure-based control of complex networks with nonlinear dynamics. \newblock {\em\protect{Proceedings of the National Academy of Sciences}} \textbf{114}, 7234-7239 (2017).

\bibitem{menara2018structural}
\reviseSix{
T. Menara, D.S. Bassett, F. Pasqualetti, Structural controllability of symmetric networks.
\newblock {\em\protect{IEEE Trans. Automat. Contr.}}
  \textbf{64}, 3740--3747 (2018).}



\bibitem{montanari2021functional}
A.N. Montanari, C. Duan, L.A. Aguirre, A.E. Motter, Functional observability and target state estimation in large-scale networks. \newblock {\em\protect{Proceedings of the National Academy of Sciences}}  \textbf{119}, e2113750119 (2022).
 
 

\bibitem{kalman1960general}
R.E. Kalman, ``On the general theory of control systems'' in {\em Proc. First
  International Conference on Automatic Control}
\newblock (Moscow, USSR, 1960), pp. 481--492.

\bibitem{hautus1970stabilization}
M. Hautus, Stabilization controllability and observability of linear autonomous
  systems.
\newblock {\em\protect{Indagationes Mathematicae}} \textbf{73}, 448--455
  (1970).

\bibitem{yan2012controlling}
G. Yan, J. Ren, Y.C. Lai, C.H. Lai, B. Li, Controlling complex networks: How much
  energy is needed?
\newblock {\em\protect{Physical Review Letter}} \textbf{108},
  218703 (2012).

\bibitem{sun2013controllability}
J. Sun, A.E. Motter, Controllability transition and nonlocality in network
  control.
\newblock {\em\protect{Physical Review Letter}} \textbf{110},
  208701 (2013).

\bibitem{yan2015spectrum}
G. Yan and G. Tsekenis and B. Barzel and J.-J. Slotine and Y.-Y. Liu and A. Barab{\'a}si, Spectrum of controlling and observing complex networks.
\newblock {\em\protect{Nat. Phys.}} \textbf{11}, 779--786
  (2015).
  
  
\bibitem{pasqualetti2014controllability}
F. Pasqualetti, S. Zampieri, F. Bullo, Controllability metrics, limitations and
  algorithms for complex networks.
\newblock {\em\protect{IEEE Trans. Control. Netw. Syst.}} \textbf{1}, 40--52 (2014).

\bibitem{gao2014target}
J. Gao, Y.Y. Liu, R.M. D'Souza, A.L. Barab{\'a}si, Target control of complex networks.
\newblock {\em\protect{Nat. Commun.}} \textbf{5}, 5415
  (2014).



\bibitem{klickstein2017energy}
I. Klickstein, A. Shirin, F. Sorrentino, Energy scaling of targeted optimal
  control of complex networks.
\newblock {\em\protect{Nature Communications}} \textbf{8}, 15145
  (2017).

\bibitem{li2018enabling}
G. Li, L. Deng, G. Xiao, P. Tang, C. Wen, W. Hu, J. Pei, L. Shi, H.E. Stanley, Enabling controlling complex networks with local topological
  information.
\newblock {\em\protect{Scientific Reports}} \textbf{8}, 4593
  (2018).
  

\bibitem{sanchez2021nonlinear}
\reviseSix{E.N Sanchez, C.J. Vega, O.J. Suarez, G. Chen, {\em Nonlinear Pinning Control of Complex Dynamical Networks: Analysis and Applications}
\newblock (CRC Press, 2021).
}  
  
\bibitem{girvan2002community}
M. Girvan, M.E. Newman, Community structure in social and biological networks. \newblock {\em\protect{Proceedings of the National Academy of Sciences}} \textbf{99}, 7821--7826 (2002).

\bibitem{watts1998collective}
D.J. Watts, S.H. Strogatz, Collective dynamics of `small-world' networks.
\newblock {\em\protect{Nature}} \textbf{393}, 440--442 (1998).

\bibitem{controltheory}
G.E. Dullerud, F. Paganini, {\em A Course in Robust Control Theory: a Convex Approach}
\newblock (Springer Science \& Business Media, 2013).

\bibitem{grochenig2006symmetry}
K. Gr{\"o}chenig, M. Leinert, Symmetry and inverse-closedness of matrix algebras
  and functional calculus for infinite matrices.
\newblock {\em\protect{Trans. of the American Mathematical
  Society}} \textbf{358}, 2695--2711 (2006).

\bibitem{motee2008decentralized}
N. Motee, A. Jadbabaie, B. Bamieh, ``On decentralized optimal control and
  information structures'' in {\em 2008 American Control Conference}
\newblock (IEEE, 2008), 4985--4990.

\bibitem{curtain2011riccati}
R. Curtain, Riccati equations on noncommutative Banach algebras.
\newblock {\em\protect{SIAM Journal on Control and Optimization}}
  \textbf{49}, 2542--2557 (2011).

\bibitem{dijkstra1959note}
E.W. Dijkstra, A note on two problems in connexion with graphs.
\newblock {\em\protect{Numerische Mathematik}} \textbf{1},
  269--271 (1959).

\bibitem{felner2011position}
A. Felner, ``Position paper: Dijkstra's algorithm versus uniform cost search or a
  case against dijkstra's algorithm'' in {\em Proceedings of the Fourth Annual Symposium on Combinatorial Search (Association for the Advancement of Artificial Intelligence, Palo Alto, CA, 2011),}
\newblock (2011), pp. 47--51.

\bibitem{konect}
J. Kunegis, ``KONECT: The Koblenz network collection'' in {\em WWW’13: Proceedings of the 22nd International Conference on World Wide Web} (International WWW Conference Committee, Geneva, Switzerland, 2013), pp. 1343--1350.

\bibitem{erdHos1960evolution}
P. Erd{\H{o}}s, A. R{\'e}nyi, On the evolution of random graphs.
\newblock {\em\protect{Publ. Math. Inst. Hung. Acad. Sci}}
  \textbf{5}, 17--60 (1960).

\bibitem{barabasi1999emergence}
A.L. Barab{\'a}si, R. Albert, Emergence of scaling in random networks.
\newblock {\em\protect{Science}} \textbf{286}, 509--512 (1999).


\bibitem{morse1971output}
A.S. Morse, Output controllability and system synthesis.
\newblock {\em\protect{SIAM Journal on Control}} \textbf{9}, 143--148 (1971).


\bibitem{zhou1996robust}
K. Zhou, J.C. Doyle, K. Glover, {\em Robust and Optimal Control}
\newblock (Pearson, 1995).


\bibitem{whalen2015observability}
\reviseSix{
A.J. Whalen, S.N. Brennan, T.D. Sauer, S.J. Schiff, Observability and controllability of nonlinear networks: The role of symmetry.
\newblock {\em\protect{Physical Review X.}}
  \textbf{5}, 011005 (2015).}
  

\bibitem{summers2015submodularity}
T.H. Summers, F.L. Cortesi, J. Lygeros, On submodularity and controllability in
  complex dynamical networks.
\newblock {\em\protect{IEEE Trans. Control. Netw. Syst.}} \textbf{3}, 91--101 (2015).

\bibitem{wang2016system}
Y.S. Wang, N. Matni, J.C. Doyle, A system-level approach to controller synthesis.
\newblock {\em\protect{IEEE Trans. Automat. Contr.}}
  \textbf{64}, 4079--4093 (2019).

\bibitem{anderson2019system}
J. Anderson, J.C. Doyle, S.H. Low, N. Matni, System level synthesis.
\newblock {\em\protect{Annual Reviews in Control}} \textbf{47},
  364--393 (2019).


\bibitem{arenas2008synchronization}
\reviseSix{
A. Arenas, A. D{\'\i}az-Guilera, J. Kurths, Y. Moreno, C. Zhou, Synchronization in complex networks.
\newblock {\em\protect{Physics Reports}} \textbf{469},
  93--153 (2008).}

  
\bibitem{dorfler2013synchronization}
F. D{\"o}rfler and M. Chertkov and F. Bullo, Synchronization in complex oscillator networks and smart grids.
\newblock {\em\protect{Proceedings of the National Academy of Sciences}} \textbf{110},
  2005--2010 (2013).  


\bibitem{machowski2011power}
J. Machowski, J. Bialek, J. Bumby, {\em Power System Dynamics: Stability and
  Control}
\newblock (John Wiley \& Sons, 2011).


\bibitem{colizza2006role}
V. Colizza, A. Barrat, M. Barth{\'e}lemy, A. Vespignani, The role of the airline
  transportation network in the prediction and predictability of global
  epidemics.
\newblock {\em\protect{Proc. Natl. Acad. Sci. U.S.A.}} \textbf{103}, 2015--2020 (2006).


\bibitem{sanchez2018design}
L.M. Sanchez-Rodriguez, et~al., Design of optimal nonlinear network controllers
  for Alzheimer's disease.
\newblock {\em\protect{PLoS Computational Biology}} \textbf{14},
  e1006136 (2018).
  
\bibitem{scheid2021time}
\reviseSix{
B.H. Scheid, et al., Time-evolving controllability of effective connectivity networks during seizure progression.
\newblock {\em\protect{Proc. Natl. Acad. Sci. U.S.A.}} \textbf{118},
  e2006436118 (2021).}

\bibitem{taylor2015optimal}
P.N. Taylor, et~al., Optimal control based seizure abatement using patient
  derived connectivity.
\newblock {\em\protect{Frontiers in Neuroscience}} \textbf{9}, 202
  (2015).

  
\bibitem{schiff2010towards}
S.J. Schiff, Towards model-based control of Parkinson's disease.
\newblock {\em\protect{Philosophical Transactions of the Royal Society A: Mathematical, Physical and Engineering Sciences}} \textbf{368},
  2269--2308 (2010).
  
  
\bibitem{githubrepo}
Codes for the localized control of network systems (2021). Available at \href{https://github.com/cduan2020/LocalizedControl}{https://github.com/cduan2020/LocalizedControl}.


\end{thebibliography}

\begin{thebibliography}{99}

\setcounter{enumiv}{0} 



\bibitem{curtain2011riccati_SI}
R. Curtain, Riccati equations on noncommutative {B}anach algebras.
\newblock {\em\protect{SIAM J. Control Optim.}} \textbf{49},
  2542--2557 (2011).

\bibitem{zhou1996robust_SI}
K. Zhou, J.C. Doyle, K. Glover, {\em Robust and Optimal Control}
\newblock (Prentice Hall, New Jersey, 1996).

\bibitem{bartolucci2020emerging_SI}
S. Bartolucci, F. Caravelli, F. Caccioli, P. Vivo, Emerging locality of network influence.
\newblock arXiv:2009.06307 (2020).

\bibitem{favero2021locality_SI}
A. Favero, F. Cagnetta, M. Wyart, Locality defeats the curse of dimensionality in convolutional teacher-student scenarios in
\newblock {\em\protect{Advances in Neural Information Processing Systems Conf.}} (2021).


\bibitem{konect_url}
J. Kunegis, {KONECT}: The {Koblenz} network collection in {\em Proceedings of
  the 22nd International Conference on World Wide Web}.
\newblock (2013). Available at \href{http://konect.cc/}{http://konect.cc}.

\bibitem{FERC}
{Federal Energy Regulatory Commission Form 715} (2017). Data obtained under a non-disclosure agreement by
  following the procedure described at
  \href{https://www.ferc.gov/legal/ceii-foia/ceii.asp}{https://www.ferc.gov/legal/ceii-foia/ceii.asp}.

\bibitem{OpenFlights}
Openflights (2014). Available at
  \href{https://openflights.org/data.html}{https://openflights.org/data.html}.

\bibitem{Crossley2013_SI}
N.A. Crossley, et~al., Cognitive relevance of the community structure of the
  human brain functional coactivation network.
\newblock {\em\protect{Proc. Natl. Acad. Sci. U.S.A.}}
  \textbf{110}, 11583--11588 (2013). Available from the Brain
  Connectivity Toolbox website at
  \href{https://sites.google.com/site/bctnet/datasets-and-demos}{https://sites.google.com/site/bctnet/datasets-and-demos}.
  
\bibitem{sanchez2018design_SI}
L.M. Sanchez-Rodriguez, et~al., Design of optimal nonlinear network controllers
  for Alzheimer's disease.
\newblock {\em\protect{PLoS Computational Biology}} \textbf{14},
  e1006136 (2018).

\bibitem{zimmerman2010matpower_SI}
R.D. Zimmerman, C.E. Murillo-S{\'a}nchez, R.J. Thomas, {MATPOWER}: Steady-state
  operations, planning, and analysis tools for power systems research and
  education.
\newblock {\em\protect{IEEE T. Power Syst.}} \textbf{26}, 12--19
  (2010).

\bibitem{machowski2011power_SI}
J. Machowski, J. Bialek, J. Bumby, {\em Power System Dynamics: Stability and
  Control}
\newblock (John Wiley \& Sons, 2011).

\bibitem{1994power_SI}
P. Kundur, {\em Power System Stability and Control}
\newblock (McGraw-Hill, 1994).

\bibitem{gao2014target_SI}
J. Gao, Y.Y. Liu, R.M. D'Souza, A.L. Barab{\'a}si, Target control of complex networks.
\newblock {\em\protect{Nat. Commun.}} \textbf{5}, 5415 (2014).

\bibitem{zhao2016non_SI}
B. Zhao, Y. Guan, L. Wang, Non-fragility and partial controllability of
  multi-agent systems.
\newblock arXiv:1607.07753 (2016).

\bibitem{horn2012matrix_SI}
R.A. Horn, C.R. Johnson, {\em Matrix Analysis}
\newblock (Cambridge University Press, 2012).

\bibitem{bartels1972solution_SI}
R.H. Bartels, G.W. Stewart, Solution of the matrix equation {$AX+XB=C$}.
\newblock {\em\protect{Communications of the ACM}} \textbf{15},
  820--826 (1972).

\bibitem{wang2016system_SI}
Y.S. Wang, N. Matni, J.C. Doyle, A system-level approach to controller synthesis.
\newblock {\em\protect{IEEE T. Automat. Contr.}} \textbf{64},
  4079--4093 (2019).

\bibitem{dullerud2013course_SI}
G.E. Dullerud, F. Paganini, {\em A Course in Robust Control Theory: A Convex Approach}
\newblock (Springer Science \& Business Media, 2013).

\bibitem{anderson2019system_SI}
J. Anderson, J.C. Doyle, S.H. Low, N Matni, System level synthesis.
\emph{Annual Reviews in Control} \textbf{47},
  364--393 (2019).

\bibitem{zhu2015stability_SI}
J. Zhu, J. Chen, Stability of systems with time-varying delays: An
  $\mathcal{L}_1$ small-gain perspective.
\newblock {\em\protect{Automatica}} \textbf{52}, 260--265 (2015).


\bibitem{aastrom1995pid_SI}
K.J. {\AA}str{\"o}m, T. H{\"a}gglund, {\em PID Controllers: Theory, Design, and
  Tuning}
\newblock (Instrument society of America Research Triangle Park, 1995).

\bibitem{liang2013analysis_SI}
Y.W. Liang, L.G. Lin, Analysis of SDC matrices for successfully implementing the
 SDRE scheme.
\newblock {\em\protect{Automatica}} \textbf{49}, 3120--3124
  (2013).

\bibitem{lin2015analytical_SI}
L.G. Lin, J. Vandewalle, Y.W. Liang, Analytical representation of the
  state-dependent coefficients in the SDRE/SDDRE scheme for multivariable
  systems.
\newblock {\em\protect{Automatica}} \textbf{59}, 106--111 (2015).

\bibitem{ccimen2008state_SI}
T. {\c{C}}imen, State-dependent Riccati equation (SDRE) control: A survey.
\newblock {\em\protect{IFAC Proceedings}} \textbf{41},
  3761--3775 (2008).

\bibitem{cloutier2002capabilities_SI}
J.R. Cloutier, D.T. Stansbery, ``The capabilities and art of state-dependent Riccati equation-based design'' in {\em Proc. American Control
  Conference}
\newblock (IEEE, 2002), pp. 86--91.

\bibitem{ccimen2010systematic_SI}
T. {\c{C}}imen, Systematic and effective design of nonlinear feedback
  controllers via the state-dependent Riccati equation (SDRE) method.
\newblock {\em\protect{Annual Reviews in Control}} \textbf{34},
  32--51 (2010).



\end{thebibliography}
\end{document}